\documentclass[a4paper,11pt]{article}
\pdfoutput=1 

\usepackage{amssymb}
\usepackage{dsfont}
\usepackage[T1]{fontenc} 
\usepackage{physics}
\usepackage{comment}
\usepackage{amsmath}
\usepackage{graphicx}
\usepackage{array, makecell}
\usepackage{float}
\setcellgapes{3pt}
\usepackage{xcolor}
\usepackage{soul}
\usepackage[numbers,sort&compress]{natbib}

\definecolor{azure}{rgb}{0.0, 0.5, 1.0}
\definecolor{darkblue}{rgb}{0.15,0.35,0.7}
\definecolor{reddish}{rgb}{0.65, 0.2, 0.2}
\definecolor{brandeisblue}{rgb}{0.0, 0.44, 1.0}
\definecolor{ceruleanblue}{rgb}{0.16, 0.32, 0.75}
\definecolor{indigo(dye)}{rgb}{0.0, 0.25, 0.42}

\usepackage[linktocpage=true]{hyperref}
\hypersetup{
colorlinks=true,
citecolor=ceruleanblue,
linkcolor=ceruleanblue,
urlcolor=ceruleanblue,
pdfauthor={},
pdftitle={},
pdfsubject={}
}

\usepackage{cleveref}
\usepackage{bm}
\crefname{lem}{lemma}{lemmas}
\crefname{thm}{theorem}{theorems}
\crefname{cor}{corollary}{corollaries}
\crefname{rem}{remark}{remarks}
\crefname{prop}{proposition}{propositions}

\makeatletter
\renewcommand\section{\@startsection {section}{1}{\z@}%
                               {-3.5ex \@plus -1ex \@minus -.2ex}
                               {2.3ex \@plus.2ex}%
                               {\normalfont\large\bfseries}}
\renewcommand\subsection{\@startsection{subsection}{2}{\z@}%
                                 {-3.25ex\@plus -1ex \@minus -.2ex}%
                                 {1.5ex \@plus .2ex}%
                                 {\normalfont\bfseries}}
\makeatother

\let\non\nonumber

\newfont{\goth}{ygoth.tfm scaled 1200}                   

\numberwithin{equation}{section}

\usepackage{graphicx,tikz}
\usepackage[framemethod=TikZ]{mdframed} 
\usepackage{tikz-cd} 
\usetikzlibrary{arrows,snakes,shapes.arrows,decorations.markings}
     \tikzset{>=triangle 90}
     \tikzstyle{bbc}=[draw,circle,fill=black,scale=.75]
     \tikzstyle{rc}=[circle,fill=red,scale=.6]
     \tikzstyle{wc}=[draw,circle,scale=.75]

\newcommand{\be}{\begin{equation}}
\newcommand{\ee}{\end{equation}}
\newcommand{\bee}{\begin{equation} \begin{aligned}}
\newcommand{\eee}{\end{aligned} \end{equation}}

\newcommand{\CA}{\mathcal{A}}

\newcommand{\CDS}{\mathcal{DS}}

\newcommand{\CH}{\mathcal{H}}
\newcommand{\CL}{\mathcal{L}}

\newcommand\doubleC{\mathbb{C}}

\newcommand\doubleZ{\mathbb{Z}}

\newcommand\scriptC{\mathcal{C}}

\newcommand\scriptH{\mathcal{H}}

\newcommand\scriptK{\mathcal{K}}
\newcommand\scriptL{\mathcal{L}}
\newcommand\scriptM{\mathcal{M}}
\newcommand\scriptN{\mathcal{N}}

\newcommand\scriptT{\mathcal{T}}

\newcommand{\VEC}{\operatorname{Vec}}

\newcommand{\dsi}{\mathds{1}}
\newcommand{\IZ}{\mathbb{Z}}
\newcommand{\otau}{\overline{\tau}}
\newcommand{\ochi}{\overline{\chi}}

\newcommand{\oQ}{\overline{Q}}
\newcommand{\oq}{\overline{q}}
\newcommand{\oL}{\overline{L}}
\newcommand{\oj}{\overline{j}}
\newcommand{\oV}{\overline{V}}
\newcommand{\oT}{\overline{T}}
\newcommand{\oeta}{\overline{\eta}}

\newcommand{\llq}{\mathcal{L}_Q}
\newcommand{\llqb}{\mathcal{L}_{\overline{Q}}}
\newcommand{\ii}{\mathsf{i}}
\newcommand{\opartial}{\overline{\partial}}
\newcommand{\oh}{\overline{h}}

\newcommand\Rep{\operatorname{Rep}}
\newcommand\Hom{\operatorname{Hom}}

\newcommand{\asym}{\sigma_\text{sym}}
\newcommand{\asymi}{\sigma^{-1}_\text{sym}}
\newcommand{\splv}[3]{V_{\CL_{#3}}^{\CL_{#1},\CL_{#2}}}
\newcommand{\figref}[1]{Figure \ref{#1}}
\newcommand{\tabref}[1]{Tab.\,\ref{#1}}

\newcommand\thetatilde{\widetilde{\theta}}
\newcommand\phitilde{\widetilde{\phi}}

\begin{document} 

\begin{titlepage}
\begin{center}

\hfill         \phantom{xxx}  

\vskip 2 cm {\Large \bf On Triality Defects in 2d CFT} 

\vskip 1.25 cm {\bf Da-Chuan Lu and Zhengdi Sun}\non\\

\vskip 0.2 cm
 {\it Department of Physics, University of California, San Diego, CA 92093, USA}
\end{center}
\vskip 1.5 cm

\begin{abstract}
\noindent We consider the triality fusion category discovered in the $c = 1$  Kosterlitz-Thouless theory \cite{Thorngren:2021yso}. We analyze this fusion category using the tools from the group theoretical fusion category and compute the simple lines, fusion rules and $F$-symbols. We then studied the physical implication of this fusion category including deriving the spin selection rule, computing the asymptotic density of states of irreducible representations of the fusion category symmetries, and analyzing its anomaly and constraints under the renormalization group flow. There is another set of $F$-symbols for the fusion categories with the same fusion rule known in the literature \cite{teo2015theory}. We find these two solutions are different as they lead to different spin selection rules. This gives a complete list of the fusion categories with the same fusion rule by the classification result in \cite{jordan2009classification}.

\baselineskip=18pt

\end{abstract}
\end{titlepage}

\tableofcontents

\flushbottom

\section{Introduction}
Global symmetry is an important guideline for constructing and analyzing quantum field theories. In modern language, the global symmetries, whether continuous or discrete, can be represented as invertible topological operators supported on a codimensions-1 (codim-1) surface. Any correlation functions with the topological surface operator insertion are invariant under the small deformation of the codim-$1$ surface where the topological operator is supported. As the result, the topological operator commute with the stress-energy tensor. For a continuous symmetry given by a conserved current $j_\mu(x)$, the corresponding topological surface operator is constructed as $e^{i\alpha \oint_{\Sigma_{d-1}}d^{d-1}x \, n^\mu(x) j_\mu(x)}$ and the invariance under the small deformation then follows from the conservation equation $\partial_\mu j^\mu(x) = 0$.

Several generalizations have been made from this point of view, which leads to the concept of generalized global symmetries \cite{Gaiotto:2014kfa}. For instance, the support of the invertible topological operator can be generalized to codim-$p$ surfaces for $p > 1$, and the corresponding symmetries are called $(p-1)$-form symmetries \cite{Gaiotto:2014kfa,Kapustin:2014gua}. The standard $0$-form symmetries can interact non-trivially with the $1$-form symmetries, and this leads to the structure of $2$-group symmetries \cite{Cordova:2018cvg,Cordova:2020tij,Cordova:2022ruw}. Another direction of generalization is to study the topological operators which are not invertible. For instance, the non-invertible $0$-form symmetries in 2-dimensional field theory can be described by the fusion categories, and the topological operators correspond to the objects in the fusion categories. Therefore, these non-invertible symmetries are also called categorical symmetries. There is a lot of progress in the study of non-invertible symmetries in dimension $d\geq 3$ recently \cite{Arias-Tamargo:2022nlf,Hayashi:2022fkw,Choi:2022jqy,Cordova:2022ieu,Bashmakov:2022jtl,Damia:2022rxw,Damia:2022bcd,Kaidi:2022uux}. Of course, the categorical $0$-form symmetries can interact non-trivially with the categorical higher form symmetries, this leads to the concept of higher categorical symmetries \cite{Bhardwaj:2022yxj}.

For a more detailed review and a more complete list of references on the development of the generalized global symmetries and their applications from the high energy physics perspective and the condensed matter physics perspective, we refer the readers to the two reviews \cite{Cordova:2022ruw,McGreevy:2022oyu}.

There are two classes of non-invertible symmetries \cite{Kaidi:2022uux, Kaidi:2022cpf}. Consider a field theory $\scriptT$ with non-invertible symmetries $\scriptC$, if there exists some topological manipulation $\phi$ such as finite gauging on the theory $\scriptT$ such that the theory $\phi(\scriptT)$ contains only invertible symmetries $\phi(\scriptC)$ and the constraints on the RG flows of the theory $\scriptT$ from the non-invertible symmetries $\scriptC$ can be completely determined from the constraints on the RG flows from the invertible symmetries $\phi(\scriptC)$ of the theory $\phi(\scriptT)$, we refer the non-invertible symmetries $\scriptC$ as non-intrinsic \cite{scitalks_22060007}. Otherwise, the non-invertible symmetries are called intrinsic. A simple example of non-intrinsic invertible symmetries is the $\Rep(S^3)$ symmetries in the 2-dimensional field theory, acquired from the topological manipulation of gauging the $S^3$ global symmetries. (For example, see \cite{Bhardwaj:2017xup}.)

The study of codim-$1$ non-invertible topological operators (also known as topological defect lines, or TDLs) in 2-dimensional CFT has a long history \cite{Cardy:1986gw,Oshikawa:1996dj,Oshikawa:1996ww,Petkova:2000ip,Fuchs:2002cm,Fuchs:2003id,Fuchs:2004dz,Fuchs:2004xi, Fjelstad:2005ua}, focusing on their connection to boundary CFT, twisted partition function on various 2-manifold, orbifolds and symmetry TFT. Recent studies not only focus on searching for interesting non-invertible TDLs \cite{Burbano:2021loy, Thorngren:2019iar, Thorngren:2021yso, Chang:2020imq, Huang:2021nvb, Chang:2022hud}, but also extending the applications of the TDLs, including their lattice construction \cite{Aasen:2016dop,Aasen:2020jwb}, constraints for 2d modular bootstrap \cite{Lin:2019hks, Lin:2021udi, Pal:2020wwd, Collier:2021ngi, Kaidi:2021ent, Lanzetta:2022lze}, constraints on the RG flow \cite{Thorngren:2019iar,Chang:2018iay, Kikuchi:2021qxz,Kikuchi:2022gfi}, gauging non-invertible symmetries \cite{Bhardwaj:2017xup,barter2021computing} etc.

Perhaps the most well-studied non-invertible TDL is the duality line $N$ under the $\doubleZ_2$ gauging. The duality line $N$ and two lines $\dsi,\eta$ in the $\doubleZ_2$ symmetries form the Ising fusion category with the fusion rule
\begin{equation}\label{eq:duality_fusion_rule}
    \eta N = N\eta = N, \quad N^2 = I + \eta.
\end{equation}
The Ising fusion category has been extensively studied in the literature \cite{Burbano:2021loy,Gaiotto:2020iye, Lin:2019hks,Aasen:2016dop,Aasen:2020jwb,Oshikawa:1996dj,Oshikawa:1996ww,Thorngren:2019iar,Thorngren:2021yso}. The existence of the duality line $N$ implies the conformal field theory is invariant under the $\doubleZ_2$ gauging. One possible generalization is of course to consider the duality line $N$ under $A$-gauging where $A$ is a finite Abelian group. The corresponding fusion category is known as the Tambara-Yamagami Fusion category $\scriptT(A,\chi,\epsilon)$ \cite{tambara1998tensor,tambara2000representations} where $\chi$ is a symmetric bicharacter of the group $A$ and $\epsilon = \pm 1$ is the Frobenius-Schur indicator. The $\scriptT(A,\chi,\epsilon)$ with the same $A$ but different $\chi$ and $\epsilon$ satisfies the same fusion rules, yet they are different fusion categories distinguished by the $F$-symbols (also known as the crossing kernels $\scriptK$ in \cite{Chang:2018iay}), which measures the difference between two different ways of resolving the crossing of two TDLs. The $F$-symbols reduce to the familiar group anomaly measured by $H^3(G,U(1))$ when considering only invertible TDLs. For instance, there are two types of Ising fusion categories with different FS indicators $\epsilon = \pm 1$ and they can also be distinguished from the so-called spin selection rules \cite{Chang:2018iay}. The studies of the Tambara-Yamagami fusion category symmetries in the physics literature include \cite{Burbano:2021loy,Lin:2019hks,Chang:2018iay,Thorngren:2019iar,Thorngren:2021yso}.

Another generalization of the Ising fusion category is the triality fusion category and recently is studied in \cite{Thorngren:2019iar,Thorngren:2021yso}. The simple TDLs contain symmetry operators which generate $\doubleZ_2\times \doubleZ_2$ global symmetries, as well as a triality line $\llq$ and its orientation reversal $\llqb$, satisfying the fusion rule \footnote{In general, the invertible symmetries in a triality fusion category does not have to be $\doubleZ_2\times \doubleZ_2$. For simplicity, however, the notion of the triality fusion category would specifically mean the case where the invertible symmetries are $\doubleZ_2\times \doubleZ_2$.} generalizing \eqref{eq:duality_fusion_rule}
\begin{equation}\label{eq:triality_fusion_rule}
    \llq \times \llqb = \sum_{g\in \doubleZ_2\times \doubleZ_2} g, \quad \llq \times \llq = 2\llqb, \quad g\times \llq = \llq \times g = \llq, \quad g \in \doubleZ_2\times\doubleZ_2.
\end{equation}
The triality fusion category is obtained by gauging the $\doubleZ_2$ global symmetry in $A_4$ global symmetry of the $SU(2)_1$ theory. In general, consider a 2-dimensional theory $\scriptT$ with the global symmetry $G$ and anomaly $\omega \in H^3(G,U(1))$. One can gauge a non-anomalous subgroup $H \subset G$ and additional data which is the discrete torsion $\psi \in H^2(H,U(1))$ needs to be specified in this gauging process. The fusion category that describes the categorical symmetries of the gauged theory $\scriptT/H$ is called the group theoretical fusion category, and it is denoted by $\scriptC(G,\omega,H,\psi)$. In this language, the triality fusion category in the KT theory is $\scriptC(A_4,1,\doubleZ_2,1)$.

In this paper, we study this triality fusion category using the tools from the group theoretical fusion category. We compute the spectrum of simple TDLs, their fusion rules, and the $F$-symbols by using the description of the group theoretical fusion category in terms of bimodules \cite{ostrik2002module}.

We then study the physical implication of the triality fusion category. We derive the spin selection rules from the $F$-symbols we acquire. We also derive the Cardy formula for densities of states using the result in \cite{Pal:2020wwd}. Since these triality fusion categories are group-theoretical, their constraints on the RG flow can be determined by the group $G$ and the anomaly $\omega$ as well as the anomaly free subgroup $H$ (with the discrete torsion $\psi$) that we are gauging. An interesting feature is that, even if the finite symmetry $G$ is anomaly free (that is, $\omega$ is trivial), the resulting group theoretical fusion categories can be anomalous.\footnote{The authors thank for the referee for pointing out a mistake on this in the draft.} We will describe a general criteria for the group theoretical fusion category to be anomaly free in the section \ref{sec:constraints_on_RG_flows}. In general, this means that finite non-intrinsic non-invertible 0-form symmetries are completely characterized by group theoretical fusion categories in a 2-dimensional bosonic theory. We then consider the $c=1$ KT theory as an example, and compute its twisted partition function explicitly and show the results indeed agree with our general analysis. 

It is a natural question to ask if there are more allowed $F$-symbols with the same fusion relations. Just like the TY-categories characterizing the duality can have distinct FS indicators with $\epsilon = \pm 1$, the possible FS indicators of the triality fusion category are given by $\alpha = e^{2\pi i k/3}\in \doubleZ_3$ with $k = 0,1,2$. One can see this from the fact that there are $\doubleZ_3$ phase rotations of the $F$-symbols which preserve the pentagon equations and are not gauge transformations. Physically, this means one can construct triality fusion category with different FS indicator $\alpha'$ from a known category with FS indicator $\alpha$ by taking the theory $\scriptT$ with the triality defect $\mathcal{L}_Q$, staking a decoupled theory $\scriptT'$ with an anomalous $\doubleZ_3$ global symmetry generated by $\eta$, and defining a new triality line $\scriptL_Q' \equiv \scriptL_Q \eta$. From the group theoretical fusion category point of view, to acquire triality fusion categories with the FS indicator $\alpha = e^{2\pi i k/3}$ means we are gauging the non-anomalous $\doubleZ_2$ symmetries in $A_4$ but now with non-trivial 't Hooft anomaly for $A_4$. The anomaly for $A_4$ is classified by $H^3(A_4,U(1))\simeq \doubleZ_6$ and let's denote its generator as $\omega_0$. The triality fusion categories with different FS indicators are $\scriptC(A_4,\omega_0^{2k},\doubleZ_2,1)$ for $k = 0,1,2$.

However, just taking into account the FS indicators does not enumerate all the possible triality fusion categories. Indeed, another set of $F$-symbols is computed in the condensed matter literature \cite{teo2015theory} and it is natural to ask if these $F$-symbols lead to the same fusion categories as ours. We show that the two sets of fusion categories are different and can be distinguished using spin selection rules. Then this enumerates all the possible inequivalent $F$-symbols for the triality fusion categories according to the classification result in \cite{jordan2009classification}.

\subsection{Outline of the Paper}
The goal of the paper is to provide a description of group theoretical triality fusion categories $\scriptC(A_4,\omega_0^{2k},\doubleZ_2,1)$ in terms of the bimodules of the $\doubleZ_2$ group algebra, and use it to compute the fusion rule and $F$-symbols. The simplicity of the group theoretical fusion category is that the objects naturally have a $\doubleC$-vector spaces structure, therefore every data we need can be described using linear algebras.\footnote{Notice that there is no $\doubleC$-vector spaces structure on objects in a generic fusion category $\scriptC$. For finite $0$-form symmetries in a bosonic theory, non-intrinsic non-invertible symmetries form group theoretical fusion categories. This means non-intrinsic non-invertible symmetries naturally have $\doubleC$-vector space structure while the intrinsic non-invertible symmetries do not. It would be interesting to check if there's any physical understanding or implication of this difference.} We also pointed out there are triality fusion categories that are not group theoretical, therefore are intrinsic. We show that whether the triality fusion category is intrinsic or non-intrinsic can be determined from the spins of the defect Hilbert space of the triality line.

In section \ref{sec:review}, we briefly review the TDLs in CFT. In section \ref{sec:tdl_review}, we introduce the basic notions relate to the TDLs in 2d CFT. In section \ref{sec:review_bootstrap}, we review the modular bootstrap program, and describe how to relate the twisted partition function computes the action of TDL $\scriptL$ on Hilbert space $\scriptH$ and to the twisted partition function computes the states of the defect Hilbert space $\scriptH_\scriptL$. Then, in section \ref{sec:review_asymptotic_density_of_states}, we briefly review the result in \cite{Pal:2020wwd} on the asymptotic density of states in various Hilbert spaces, which will be useful for us later.

In section \ref{sec:group_theoretical_triality}, we describe how to understand the triality fusion categories discovered in \cite{Thorngren:2019iar,Thorngren:2021yso} as group theoretical fusion category $\scriptC(A_4,1,\doubleZ_2,1)$. In section \ref{sec:group_theoretical_triality_review}, we briefly review the triality fusion category, which is acquired from gauging $\doubleZ_2$ subgroup in $A_4$. Then, we introduce the notion of group theoretical fusion category in section \ref{sec:group_theroretical_fusion_category_more} and show the triality fusion category in the KT theory can be described as $\scriptC(A_4,1,\doubleZ_2,1)$. We then begin to describe the data of this triality fusion category from the $\scriptC(A_4,1,\doubleZ_2,1)$ using the language of bimodules. In the rest of this section, we give an explicit calculation of the spectrum of simple TDLs, their fusion rules, and the $F$-symbols for $\scriptC(A_4,\omega_0^{2k},\doubleZ_2,1)$ using bimodules.

In section \ref{sec:Physical_Implications}, we use the above result to derive physical consequences. In section \ref{sec:group_theoretical_spin_selection_rules}, we compute the spin selection rules following the techniques in \cite{Chang:2018iay}. Since $\scriptC(A_4,\omega_0^{2k},\doubleZ_2,1)$ is acquired from gauging the $\doubleZ_2$ subgroup in a theory with global symmetry $A_4$, we can match the irreducible representations (irreps) of the fusion ring of $\scriptC(A_4,\omega_0^{2k},\doubleZ_2,1)$ to different sectors in $\scriptT$. This allows us to derive the Cardy-like formulas for different irreps of $\scriptC(A_4,\omega_0^{2k},\doubleZ_2,1)$ in the section \ref{sec:asymptotic_density_of_states}. Finally in the section \ref{sec:constraints_on_RG_flows}, we study the anomaly of the group theoretical fusion categories $\scriptC(G,\omega,H,\psi)$ by analyzing its possible symmetric gapped phases\footnote{Here, by $\scriptC$-symmetric gapped phase we mean the Hamiltonian or Lagrangian preserves the $\scriptC$-symmetry while it could be spontaneously broken etc. By $\scriptC$-symmetric trivial gapped phase, we mean there's a unique ground state in this phase.}. We find that while the symmetric gapped phases of $\scriptC(G,\omega,H,\psi)$ are in canonical bijection with the symmetric gapped phases of $\VEC_G^\omega$, this bijection does not preserve the number of the ground states. From the analysis of the possible symmetric gapped phases of $\scriptC(A_4,\omega_0^{2k}, \doubleZ_2,1)$ we conclude that they are all anomalous. We then give an anomaly free condition for the group theoretical fusion category in general. This means that the finite non-intrinsic non-invertible symmetries are completely characterized by group theoretical fusion categories for 2-dimensional bosonic field theory.

In section \ref{sec:KT_theory}, we first review the Kosterlitz-Thouless (KT) theory and its triality fusion categories discovered in \cite{Thorngren:2019iar,Thorngren:2021yso}. We then compute the twisted partition function of the triality defect and show the spins in the defect Hilbert space $\scriptH_{\scriptL_Q}$ indeed match the spin selection rules in the section \ref{sec:KT_twisted_partition_function}. Since the KT theory has $\doubleZ_3\times \doubleZ_3 \subset (U(1)^{\tilde{\theta}}\times U(1)^{\tilde{\phi}})\rtimes \doubleZ_2$ global symmetries, it is interesting to check if one can construct new triality line via $\llq \eta$, where $\eta$ generates the one of the $\doubleZ_3$ symmetry. We show explicitly this is not possible by checking that the line $\llq \eta$ does not satisfy the fusion rule in the section \ref{sec:KT_new_trialities}.

In section \ref{sec:more_trialities}, we consider other triality fusion categories. We first review the classification of the triality fusion categories in \cite{jordan2009classification} in the section \ref{sec:classifications}. We then list the $F$-symbols for the intrinsic triality fusion categories computed in \cite{teo2015theory} in the section \ref{sec:instric_triality_F_symbol}. In the section \ref{sec:intrinsic_spin_selection_rule}, We then compute the spin selection for the triality line $\llq$ for the intrinsic triality fusion categories and show that the triality fusion categories can be distinguished by the spins of the states in the $\scriptH_{\llq}$. We conclude the paper with a comment on when the spin selection rules should be saturated.

\subsection{Future Directions}
We outline some of the future directions to explore.

\subsubsection*{\ul{\it Exploring generic group theoretical fusion categories}}
In this paper, we mainly focus on describing the group theoretical fusion categories $\scriptC(A_4,1,\doubleZ_2^\sigma,1)$. Since our approach can be easily generalized to other group theoretical fusion categories, it would be interesting to systematically explore the group theoretical fusion categories using Mathematica, for example, computing the simple TDLs, their fusion rules, and the $F$-symbols, as well as the physical implications such as the spin selection rules, classification of $\scriptC$-symmetric gapped phases, and solutions of general modular bootstrap equations, etc. Group theoretical fusion categories provide a systematic way of constructing $N$-ality fusion categories in 2-dimension. Some of these will be done in the upcoming work by the authors \cite{upcoming}. 

\subsubsection*{\ul{\it Intrinsic triality defects}}
In this paper, we only study the spin selection rules of the intrinsic triality fusion categories, which is sufficient to distinguish them from the non-intrinsic ones. Unlike the group theoretical fusion category symmetries, it's harder to find examples of CFT with non-intrinsic triality symmetries. A possible candidate is the bosonization of the theory of 8 Majorana fermions discussed in \cite{Tong:2019bbk}. Furthermore, it would be interesting to understand the anomaly and in general, the possible symmetric gapped phases, which are described to module categories $\scriptM$ over the fusion category $\scriptC$. This can be done using the techniques developed in \cite{meir2012module}. More generally, a class of intrinsic $N$-ality fusion categories is constructed in \cite{jordan2009classification} and it would be interesting to answer the above questions for these as well.

\hfill \break

\textbf{Note added:} By the time we are about to post this paper, a nice paper \cite{Lin:2022dhv} generalizing the result in \cite{Pal:2020wwd} appears and their general result can be used to produce our results in the section \ref{sec:asymptotic_density_of_states}. Also, the three papers \cite{Lin:2022xod,Bhardwaj:2022lsg,Bartsch:2022mpm} appear on the arXiv the same day as the authors post the first version of the draft on arXiv, which generalizes the idea of gauging non-normal or non-Abelian subgroup will lead to non-invertible symmetries to higher form and higher group symmetries in higher dimensions.\footnote{For understanding the non-invertible symmetries from holographic point of view, see recent papers \cite{Apruzzi:2022rei,GarciaEtxebarria:2022vzq,Heckman:2022muc,Antinucci:2022vyk}.}

\section{Review on TDLs and their applications in 2d CFT}\label{sec:review}
In this section, we give a brief review of TDLs in 2d CFT to fix the convention, and readers who are interested in a more detailed review should look at \cite{Chang:2018iay,Bhardwaj:2017xup}. We also review results on modular bootstraps and Cardy-like formulas which will be useful later.

\subsection{TDLs in 2d CFT}\label{sec:tdl_review}
In 2-dimensional conformal field theory, the symmetry defects of the ordinary symmetry are line operators and they are the very first examples of TDLs. For symmetry $G$ of a CFT, the corresponding TDLs are denoted as $\CL_g, g\in G$. The juxtaposition of two such TDLs satisfies the group algebra, $\CL_g \times \CL_h = \CL_{gh}$. The TDL corresponds to the identity $\dsi$ in the group $G$ is the identity line which we will also denote as $\dsi \equiv \scriptL_\dsi$. Since every group element has its inverse, these TDLs also have their inverse in the sense that there exists another line such that they fuse into a single identity line $\dsi$, i.e. they are invertible TDLs. 

However, TDLs are ubiquitous in the CFT, and the way they fuse is beyond the group algebra, more generally satisfying the fusion algebra. These TDLs then generate the non-invertible symmetries or the fusion category symmetries, since the TDLs may not have their inverse. The Kramer-Wannier duality line in the Ising CFT and most of the Verlinde lines in rational CFT are examples of the non-invertible TDLs. 

The TDL $\scriptL$ can be deformed locally without changing the correlation functions $\langle \scriptL \cdots\rangle$ where $\cdots$ denotes any other operator insertions. The topological nature of these TDLs imply that they commute with stress energy tensor, therefore can be algebraically expressed as,
\begin{equation}
    [L_n,\CL_a] = [\bar{L}_n,\CL_a] = 0,
\end{equation}
where $L_n,\bar{L}_n$ are the generators of the Virasoro algebra. When the TDL $\CL_a$ acts on a state $\ket{\phi}$ in Hilbert space $\CH$, the resulting state $\CL_a \ket{\phi}$ is still in $\CH$. 

When two TDLs are brought close to each other, their juxtaposition satisfies the fusion rule,
\begin{equation}
    \CL_a \times \CL_b =\bigoplus_c N_{ab}^c \CL_c
\end{equation}
where $N_{ab}^c$ is the fusion multiplicity which can only be non-negative integers. More specifically, the TDLs can join at the point-like junction, which is equipped with a Hilbert space. For example, the fusion of TDLs $\CL_1,\CL_2$ into $\CL_3$ corresponds to a vector space with dimension depending on the fusion multiplicity, we dub the vector space as fusion space and denote by $V_{\CL_1,\CL_2}^{\CL_3}$. Correspondingly, the TDL $\CL_3$ can split into $\CL_1,\CL_2$, whose vector space is split space and denoted by $V_{\CL_3}^{\CL_1,\CL_2}$.

\begin{figure}[h]
    \centering
    \includegraphics[width=0.7\textwidth]{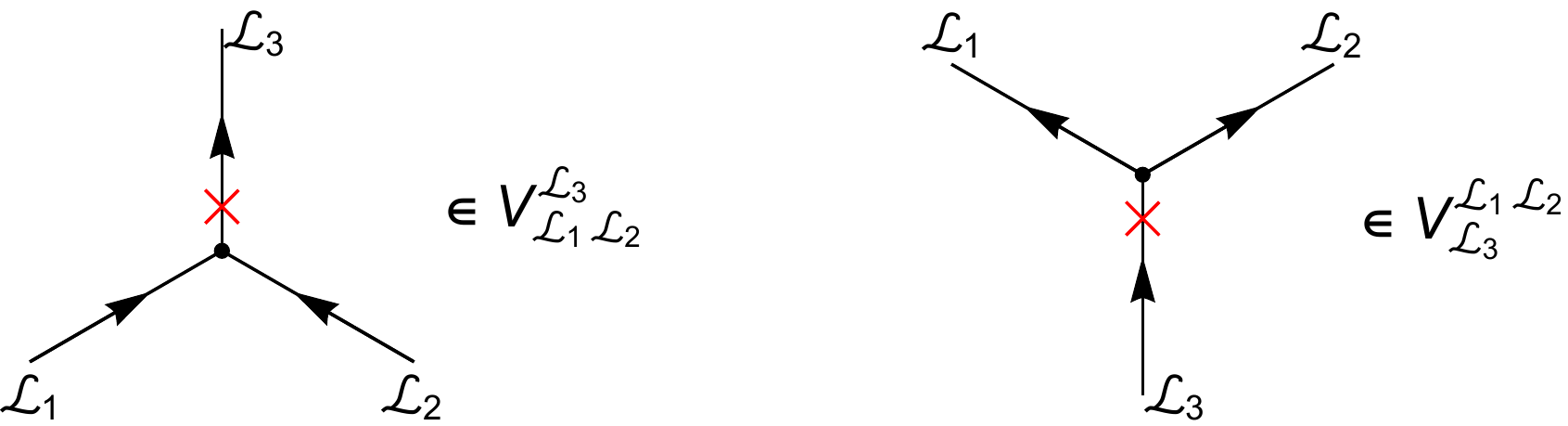}
    \caption{The fusion vertex of TDLs $\CL_1,\CL_2$ fuse into $\CL_3$ and split vertex of TDLs $\CL_3$ splits into $\CL_1,\CL_2$. The red cross marks the last leg, which determines the ordering of each vertex.}
    \label{fig:vert}
\end{figure}

\begin{figure}[h]
    \centering
    \includegraphics[width=0.7\textwidth]{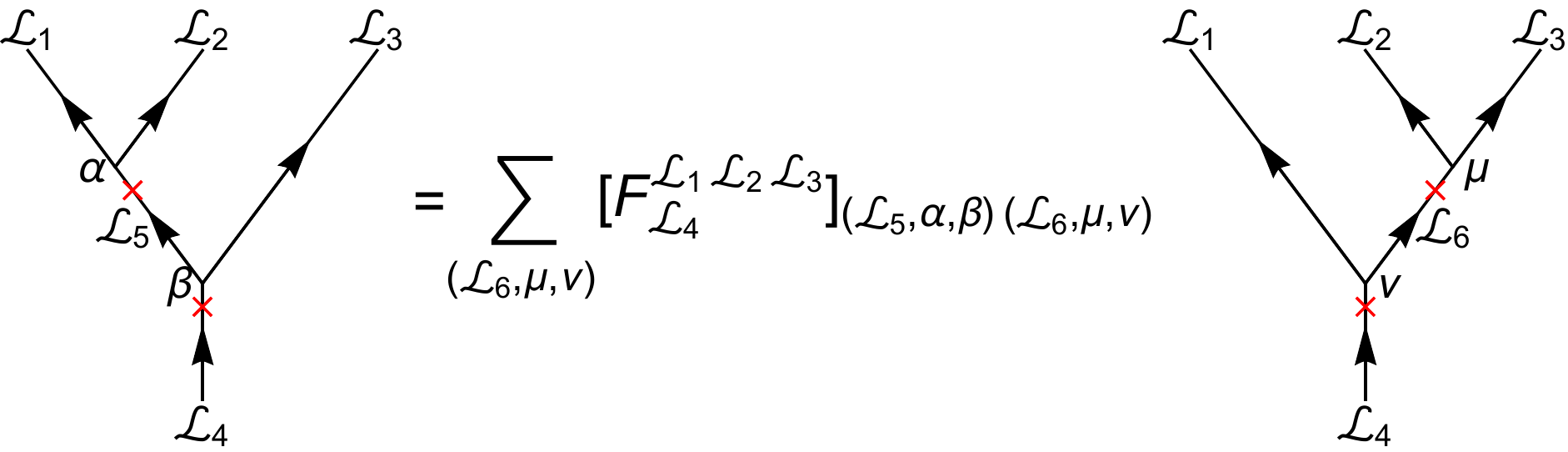}
    \caption{The fusion of three TDLs has two different ways, but they are related by the $F$-move and characterized by the $F$-symbol defined in this diagram.}
    \label{fig:fmove}
\end{figure}

More complicated fusion/split process can be decomposed into the fusion/split space with 3 TDLs. However, the decomposition is not unique but under the isomorphism, $\bigoplus_{\CL_5} \splv{1}{2}{5}\otimes \splv{5}{3}{4}\cong \bigoplus_{\CL_6} \splv{2}{3}{6}\otimes \splv{1}{6}{4}$. This is called $F$-move and can be written diagrammatically as shown in \figref{fig:fmove}. Notice that here we use the notion of $F$-symbol, which is same as the crossing kernels $\scriptK$ in \cite{Chang:2018iay} up to the flipping of some of the orientation of the TDLs. 

\begin{figure}[h]
    \centering
    \includegraphics[width=0.5\textwidth]{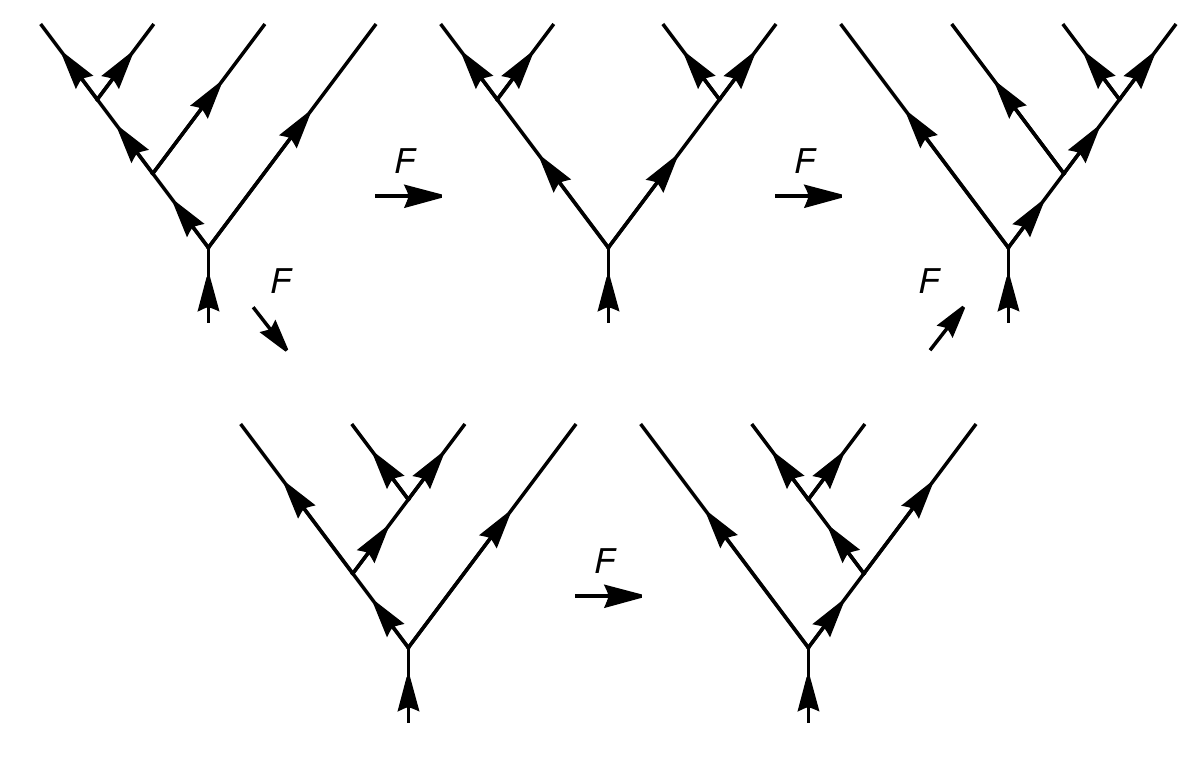}
    \caption{The upper 2 $F$-moves and the lower 3 $F$-moves yield the same diagram, which gives the pentagon equation in \eqref{eq:pentaequ}.}
    \label{fig:pentaequ}
\end{figure}

The $F$-symbols are constrained by the self-consistent conditions when applying $F$-moves to the split process with $5$ TDLs. Two sequences of $F$-moves start with the same state and end with the same state should be equivalent. These consistent conditions diagrammatically shown in \figref{fig:pentaequ} yield the pentagon equations on the $F$-symbols,
\begin{align}\label{eq:pentaequ}
&\sum_{\delta }\left[ F_{e}^{fcd}\right] _{\left( g,\beta ,\gamma \right)
\left( l,\nu, \delta \right) }\left[ F_{e}^{abl}\right]
_{\left( f,\alpha ,\delta \right) \left( k,\mu, \lambda \right) }\nonumber\\
=&\sum_{h,\sigma ,\psi ,\rho }\left[ F_{g}^{abc}\right] _{\left( f,\alpha
,\beta \right) \left( h,\psi,\sigma \right) }\left[ F_{e}^{ahd}\right]
_{\left( g,\sigma ,\gamma \right) \left( k,\rho, \lambda \right) }%
\left[ F_{k}^{bcd}\right] _{\left( h,\psi ,\rho \right) \left(
l,\nu ,\mu \right) }.
\end{align}
For example, the invertible symmetry $G$ with anomaly $\omega$ is described by the category of $G$-graded vector spaces, denoted by $\VEC_G^\omega$. The simple objects are $1-$dimensional $\doubleC$-vector space labeled by $g \in G$, and physically they correspond to the TDLs which generates the $g$-action. The fusion rule is $\CL_g \times \CL_h = \CL_{gh}$. The $F$-symbols of this category are $U(1)$ phase factors (rather than generic complex numbers in order to preserve the normalization), $\omega(g,h,k) \equiv F^{\CL_g,\CL_h,\CL_k}_{\CL_{ghk}}$ for $\CL_{1,2,3}=\CL_{g,h,k}$ as in \figref{fig:fmove}. These $U(1)$ phase factors satisfy the pentagon equations,
\begin{equation}
    \omega(g,h,kl)\omega(gh,k,l) = \omega(h,k,l) \omega(g,hk,l)\omega(g,h,k),
\end{equation}
which is the cocycle condition for the 3-cocycle $\omega: G\times G\times G \rightarrow U(1)$.

Notice that one can consider shifting the basis in $V^{g,h}_{gh}$ by a phase $\beta(g,h)$, which is an element in $C^2(G,U(1))$. This basis change does not change the physics and should be understood as a gauge transformation of the $F$-symbols. The $F$-symbol changes as
\begin{equation}
    \omega(g,h,k) \rightarrow \omega(g,h,k) \frac{\beta(g,hk)\beta(h,k)}{\beta(g,h)\beta(gh,k)},
\end{equation}
and is interpreted as changing $\omega$ by an exact element in the set of 3-coboundaries $Z^3(G,U(1))\subset C^3(G,U(1))$. Therefore, inequivalent $\VEC_G^\omega$'s are labelled by a finite group $G$ and its anomaly $\omega \in H^3(G,U(1)) = C^3(G,U(1))/Z^3(G,U(1))$.

\subsection{Modular bootstrap with TDLs}\label{sec:review_bootstrap}
\begin{figure}[htbp]
    \centering
    \includegraphics{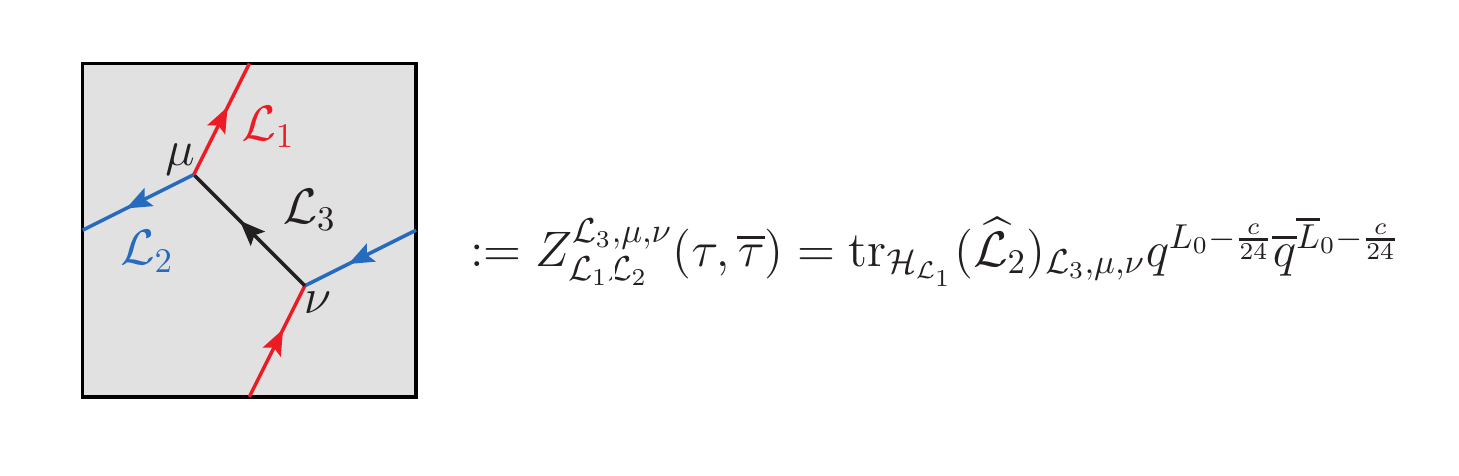}
    \caption{Convention on the twisted partition function $Z_{\mathcal{L}_1 \CL_2}^{\CL_3}(\tau,\otau)$.}
    \label{fig:deftwistedpartitionfunc}
\end{figure}

By utilizing the modular transformation properties of the partition functions or correlation functions of a CFT on a torus with complex structure $\tau$, one can extract many useful information of the CFT \cite{Cardy:1986ie,Cardy:2017qhl,Pal:2019zzr,Pal:2020wwd,Lin:2019hks, Lin:2021udi, Collier:2021ngi, Kaidi:2021ent, Lanzetta:2022lze, Mukhametzhanov:2019pzy, Mukhametzhanov:2020swe}. Here, we are interested in studying the modular properties of the CFT partition functions with networks of TDLs inserted on the torus, which is called the twisted partition function on the torus. We will adopt the convention from \cite{Thorngren:2021yso} on the twisted partition function over torus, as in Figure \ref{fig:deftwistedpartitionfunc}.
To reveal the physical meaning of $Z_{\scriptL_1,\scriptL_2}^{\scriptL_3,\mu,\nu}(\tau,\otau)$, we first consider the simple case where $\scriptL_2 = \dsi$ and $\scriptL_3 = \scriptL_1 = \scriptL$. For convenience, we will sometimes abbreviate this partition function as $Z_{\scriptL}(\tau,\otau)$. Because the topological defect $\scriptL$ is inserted along the time direction, it should be interpreted as a defect along the spatial direction, leading to a different Hilbert space on the spatial circle $S^1$, and we will denote this Hilbert space as the defect Hilbert space $\scriptH_{\scriptL}$.
\begin{equation}
    Z_{\scriptL}(\tau,\otau) = Z^{\scriptL,1,1}_{\scriptL,\dsi}(\tau,\otau) = \Tr_{\scriptH_\scriptL} q^{L_0 - \frac{c}{24}} \oq^{\oL_0 - \frac{c}{24}}.
\end{equation}
For instance, if $\scriptL$ is a $\doubleZ_2$ symmetry defect $\eta$, then the defect Hilbert space $\scriptH_{\scriptL}$ is the $\doubleZ_2$ twisted Hilbert space acquired by imposing the twisted boundary condition.

\begin{figure}
    \centering
    \includegraphics[scale = 0.8]{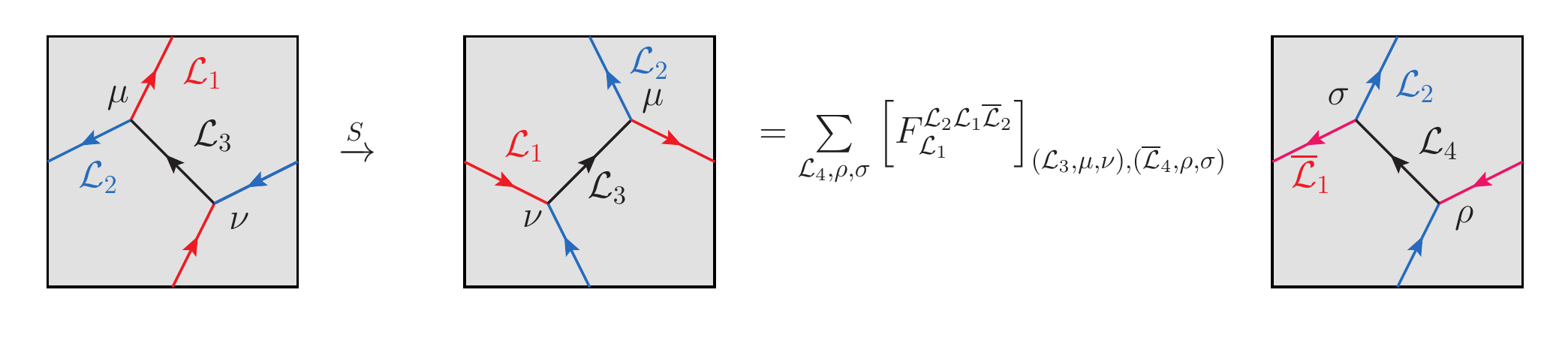}
    \caption{Generalized modular bootstrap equations from $S$ modular transformation of the twisted partition function $Z^{\scriptL_3,\mu,\nu}_{\scriptL_1,\scriptL_2}(\tau,\overline{\tau})$.}
    \label{fig:modularbootstrap}
\end{figure}

As an another simple example, let's consider instead $\scriptL_1 = I$ and $\scriptL_2 = \scriptL_3 = \scriptL$. Similarly, we will sometimes abbreviate this partition function as $Z^{\scriptL}(\tau,\otau)$. In this configuration, the TDL $\scriptL$ is inserted on an equal time slice, therefore should be interpreted as a quantum operator $\widehat{\scriptL}$ on the Hilbert space $\scriptH$:
\begin{equation}
    Z^{\scriptL}(\tau,\otau) \equiv Z^{\scriptL,1,1}_{\dsi,\scriptL}(\tau,\otau) = \Tr_{\scriptH} \widehat{\scriptL} q^{L_0 - \frac{c}{24}} \oq^{\oL_0 - \frac{c}{24}}.
\end{equation}
More generally, if we consider the twisted partition function $Z_{\scriptL_1,\scriptL_2}^{\scriptL_3,\mu,\nu}(\tau,\otau)$, we should interpret $\scriptL_2$ as a quantum operator $(\widehat{\scriptL}_2)_{\scriptL_3,\mu,\nu}$ acting on the defect Hilbert space $\scriptH_{\scriptL_1}$, where the subscript $(\scriptL_3,\mu,\nu)$ indicates that for different intermediate lines $\scriptL_3$ and different vertex labels $\mu,\nu$ when the fusion multiplicities are greater than $1$, we will have a different operator in general. 
\begin{equation}\label{eq:general_twisted_partition_function}
    Z_{\scriptL_1,\scriptL_2}^{\scriptL_3,\mu,\nu}(\tau,\otau) = \Tr_{\scriptH_{\scriptL_1}} (\widehat{\scriptL}_2)_{\scriptL_3,\mu,\nu} q^{L_0 - \frac{c}{24}}\oq^{\oL_0 - \frac{c}{24}}.
\end{equation}
Under the modular transformation, the twisted partition function $Z^{\scriptL_3,\mu,\nu}_{\scriptL_1,\scriptL_2}(\tau,\otau)$ transforms as the Figure \ref{fig:modularbootstrap},
\begin{equation}
    Z^{\scriptL_3,\mu,\nu}_{\scriptL_1,\scriptL_2}\left(-\frac{1}{\tau},-\frac{1}{\otau}\right) = \sum_{\scriptL_4,\rho,\sigma} \left[F^{\scriptL_2 \scriptL_1 \overline{\scriptL}_2}_{\scriptL_1}\right]_{(\scriptL_3,\mu,\nu),(\overline{\scriptL}_4,\rho,\sigma)} Z^{\scriptL_4,\sigma,\rho}_{\scriptL_2,\overline{\scriptL}_1}(\tau,\overline{\tau}).
\end{equation}

As a simple example, we may consider taking $\scriptL_1$ to be the identity line. Then the $F$-symbol is always trivial in this case (for example, see \cite{Chang:2018iay}) and $\mu,\nu = 1$ as the fusion is always one dimensional. Then we have
\begin{equation}
    Z_{\scriptL}\left(-\frac{1}{\tau},-\frac{1}{\overline{\tau}}\right) \equiv Z_{\dsi,\scriptL}^{\scriptL,1,1}\left(-\frac{1}{\tau},-\frac{1}{\overline{\tau}}\right) = Z^{\scriptL}_{\scriptL,\dsi}(\tau,\overline{\tau}) \equiv Z^{\scriptL}(\tau,\overline{\tau}).
\end{equation}
As one can see, the $S$ modular transformation relates the partition functions that count states the defect Hilbert space $\scriptH_\scriptL$ to the partition functions that compute the action of the TDL $\scriptL$ on the Hilbert space $\scriptH$. This allows us to derive interesting Cardy-like formula which we will now review in the next subsection. 

\subsection{Asymptotic density of states}\label{sec:review_asymptotic_density_of_states}
As an application of the modular bootstrap reviewed above, one can derive the asymptotic density of states of a CFT \cite{Cardy:1986ie, Mukhametzhanov:2019pzy}. Intuitively, the modular bootstrap equation,
\begin{equation}\label{eq:bootstrap}
    Z^{\scriptL}\left(-\frac{1}{\tau}, -\frac{1}{\otau}\right) = Z_{\scriptL}(\tau,\overline{\tau}),
\end{equation}
relates the high-temperature limit of the partition function $Z^{\scriptL}$ which computes the action of the TDL $\scriptL$ on $\scriptH$ to the low-temperature limit of the partition function $Z_{\scriptL}$ which computes the spectrum of the defect Hilbert space $\scriptH_{\scriptL}$ and vice versa. Since in the low-temperature limit, the partition function is always dominated by the ground state, \eqref{eq:bootstrap} essentially allows us to determine the partition function $Z^{\scriptL}$ and $Z_{\scriptL}$ in the high-temperature limit. In the high-temperature limit, states with different energy would contribute equally to the partition sum, knowing $Z_{\scriptL}$ in such limit would allow us to derive an approximation of the density of states for the defect Hilbert space $\scriptH_{\scriptL}$. This idea is made into a rigorous mathematical statement in \cite{Pal:2020wwd} using the techniques in \cite{Mukhametzhanov:2019pzy}, and we will only mention the result below.

Let $F_{\scriptL}(\Delta)$ denote the total number of states with scaling dimension $\Delta' < \Delta$ in the defect Hilbert space. By asymptotic density of states, we mean a continuous function $\rho_{0,\scriptL}(\Delta)$ which approximates the actual density of states in the sense that
\begin{equation}
    F_{\scriptL}(\Delta) = \int_0^\Delta d\Delta' \, \rho_{0,\scriptL}(\Delta') + O(\Delta^{-1/2}), \quad \Delta \rightarrow \infty.
\end{equation}
Let 
\begin{equation}\label{eq:asym_dos}
    \rho_0(\Delta) = \left(\frac{c}{48\Delta^3}\right)^{1/4} \exp\left[2\pi \sqrt{\frac{c\Delta}{3}}\right],
\end{equation}
it is shown in \cite{Pal:2020wwd} that the asymptotic density of states $\rho_{0,\scriptL}$ of the defect Hilbert space $\scriptH_\scriptL$ is simply given by
\begin{equation}
    \rho_{0,\scriptL}(\Delta) = \langle 0| \scriptL |0\rangle \rho_0(\Delta),
\end{equation}
where $|0\rangle$ is the ground state and $\langle 0|\scriptL |0\rangle$ is the quantum dimension of the TDL $\scriptL$.

Furthermore, let's consider the theory with finite global symmetry $G$. Then the states in the Hilbert space will organize into irreducible representations of $G$. For simplicity, we assume $G$ acts faithfully. Recall that a particular type of irreducible representation $\alpha$ can be counted by $\frac{1}{|G|}\sum_{g\in G}\chi_\alpha(g)^*\text{Tr} g$ in a reducible representation of $G$ where $\chi_\alpha(g)$ is the character function of the irrep $\alpha$ and $|G|$ is the order of the group, then the partition function counts the number of irrep $\alpha$ in the Hilbert space is given by
\begin{equation}\label{eq:projection}
    \frac{1}{|G|}\sum_{g\in G}\chi_\alpha(g)^* Z^{g}(\tau,\overline{\tau}),
\end{equation}
whose high-temperature limit is known since each $Z^g$'s high-temperature limit is known. Moreover, the assumption that $G$ acts faithfully implies that the ground state in the defect Hilbert space has $h+\overline{h}>0$, therefore \eqref{eq:projection} is dominated by $Z^{\dsi}(\tau,\otau) = Z(\tau,\otau)$ in the high temperature limit. This allows us to derive the following result. Every irreducible representation has to appear in the Hilbert space $\scriptH$ and they also have a Cardy-like growth. Specifically, we can consider the asymptotic growth $\rho_{0,\alpha}(\Delta)$ of the occurrence of a particular irrep $\alpha$ of $G$ as a function of the scaling dimension $\Delta$ takes the form,
\begin{equation}
    \rho_{0,\alpha}(\Delta) = \frac{d_\alpha}{|G|}\rho_0(\Delta).
\end{equation}
This result will be useful for us later.

\section{Non-intrinsic triality fusion category as group theoretical fusion category}\label{sec:group_theoretical_triality}
As pointed out in \cite{Thorngren:2021yso}, the origin of the triality fusion rule in the KT theory arises from gauging $\doubleZ_2$ subgroup of $A_4 \subset SO(4)$ in the $SU(2)_1$ theory. Generically, gauging a subgroup $H$ which is not normal in $G$ or not Abelian leads to non-invertible symmetries \cite{naidu2007categorical}. In this section, we first review the construction of triality acquired in \cite{Thorngren:2019iar,Thorngren:2021yso}. We then discuss how to understand the group theoretical triality fusion category using the mathematical tools of bimodule categories in \cite{ostrik2002module}. Specifically, we describe how to understand and compute simple TDLs, fusion rules and $F$-symbols.

\subsection{A brief review of the triality category discovered in \cite{Thorngren:2019iar,Thorngren:2021yso}}\label{sec:group_theoretical_triality_review}
The origin of the triality fusion category discovered in \cite{Thorngren:2019iar,Thorngren:2021yso} is a result of gauging $\doubleZ_2$ subgroup in a theory with $A_4$ global symmetry. Notice that $A_4$ is an order $12$ group and can be presented as
\begin{equation}\label{eq:defA_4}
    A_4 = \langle \sigma,\eta,q|q^3 = \sigma^2 = \eta^2, q \sigma q^{-1} = \sigma\eta = \eta\sigma, q \eta q^{-1} = \sigma\rangle.
\end{equation}
This means we can think of $A_4$ containing a $\doubleZ_2^\sigma\times \doubleZ^\eta_2$ subgroup, and the conjugation by the $\doubleZ^q_3$ generator $q$ will permute the three $\doubleZ_2$ generators in $\doubleZ^\sigma_2\times \doubleZ^\eta_2$. After gauging the $\doubleZ_2^\sigma$ symmetry, the $\doubleZ_2^\eta$ subgroup will survive since it commutes with the $\doubleZ_2^\sigma$ subgroup, together with the quantum $\doubleZ_2$ which we will denote as $\doubleZ_2^{\hat{\sigma}}$, they form the $\doubleZ_2^{\hat{\sigma}}\times \doubleZ_2^{\eta}$ invertible symmetries in the gauged theory. The symmetry operator $q$ does not commute with $\sigma$, therefore is not gauge invariant, and will not appear as a genuine topological line operator in the gauged theory. However, the linear combination
\begin{equation}
    q + \sigma q \sigma
\end{equation}
does commute with $\sigma$ and will survive as a genuine topological line operator in the gauged theory, which we will denote as $\llq$. However, this operator is not invertible, as it has quantum dimension $2$. Similarly, its orientation reversal $\llqb$ relates to the gauge-invariant linear combination $q^{-1} + \sigma q^{-1} \sigma$. It is pointed out in \cite{Thorngren:2019iar,Thorngren:2021yso} that $\llq,\llqb$ together with the $\doubleZ_2^{\hat{\sigma}}\times \doubleZ_2^{\eta}$ forms the triality fusion categories with the fusion rules,
\begin{equation}
\begin{aligned}
    & g\times\llq = \llq \times g = \llq, \quad g\times \llqb = \llqb \times g = \llqb, \\
    & \llq\times \llq = 2\llqb, \quad \llqb \times \llqb = 2\llq, \quad \llqb \times \llq =\llq \times \llqb = \sum_{g\in\doubleZ_2^{\hat{\sigma}}\times\doubleZ_2^\eta} g.
\end{aligned}
\end{equation}
The existence of the triality fusion category implies the theory is invariant under the $\doubleZ_2^{\hat{\sigma}}\times \doubleZ_2^{\eta}$ gauging, but $\doubleZ_2^{\hat{\sigma}}\times \doubleZ_2^{\eta}$ charge assignments will be permuted. This can be understood as the following. Gauging the $\doubleZ_2^{\hat{\sigma}}$ quantum symmetry gives back the original theory with $A_4$ symmetry, then gauging $\doubleZ_2^{\eta}$ symmetry will give the same theory acquired from gauging the $\doubleZ_2^{\sigma}$ symmetry, since $\doubleZ_2^{\eta}$ and $\doubleZ_2^{\sigma}$ are related by the conjugation of $q$. However, this will change the $\doubleZ_2\times \doubleZ_2$ charge assignment in the gauged theory since we are gauging different but equivalent $\doubleZ_2$'s.

\subsection{Group theoretical fusion category}\label{sec:group_theroretical_fusion_category_more}
Let's consider the fusion category $\VEC_G^\omega$, where $G$ is a finite group and $\omega \in H^3(G,U(1))$. Physically, this fusion category describes the global symmetry $G$ with anomaly $\omega$ for 2-dimensional field theory. The simple elements are labeled by group element $g \in G$ and the fusion rule is simply the product of the group elements. The $F$-symbols are given by
\begin{equation}
    F_{l}^{ghk} = \omega(g,h,k).
\end{equation}
If we consider a finite subgroup $H$ in $G$, then the anomaly-free condition for $H$ is that $\omega$ as a function from $G^3$ to $U(1)$ restricting to $H^3$ is trivial, that is,
\begin{equation}
    \omega(h_1,h_2,h_3) = 1, \quad \forall \, h_i \in H.
\end{equation}
When $H$ is anomaly free, we can consider gauging the subgroup $H$ and we have to choose a discrete torsion $\psi \in H^2(G,U(1))$. The resulting fusion category which describes the global symmetry in the gauged theory is denoted as $\scriptC(G,\omega,H,\psi)$ and this class of fusion category is called the group theoretical fusion category. In particular, $\VEC_G^\omega = \scriptC(G,\omega,\doubleZ_1,1)$. 

Generically, when $H$ is not a normal subgroup of $G$ or is not Abelian, then the resulting group theoretical fusion category is not of the form $\VEC_G^\omega$, meaning it contains non-invertible simple lines. For example, gauging the $\doubleZ_2$ symmetry of $S^3$ with $\omega = 1$ leads to the fusion category $\Rep(S^3)$ \cite{Bhardwaj:2017xup} and gauging the $\doubleZ_2$ symmetry of $A_4$ with $\omega = 1$ leads to the triality fusion category \cite{Thorngren:2019iar,Thorngren:2021yso}.

Below, we will describe how to understand and compute the simple TDLs, fusion rules, and $F$-symbols of the triality fusion category from the data of the group theoretical fusion category. It would be helpful to point out that although the $F$-symbols for the group theoretical fusion category can be computed by the more general method given in \cite{barter2021computing} by considering the tube algebra $\mathbf{Tub}_\scriptC(\scriptM)$, it is more convenient to compute the $F$-symbols using our approach. For instance, the calculation of the $F$-symbol for $\scriptC(A_4,1,\doubleZ_2,1)$ using \cite{barter2021computing} requires to construct an explicit Artin-Wedderburn isomorphism from the tube algebra $\mathbf{Tub}_{\VEC_{A_4}}(\scriptM)$ to a direct sum of several matrix algebras, but since the tube algebra $\mathbf{Tub}_{\VEC_{A_4}}(\scriptM)$ has a very large dimension $432$ in this case, it's computationally hard to explicitly construct the Artin-Wedderburn isomorphism. 

\subsection{Simple lines}\label{sec:simplelines}
As pointed out in \cite{ostrik2002module}, the simple TDLs in the gauged theory are described by indecomposable $A-A$ bimodules in ${\VEC}_G^\omega$, where $A$ is the group algebra of $H$ twisted by some 2-cocycle of $\psi \in H^2(H,U(1))$, corresponding to a choice of discrete torsion. In the case of $SU(2)_1$ theory, we take $G = A_4$ and $G$ is anomaly free so that $\omega = 1 \in H^3(G,U(1))$. $H = \doubleZ_2 \subset G$ and there's no non-trivial discrete torsion for $\doubleZ_2$, so we can choose $\psi$ to be trivial as well. Hence, $A$ is simply the group algebra of $\doubleZ_2$, namely a 2-dimensional vector space over $\doubleC$ with a basis $\{1,\sigma\}$ equipped with multiplication $1^2 = \sigma^2 = 1$ and $\sigma 1 =1 \sigma =\sigma$ which is the group multiplication of $\doubleZ_2$.

In the special case of group theoretical fusion category, by $A-A$ bimodule $M$, we mean a $\doubleC$-vector space $M$ together a left multiplication $A\times M \rightarrow M$ denoted as $am$ for $a \in A, m\in M$ and a right multiplication $M\times A \rightarrow M$ denoted as $ma$ for $a\in A, m\in M$ such that $\forall a_i \in A$ and $m\in M$:
\begin{equation}
    a_1 (a_2 m) = (a_1 a_2) m \equiv a_1 a_2 m, \quad (m a_1) a_2 = m (a_1 a_2) \equiv m a_1 a_2, \quad (a_1 m)a_2 = a_1 (m a_2) \equiv a_1 m a_2.
\end{equation}
The indecomposable $A-A$ bimodule $M$ is of the following form. Consider a double coset $HgH$ of $H$ in $G$, and let $H^g$ denote the little group of $HgH$:
\begin{equation}
    H^g = \{h \in H|\exists h' \in H \, s.t. \, hgh' = g \} \subset H.
\end{equation}
As one can check from the above definition, the little group $H^{g'}$ does not depend on the choice of representative $g' \in HgH$ up to isomorphism.

Each indecomposable $A-A$ bimodule is labelled by a double coset $HgH$ and an irreducible representation $\rho$ (potentially twisted by some 2-cocycle of the little group) of the little group of an arbitrary element in the double coset $HgH$. More specifically, given $(HgH,\rho)$, an indecomposable $A-A$ bimodule $M_{HgH}^{\rho}$ is a vector space over $\doubleC$ such that
\begin{equation}
    M_{HgH}^\rho = \bigoplus_{g'\in HgH} M_{g'}^{\rho}.
\end{equation}
To describe the action of $A$ on $M_{HgH}^\rho$, we only need to describe the action of the basis $\{h\}_{h\in H}$ on $M_{HgH}^\rho$. And multiplication of $h$ on the left induces an isomorphism from $M_{g'}^\rho$ to $M_{hg'}^\rho$ while the multiplication of $h$ on the right induces an isomorphism from $M_{g'}^\rho$ to $M_{g'h}^\rho$. This implies $M_{g'}^\rho$'s are isomorphic to each other. Consider $M_{g}^\rho$ and we find for $h \in H^g$, multiplying $h$ on the left and multiplying the corresponding $h'$ on the right leads to an isomorphism between $M_g^\rho$ and itself (since $hgh' = g$), this is why $H^g$ is called the little group. $M_g^\rho$ is the vector space of the irreducible representation $\rho$ of $H^g$. Notice that in the case where the group algebra $A$ is twisted by non-trivial 2-cocycle $\psi$, the composition of two left(right) multiplications on $M_g^\rho$ can composite non-trivially. But in the case of the gauging $\doubleZ_2$ symmetry in $A_4$ with no anomaly, such data is trivial. The fusion of the TDLs in the gauged theory is described by the tensor product of $A-A$ bimodule over the algebra $A$. 

We now list the indecomposable bimodule in $\scriptC(A_4,1,\doubleZ_2^\sigma,1)$ and compute its fusion rule, and show indeed we reproduce the fusion rule of the triality fusion category. There are 4 double cosets of $H = \doubleZ_2 = \{(), (12)(34)\} \equiv \{1,\sigma\}$ in $A_4$:\footnote{The $q$ used in \eqref{eq:defA_4} is the cycle $(143)$.}
\begin{equation}
\begin{aligned}
    & I = \{(), (12)(34)\} \equiv \{1,\sigma\}, \quad J = \{(13)(24),(14)(23)\} \equiv \{\eta,\eta\sigma\}, \quad \\
    & Q = \{(143),(124),(132),(234)\}, \quad \oQ = \{(134),(142),(123),(243)\}.
\end{aligned}
\end{equation}
The little group for the first two double cosets is $H$ itself while for the last two double cosets are trivial. Hence, for the first two double cosets $I,J$, we need to label the irreducible representation $\pm$ of $H = \doubleZ_2$ as well. Hence, we find the following $6$ indecomposable $A-A$ bimodules,
\begin{equation}
\begin{aligned}
    & M_I^\pm = M_1^\pm \oplus M_\sigma^\pm, \quad M_J^\pm = M_{\eta}^{\pm} \oplus M_{\sigma \eta}^{\pm}, \\
    & M_{Q} = M_{(143)} \oplus M_{(124)} \oplus M_{(132)} \oplus M_{(234)}, \\
    & M_{\oQ} = M_{(134)}\oplus M_{(142)} \oplus M_{(123)} \oplus M_{(243)}.
\end{aligned}
\end{equation}
Notice that each $M_g^\rho$ appears in the direct sum is a 1-dimensional vector space, so we will choose a basis vector $m_g^\rho$ for each $M_g^\rho$. We choose the action of $h \in A$ on $m_g^\rho$'s as,
\begin{equation}
    1 m_g^\rho = m_g^\rho 1 = m_g^\rho, \quad \sigma m_g^\pm = \pm m^{\pm}_{\sigma g}, \quad m_g^\pm \sigma = m_{g\sigma}^\pm.
\end{equation}
As one can check, for $M_I^-$ and $M_J^-$, the multiplication of $\sigma$ on left and right simultaneously does generate non-trivial action of $\doubleZ_2$ on $M_g^-$ for $g \in I, J$. 
The bimodules $M_I^\pm$ correspond to the quantum $\doubleZ_2$ symmetry in the gauged theory, while $M_J^+$ generates the unbroken $\doubleZ_2$ symmetry in $A_4$. Together they form the $\doubleZ_2\times \doubleZ_2$ symmetry in the gauged theory. $M_Q$ and $M_{\oQ}$ are identified with the triality line $\CL_Q$ and $\CL_{\oQ}$.

To conclude, we point it out there is a natural dual module $\Hom_A(M_{HgH}^\rho,A)$ for each bimodule $M_{HgH}^{\rho}$, which is isomorphic to $M_{Hg^{-1}H}^{\rho^\dagger}$.\footnote{Generically, for an $A-B$ bimodule $M$, one can either consider the dual being $\Hom_A(M,A)$ or $\Hom_B(M,B)$, which are both $B-A$ bimodules. Here, since we are considering $A-A$ bimodule, we can consider either choice and the results should be isomorphic.} Physically this corresponds to taking the orientation reversal of the TDL. 

\subsection{Fusion rules}
The fusion of the TDLs in the gauged theory corresponds to the tensor product of bimodules over the algebra $A$. Generically, let $M$ and $N$ be $A-A$ bimodules, then 
\begin{equation}
    M \otimes_A N = M\otimes N/\sim
\end{equation}
where the first tensor product is the tensor product of vector spaces and $\sim$ is the equivalence relation
\begin{equation}
    (ma)\otimes n \sim m \otimes (an), \quad a \in A, m \in M, n \in N.
\end{equation}
The $M\otimes_A N$ is naturally an $A-A$ bimodule and we can then decompose it as a direct sum over indecomposable $A-A$ bimodules.

Using this, one can compute the tensor product of bimodules explicitly and acquire the fusion rule. There is an additional rule one needs to know is that the grading of $M_g^\rho \otimes_A M_{g'}^{\rho'}$ is $gg'$.

We will not list all the calculations but provide several examples to help the reader understand the procedure.

To start, let's consider the fusion between $M_I^+$ with a generic $M_{HgH}^\rho$. The tensor product $M_I^+\otimes M_{HgH}^\rho$ has a basis $\{m_1^+\otimes m_{g'}^\rho, m_\sigma^+ \otimes m_{g'}^\rho\}_{g'\in HgH}$. After the identification with the equivalence relation, we denote the equivalence class as $m_1^+\otimes_A m_{g'}^\rho = \rho(\sigma) m_{\sigma}^+\otimes_A m_{\sigma g'}^\rho$. The resulting bimodule $M_I^+ \otimes_A M_g^\rho$ is isomorphic to $M_g^\rho$ itself where $m_1^+\otimes_A m_g^\rho \simeq m_g^\rho$. We can check the left action of $\sigma$ on $m_1^+\otimes_A m_g^\rho$ gives the correct sign for $\rho = -$:
\begin{equation}
    \sigma (m_1^+\otimes_A m_g^\rho) = (\sigma m_1^+) \otimes_A m_g^\rho = m_\sigma^+\otimes_A m_g^\rho = (m_1^+ \sigma) \otimes_A m_g^\rho = m_1^+ \otimes_A (\sigma m_g^\rho) = \rho(\sigma) m_1^+\otimes_A m_{\sigma g}^\rho, 
\end{equation}
where $\rho(g)$ will produce the desired sign when $\rho = -$. Hence, we find the fusion rule $M_I^+ \otimes_A M_{HgH}^\rho = M_{HgH}^\rho $.

As another example, let's consider the fusion between $M_Q$ and $M_Q$. There are 16 basis vectors in $M_Q\otimes M_Q$ and after modding out the equivalence relation we are left with 8 basis vectors. The grading suggests there are two copies of $M_{\oQ}$ and we take them to be
\begin{equation}\label{eq:MQMQtensorproduct}
\begin{aligned}
    & m_{(143)} \otimes_A m_{(143)} \simeq m_{(134),1}, \quad m_{(143)} \otimes_A m_{(124)} \simeq m_{(123),1}, \\
    & m_{(132)} \otimes_A m_{(143)} \simeq m_{(142),1}, \quad m_{(132)}\otimes_A m_{(124)} \simeq m_{(243),1}, \\
    & m_{(134)}\otimes_A m_{(234)} \simeq m_{(142),2}, \quad  m_{(143)} \otimes_A m_{(132)} \simeq m_{(243),2}, \\
    & m_{(132)} \otimes_A m_{(132)} \simeq m_{(123),2}, \quad m_{(132)} \otimes_A m_{(234)} \simeq m_{(134),2},
\end{aligned}
\end{equation}
where the subscript $m_{g,i}, i=1,2$ indicates which copies of $M_{\oQ}$.

As the final example, let's consider the fusion between $M_Q$ and $M_{\oQ}$. There are 16 basis vectors in $M_Q\otimes M_{\oQ}$ and after modding out the equivalence relation there are only 8 left. After rewrite them as different linear combinations, we find they generate $M_I^+ \oplus M_I^- \oplus M_J^+ \oplus M_J^-$ where
\begin{equation}\label{eq:MQMQbtensorproduct}
\begin{aligned}
    & \frac{m_{(132)}\otimes_A m_{(123)} + m_{(143)}\otimes_A m_{(134)}}{\sqrt{2}} \simeq m_1^+, \quad \frac{m_{(132)}\otimes_A m_{(134)} + m_{(143)}\otimes_A m_{(123)} }{\sqrt{2}} \simeq m_\sigma^+, \\
    & \frac{m_{(132)}\otimes_A m_{(123)} - m_{(143)}\otimes_A m_{(134)}}{\sqrt{2}} \simeq m_1^-, \quad \frac{m_{(132)}\otimes_A m_{(134)} - m_{(143)}\otimes_A m_{(123)} }{\sqrt{2}} \simeq m_\sigma^-, \\
    & \frac{m_{(132)}\otimes_A m_{(243)} + m_{(143)}\otimes_A m_{(142)}}{\sqrt{2}} \simeq m_\eta^+, \quad \frac{m_{(132)}\otimes_A m_{(142)} + m_{(143)}\otimes_A m_{(243)}}{\sqrt{2}} \simeq m_{\sigma\eta}^+, \\
    & \frac{m_{(132)}\otimes_A m_{(243)} - m_{(143)}\otimes_A m_{(142)}}{\sqrt{2}} \simeq m_\eta^-, \quad \frac{m_{(132)}\otimes_A m_{(142)} - m_{(143)}\otimes_A m_{(243)}}{\sqrt{2}} \simeq m_{\sigma\eta}^-.
\end{aligned}
\end{equation}
One can check the above identification is consistent with the action of $\sigma$. For instance,
\begin{equation}
    \sigma m_1^- \simeq \sigma \frac{m_{(132)}\otimes_A m_{(123)} - m_{(143)}\otimes_A m_{(134)}}{\sqrt{2}} = \frac{m_{(143)}\otimes_A m_{(123)} - m_{(132)} \otimes_A m_{(134)}}{\sqrt{2}} \simeq - m_\sigma^-.
\end{equation}
\begin{table}[h]
\center
\makegapedcells
    \begin{tabular}{|c|c|c|c|c|c|c|}
    \hline
    $\otimes_A$ & $M_I^+$ & $M_I^-$ & $M_J^+$ & $M_J^-$ & $M_Q$ & $M_{\oQ}$ \\
    \hline
    $M_I^+$ & $M_I^+$ & $M_I^-$ & $M_J^+$ & $M_J^-$ & $M_Q$ & $M_{\oQ}$ \\
    \hline
    $M_I^-$ & $M_I^-$ & $M_I^+$ & $M_J^-$ & $M_J^+$ & $M_Q$ & $M_{\oQ}$ \\
    \hline
    $M_J^+$ & $M_J^+$ & $M_J^-$ & $M_I^+$ & $M_I^-$ & $M_Q$ & $M_{\oQ}$ \\
    \hline
    $M_J^-$ & $M_J^-$ & $M_J^+$ & $M_I^-$ & $M_I^+$ & $M_Q$ & $M_{\oQ}$ \\
    \hline
    $M_Q$ & $M_Q$ & $M_Q$ & $M_Q$ & $M_Q$ & $M_{\oQ}\oplus M_{\oQ}$ & $M_I^+\oplus M_I^- \oplus M_J^+ \oplus N_J^-$ \\
    \hline
    $M_{\oQ}$ & $M_{\oQ}$ & $M_{\oQ}$ & $M_{\oQ}$ & $M_{\oQ}$ & $M_I^+\oplus M_I^- \oplus M_J^+ \oplus N_J^-$ & $M_{Q}\oplus M_{Q}$ \\
    \hline
\end{tabular}
\caption{Result of tensor product of $A-A$ bimodules.}
\label{tab:fusionaa}
\end{table}
The rest of the tensor products can be computed as above and we list the result in \tabref{tab:fusionaa}. With the identifications between bimodules and TDLs given previously, we reproduce the fusion rule of the triality fusion category.

\subsection{$F$-symbols}\label{sec:F_symbols_group_theoretical}
The advantage of the description using bimodules is that we can easily compute the $F$-symbols for the resulting fusion category. The local fusion of two TDLs labeled by bimodules $M,N$ into the TDL labeled by bimodule $L$ can be identified as the vector space of bimodule homomorphism from $M\otimes_A N$ to $L$:
\begin{equation}
    \Hom_{A-A}(M\otimes_A N,L).
\end{equation}
Here, a $A-A$ bimodule homomorphism $\phi \in \Hom_{A-A}(M,N)$ from $A-A$ bimodule $M$ to $A-A$ bimodule $N$ is a $\doubleC$-linear map from $M$ to $N$ such that
\begin{equation}
    a\phi(m)a' = \phi(ama'), \quad \forall a,a' \in A, m \in M.
\end{equation}
We can then consider choose a basis for each vector space $\Hom_{A-A}(M\otimes_A N,L)$ and use the Greek letter $\mu,\nu,\cdots = 1,2,\cdots$ to label the basis vectors.

For instance, let's consider the fusion $M_1^+\otimes_A M_g^\rho$ studied previously. Since $M_1^+\otimes_A M_g^\rho \simeq M_g^\rho$, there is only one fusion channel and $\Hom_{A-A}(M\otimes_A N,L)$ is a 1-dim vector space. The isomorphism
\begin{equation}
    m_1^+\otimes_A m_g^\rho \simeq m_g^\rho
\end{equation}
is a basis vector for $\Hom_{A-A}(M\otimes_A N,L)$.

As an another example, let's consider $M_Q\otimes_A M_Q \simeq M_{\oQ}\oplus M_{\oQ}$. In this case, $M_{\oQ}$ appears twice in the direct sum, therefore $\Hom_{A-A}(M_Q\otimes_A M_Q, M_{\oQ})$ would be 2-dimensional. The choice of identifications in \eqref{eq:MQMQtensorproduct} leads a basis in $\Hom_{A-A}(M_Q\otimes_A M_Q, M_{\oQ})$, which are given by two $A-A$ bimodule homomorphism $\phi_1, \phi_2$ from $M_Q\otimes_A M_Q \rightarrow M_{\oQ}$ :
\begin{equation}
    \phi_1: \begin{pmatrix} m_{(143)}\otimes_A m_{(143)} \\ m_{(143)}\otimes_A m_{(124)}  \\ m_{(132)}\otimes_A m_{(124)} \\ m_{(132)}\otimes_A m_{(143)}  \end{pmatrix} \mapsto \begin{pmatrix} m_{(134)} \\ m_{(123)} \\ m_{(243)} \\ m_{(142)} \end{pmatrix},
\end{equation}
and,
\begin{equation}
    \phi_2:\begin{pmatrix} m_{(143)}\otimes_A m_{(234)}  \\ m_{(143)}\otimes_A m_{(132)}  \\ m_{(132)}\otimes_A m_{(132)} \\ m_{(132)}\otimes_A m_{(234)}  \end{pmatrix} \mapsto \begin{pmatrix} m_{(142)} \\ m_{(243)} \\ m_{(123)} \\ m_{(134)} \end{pmatrix}.
\end{equation}
Finally, let's consider the fusion $M_Q\otimes_A M_{\oQ} \simeq M_I^+ \oplus M_I^- \oplus M_J^+ \oplus M_J^-$. Since there are 4 modules that appear in the direct sum and each only appears once, there are 4 $\Hom$ spaces and each has dimension $1$. For instance, $\Hom_{A-A}(M_Q\times_A M_{\oQ}, M_I^-)$ is the space of projection maps from $M_Q\otimes_A M_{\oQ}$ to $M_I^-$. Our identification in \eqref{eq:MQMQbtensorproduct} also determines a basis vector (which is a projection map) in $\Hom_{A-A}(M_Q\times_A M_{\oQ}, M_I^-)$, given by the following $A-A$ bimodule  homomorphism,
\begin{equation}
    \begin{pmatrix} m_{(132)}\otimes_A m_{(123)} \simeq m_{(234)}\otimes_A m_{(243)} \\ m_{(143)}\otimes_A m_{(134)} \simeq m_{(124)}\otimes_A m_{(142)} \\ m_{(143)}\otimes_A m_{(123)} \simeq m_{(124)}\otimes_A m_{(243)} \\ m_{(132)}\otimes_A m_{(134)} \simeq m_{(234)}\otimes_A m_{(142)} \end{pmatrix} \mapsto \begin{pmatrix} \frac{1}{\sqrt{2}} m_1^- \\ - \frac{1}{\sqrt{2}} m_1^- \\ - \frac{1}{\sqrt{2}} m_\sigma^- \\ \frac{1}{\sqrt{2}} m_\sigma^- \end{pmatrix}
\end{equation}
where unlisted elements are mapped to $0$.

Let's consider the $F$-symbol. The local fusions of the diagrams on both sides of the diagram give $A-A$ bimodule homomorphisms from $M\otimes_A N\otimes_A L$ to $G$ in $\Hom_{A-A}(M\otimes_A N\otimes_A L, G)$, and the $F$-symbol can be interpreted as the matrix elements of the linear transformation between two sets of elements in $\Hom_{A-A}(M\otimes_A N\otimes_A L, G)$.

With a choice of basis for each $\Hom_A(M\otimes_A N, P)$, we can acquire $F$-symbols which are the matrix elements of the above transformation. Practically, let's consider the fixed bimodule $M,N,L,G$ and choose basis vectors $\phi_{M\otimes_A N \rightarrow P,\mu} \in \Hom_A(M\otimes_A N,L)$ for every junction space $\Hom_A(M\otimes_A N,L)$, then the diagram on the left hand side leads to an element $\phi_{P\otimes_A L\rightarrow G,\nu} \circ \phi_{M\otimes_A N\rightarrow P,\mu} \in \Hom_{A-A}(M\otimes_A N\otimes_A L, G)$: 
\begin{equation}
    \phi_{P\otimes_A L\rightarrow G,\nu}\circ \phi_{M\otimes_A N\rightarrow P,\mu}: m\otimes_A n \otimes_A l \mapsto \phi_{P\otimes_A L \rightarrow G, \nu}(\phi_{M\otimes_A N, \mu}(m\otimes_A n)\otimes_A l).
\end{equation}
Similarly, $\phi_{M\otimes_A Q \rightarrow G, \mu} \circ \phi_{N\otimes_A L \rightarrow Q, \nu} \in \Hom_{A-A}(M\otimes_A N\otimes_A L,G)$ is defined as
\begin{equation}
    \phi_{M\otimes_A Q \rightarrow G, \mu} \circ \phi_{N\otimes_A L \rightarrow Q, \nu}: m \otimes_A n \otimes_A l \mapsto \phi_{M\otimes_A Q \rightarrow G, \mu}(m\otimes_A \phi_{N\otimes_A L \rightarrow Q,\nu} (n\otimes_A l) ).
\end{equation}
The $F$-symbols are just $\doubleC$-numbers such that the following equations of $A-A$ bimodule homomorphisms from $M\otimes_A N\otimes_A L$ to $G$ hold
\begin{equation}
    \phi_{P\otimes_A L\rightarrow G,\nu} \circ \phi_{M\otimes_A N\rightarrow P,\mu} = \sum_{Q,\alpha,\beta} \left[F^{MNL}_G\right]_{(P,\mu,\nu),(Q,\alpha,\beta)} \phi_{M\otimes_A Q \rightarrow G,\beta} \circ \phi_{N\otimes_A L \rightarrow Q, \alpha}.
\end{equation}
To solve the above equation, we only need to evaluate the $A$-homomorphism on basis vectors of $M\otimes_A N\otimes_A L \rightarrow G$, which will produce a set of linear equations and can be solved quite easily.

As an example, let's consider taking $M = N = Q = M_Q, L = M_1^-, P = G = M_{\oQ}$. This would compute the $F$-symbol $\left[F^{\llq \llq \hat{\sigma}}_{\llqb}\right]_{(\llqb,\mu,1),(\llq,1,\beta)}$, that is, we want to solve,
\begin{equation}
\begin{aligned}
    & \phi_{M_{\oQ}\otimes_A M_1^-\rightarrow M_{\oQ}, 1} \circ \phi_{M_Q\otimes_A M_Q\rightarrow M_{\oQ},\mu} \\
    = & \sum_{\beta=1,2}\left[F^{\llq \llq \hat{\sigma}}_{\llqb}\right]_{(\llqb,\mu,1),(\llq,1,\beta)} \phi_{M_{Q}\otimes_A M_{Q} \rightarrow M_{\oQ}, \beta} \circ \phi_{M_Q\otimes_A M_1^- \rightarrow M_Q, 1}.
\end{aligned}
\end{equation}
We consider evaluate the above equation on $m_{(143)}\otimes_A m_{(143)} \otimes_A m_1^-$ and $m_{(132)}\otimes_A m_{(132)} \otimes_A m_1^-$. As an example, we do this for $m_{(143)}\otimes_A m_{(143)} \otimes_A m_1^-$. On the left-hand side, we have,
\begin{equation}
\begin{aligned}
    & \phi_{M_{\oQ}\otimes_A M_1^-\rightarrow M_{\oQ}, 1}\circ \phi_{M_Q\otimes_A M_Q\rightarrow M_{\oQ},\mu}(m_{(143)}\otimes_A m_{(143)} \otimes_A m_1^-) \\
    = & \phi_{M_{\oQ}\otimes_A M_1^-\rightarrow M_{\oQ},1}(\phi_{M_Q\otimes_A M_Q\rightarrow M_{\oQ},\mu}(m_{(143)}\otimes_A m_{(143)})\otimes_A m_1^-) \\
    = & \phi_{M_{\oQ}\otimes_A M_1^-\rightarrow M_{\oQ}, 1}(\delta_{\mu,1} m_{(134)}\otimes_A m_1^-) \\
    = & \delta_{\mu,1} m_{(134)}, 
\end{aligned}
\end{equation}
and on the right-hand side, we have,
\begin{equation}
\begin{aligned}
    & \phi_{M_{Q}\otimes_A M_{Q} \rightarrow M_{\oQ}, \beta} \circ \phi_{M_Q\otimes_A M_1^- \rightarrow M_Q, 1}(m_{(143)}\otimes_A m_{(143)} \otimes_A m_1^-) \\
    = & \phi_{M_{Q}\otimes_A M_{Q} \rightarrow M_{\oQ}, \beta}(m_{(143)}\otimes_A\phi_{M_Q\otimes_A M_1^-\rightarrow M_Q,1}(m_{(143)}\otimes_A m_1^-)) \\
    = & \phi_{M_{Q}\otimes_A M_{Q} \rightarrow M_{\oQ}, \beta}(m_{(143)}\otimes_A m_{(143)}) \\
    = & \delta_{\beta,1} m_{(134)}.
\end{aligned}
\end{equation}
Hence, we get the following 2 linear equations,
\begin{equation}
    1 = \left[F^{\llq\llq\hat{\sigma}}_{\llqb}\right]_{(\llqb,1,1),(\llq,1,1)}, \quad 0 = \left[F^{\llq\llq\hat{\sigma}}_{\llqb}\right]_{(\llqb,2,1),(\llq,1,1)}.
\end{equation}
Similarly, we get another two equations when evaluating on $m_{(132)}\otimes_A m_{(132)}\otimes_A m_1^-$:
\begin{equation}
    0 = \left[F^{\llq\llq\hat{\sigma}}_{\llqb}\right]_{(\llqb,1,1),(\llq,1,2)}, \quad -1 = \left[F^{\llq\llq\hat{\sigma}}_{\llqb}\right]_{(\llqb,2,1),(\llq,1,2)}.
\end{equation}
Solving the four equations, we found,
\begin{equation}
    \left[F^{\llq\llq\hat{\sigma}}_{\llqb}\right]_{(\llqb,\mu,1),(\llq,1,\beta)} = \begin{pmatrix} 1 & 0 \\ 0 & -1\end{pmatrix} = \sigma^3.
\end{equation}
One can solve the rest of the $F$-symbols in a similar way. 

Just as the Ising fusion category containing duality line $N$ is determined up to the FS indicator $\epsilon \in H^2(\doubleZ_2, U(1))$, the fusion category containing triality defect also has a FS indicator $\alpha \in H^3(\doubleZ_3,U(1))$. From the $F$-symbol point of view, given a solution of $F$-symbols for the pentagon equations of the triality fusion categories, one can generate a new set of solutions \cite{teo2015theory} by
\begin{equation}
\begin{aligned}
    &F^{\llq\llqb \llq}_{\llq} \rightarrow e^{2\pi m\ii/3} F^{\llq\llqb \llq}_{\llq}, \quad F^{\llqb\llq \llqb}_{\llqb} \rightarrow e^{-2\pi m\ii/3}F^{\llqb\llq \llqb}_{\llqb}, \\ & F^{\llq \llqb \llqb}_{\llqb} \rightarrow e^{-2\pi m\ii/3} F^{\llq \llqb \llqb}_{\llqb}, \quad  F^{\llqb \llq \llq}_{\llq} \rightarrow e^{2\pi m \ii/3} F^{\llqb \llq \llq}_{\llq}, \\
    & F^{\llqb \llqb \llq}_{\llqb} \rightarrow e^{2\pi m\ii/3}F^{\llqb \llqb \llq}_{\llqb},  \quad F^{\llqb \llqb \llqb}_{g} \rightarrow e^{-2\pi m \ii/3} F_g^{\llqb \llqb \llqb}.
\end{aligned}
\end{equation}
Alternatively, given a triality fusion category, we can construct the triality fusion category with different FS indicator by stacking the theory with another theory with anomalous $\doubleZ_3$ symmetry $\tilde{\eta}$ and identify the new triality line as $\widetilde{\llq} = \llq \tilde{\eta}$. If we gauge the quantum $\doubleZ_2$ symmetry, then we would recover the theory with $\widetilde{A_4}$ symmetry but now the $\widetilde{A_4}$ symmetry has an anomaly due to the anomaly of $\doubleZ_3$. This implies the triality fusion category with different FS indicators can be realized by $\doubleZ_2$ gauging of the $A_4$ global symmetry with different anomalies. Indeed, $H^3(A_4,U(1)) = \doubleZ_6$ and let $\omega_0$ denote the generator of $H^3(A_4,U(1))$. The $\doubleZ_2\times \doubleZ_2$ subgroup of $A_4$ is anomaly free only when the anomaly of $A_4$ is $\omega_0^{2k}$ for $k = 0,1,2$ (where we use the multiplicative notation for $\doubleZ_6$), and gauging one of the $\doubleZ_2$ subgroup leads to the triality fusion category with FS indicator $\alpha = e^{2\pi k \ii/3}$. \footnote{Notice that $k = 0$ (i.e. the anomaly of $A_4$ is trivial) always leads to the trivial FS indicator $\alpha =1$. We choose the generator $\omega_0$ such that $\scriptC(A_4,\omega_0^{2k},\doubleZ_2^\sigma,1)$ has FS indicator $\alpha = e^{2\pi \ii k/3}$.}

Let $\alpha = e^{2\pi \ii/3}\in U(1)$, and we list the $F$-symbols below.
\begin{equation}
\begin{aligned}
    & F^{\llq g, h}_{\llq} = F^{\llqb\llq g}_{h} = \begin{pmatrix} 1 & 1 & 1 & 1 \\ 1 & 1 & 1 & 1 \\ 1 & -1 & 1 & -1 \\ 1 & -1 & 1 & -1 \end{pmatrix}, \quad F^{g \llq h}_{\llq} = F^{g \llqb h}_{\llqb} = F^{\llqb g \llq}_{h} = \begin{pmatrix} 1 & 1 & 1 & 1 \\ 1 & 1 & -1 & -1 \\ 1 & -1 & 1 & -1 \\ 1& -1 & -1 &1\end{pmatrix}, \\
    & F^{g,h,\llq}_{\llq} = F^{\llqb,g,h}_{\llqb} = \begin{pmatrix} 1 & 1 & 1 & 1 \\ 1 & 1 & 1 & 1\\ 1 & 1 & 1 & 1 \\ 1 & 1 & 1 & 1 \end{pmatrix}, \quad F^{\llq \llqb g}_h = F^{g\llq \llqb}_h =  \begin{pmatrix} 1 & 1 & 1 & 1\\ -1 & -1 &-1 &-1 \\ 1 &1 &1 &1 \\ -1 &-1 &-1 &-1 \end{pmatrix}, \\
    & F^{g,h,\llqb}_{\llqb} = \begin{pmatrix} 1 & 1 & 1 & 1 \\ 1 & 1 & -1 & -1 \\ 1 & 1 & 1 & 1 \\ 1 & 1 & -1 & -1 \end{pmatrix}, F^{g\llqb \llq}_h = \begin{pmatrix} 1&1&1 &1 \\ 1&1&-1&-1 \\ 1& 1& 1&1 \\-1&-1&1 &1 \end{pmatrix}, F^{\llq g\llqb}_h = \begin{pmatrix} 1 & 1 & 1 &1 \\ 1& 1 &-1&-1 \\ 1&-1 &1&-1 \\-1&1 &1 &-1 \end{pmatrix}, \\
    & \left[F^{\llq \llqb \llq}_{\llq}\right]_{g,h} = \frac{\alpha}{2} \begin{pmatrix} 1 & 1 & 1 & -1 \\ -1 & -1 & 1 & -1 \\ 1 & -1 & 1 & 1 \\ -1 & 1 & 1 & 1 \end{pmatrix}, \quad \left[F^{\llqb \llq \llqb}_{\llqb}\right]_{g,h} = \frac{\alpha^{-1}}{2} \begin{pmatrix} 1 & -1 & 1 & -1 \\ 1 & -1 & -1 & 1 \\ 1 & 1 & 1 & 1 \\ 1 & 1 & -1 & -1 \end{pmatrix}.
\end{aligned}
\end{equation}
The rest of the $F$-symbols are listed in Table \ref{table:group_theoretical_F_Symbols}.
\begin{table}[h]
\center
\makegapedcells
    \begin{tabular}{|c|c|c|c|c|}
    \hline
    $g$ & $1$ & $\hat{\sigma}$ & $\eta$ & $\eta\hat{\sigma}$ \\
    \hline
    $[F^{\llq \llq g}_{\llqb}]_{(\llqb,\mu,1),(\llq,1,\nu)}$ & $\sigma^0$ & $\sigma^3$ & $\sigma^1$ & $\ii\sigma^2$ \\
    \hline
    $[F^{\llq g \llq}_{\llqb}]_{(\llq,1,\mu),(\llq,1,\nu)}$ & $\sigma^0$ & $\sigma^3$ & $\sigma^1$ & $-\ii\sigma^2$ \\
    \hline
    $[F^{g \llq \llq}_{\llqb}]_{(\llq,1,\mu),(\llqb,\nu,1)}$ & $\sigma^0$ & $\sigma^3$ & $\sigma^1$ & $-\sigma^2$ \\
    \hline
    $[F^{\llqb\llqb g}_{\llq}]_{(\llq,\mu,1),(\llqb,1,\nu)}$ & $\sigma^0$ & $\sigma^3$ & $\sigma^1$ & $\ii\sigma^2$ \\
    \hline
    $[F^{\llqb g \llqb}_{\llq}]_{(\llqb,1,\mu),(\llqb,1,\nu)}$ & $\sigma^0$ & $\sigma^3$ & $\sigma^1$ & $\ii\sigma^2$ \\
    \hline
    $[F^{g\llqb\llqb}_{\llq}]_{(\llqb,1,\mu),(\llq,\nu,1)}$ & $\sigma^0$ & $\sigma^3$ & $\sigma^1$ & $-\ii\sigma^2$\\
    \hline 
    $[F^{\llq\llq\llq}_g]_{(\llqb,\mu,1),(\llqb,\nu,1)}$ & $\sigma^0$ & $-\sigma^3$ & $\sigma^1$ & $\ii\sigma^2$ \\
    \hline
    $[F^{\llqb\llqb\llqb}_g]_{(\llq,\mu,1),(\llq,\nu,1)}$ & $\alpha^{-1}\sigma^0$ & $-\alpha^{-1}\sigma^3$ & $\alpha^{-1}\sigma^1$ & $\ii\alpha^{-1} \sigma^2$ \\ 
    \hline
    $[F^{\llq \llq \llqb}_{\llq}]_{(\llq,\mu,\nu),(g,1,1)}$ & $\frac{1}{\sqrt{2}}\sigma^0$ & $  \frac{-1}{\sqrt{2}}\sigma^3$ & $ \frac{1}{\sqrt{2}}\sigma^1$ & $\frac{-\ii}{\sqrt{2}}\sigma^2$ \\
    \hline
    $[F^{\llqb\llqb\llq}_{\llqb}]_{(\llq,\mu,\nu),(g,1,1)}$ & $\frac{\alpha}{\sqrt{2}}\sigma^0$ & $\frac{\alpha}{\sqrt{2}}\sigma^3$ & $\frac{\alpha}{\sqrt{2}}\sigma^1$ & $\frac{-\ii\alpha}{\sqrt{2}}\sigma^2$ \\
    \hline
    $[F^{\llqb\llq\llq}_{\llq}]_{(g,1,1),(\llqb,\mu,\nu)}$ & $\frac{\alpha}{\sqrt{2}}\sigma^0$ & $\frac{\alpha}{\sqrt{2}}\sigma^3$ & $\frac{\alpha}{\sqrt{2}}\sigma^1$ & $\frac{\ii\alpha}{\sqrt{2}}\sigma^2$ \\
    \hline
    $[F^{\llq\llqb\llqb}_{\llqb}]_{(g,1,1),(\llq,\mu,\nu)}$ & $\frac{\alpha^{-1}}{\sqrt{2}}\sigma^0$ & $\frac{-\alpha^{-1}}{\sqrt{2}}\sigma^3$ & $\frac{\alpha^{-1}}{\sqrt{2}}\sigma^1$ & $\frac{-\ii\alpha^{-1}}{\sqrt{2}}\sigma^2$ \\
    \hline
\end{tabular}
\caption{The $F$-symbols of group theoretical triality fusion categories $\scriptC(A_4,\omega_0^{2k},\doubleZ_2^\sigma,1)$ for fusion multiplicity $2$ where $\alpha = e^{2\pi k \ii/3}$. The $\sigma^i$ denotes the Pauli $i$ matrix and $\sigma^0$ is the $2\times 2$ identity matrix.}
\label{table:group_theoretical_F_Symbols}
\end{table}

\section{Physical implication of the group theoretical triality fusion categories}\label{sec:Physical_Implications}
In this section, we derive the physical implication of the group theoretical triality fusion categories $\scriptC(A_4,\omega_0^{2k},\doubleZ_2^\sigma, 1)$. We first derive the spin selection rules for the triality defect. Then we show how to match the states in the Hilbert space which transforms in different irreducible representations of the fusion category symmetry $\scriptC(A_4,\omega_0^{2k},\doubleZ_2^\sigma, 1)$ of the theory $\scriptT/\doubleZ_2^{\sigma}$ with the states in the Hilbert space $\scriptH$ or the defect Hilbert space $\scriptH_\sigma$ of the theory $\scriptT$. This allows us to derive the asymptotic density of states in $\scriptH$ transforms in different irreducible representations of $\scriptC(A_4,\omega_0^{2k},\doubleZ_2^\sigma, 1)$ by applying the result in \cite{Pal:2020wwd}. Finally, we study the anomaly of group theoretical triality fusion categories by analyzing its symmetric gapped phases and describe the anomaly free conditions for generic group theoretical fusion categories. 

\subsection{Spin selection rules}\label{sec:group_theoretical_spin_selection_rules}
We now derive the spin selection rules for using the $F$-symbols computed in section \ref{sec:F_symbols_group_theoretical}.

\begin{figure}[H]
    \centering
    \includegraphics{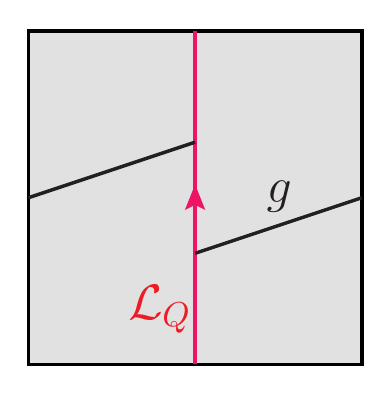}
    \caption{Symmetry operator $g\in \doubleZ_2 \times \doubleZ_2$ acts on the defect Hilbert space $\scriptH_{\llq}$, which we denote as $\hat{g}_{\llq}$.}
    \label{fig:symmetryondefectH}
\end{figure}

\begin{figure}[H]
    \centering
    \includegraphics[scale = 0.75]{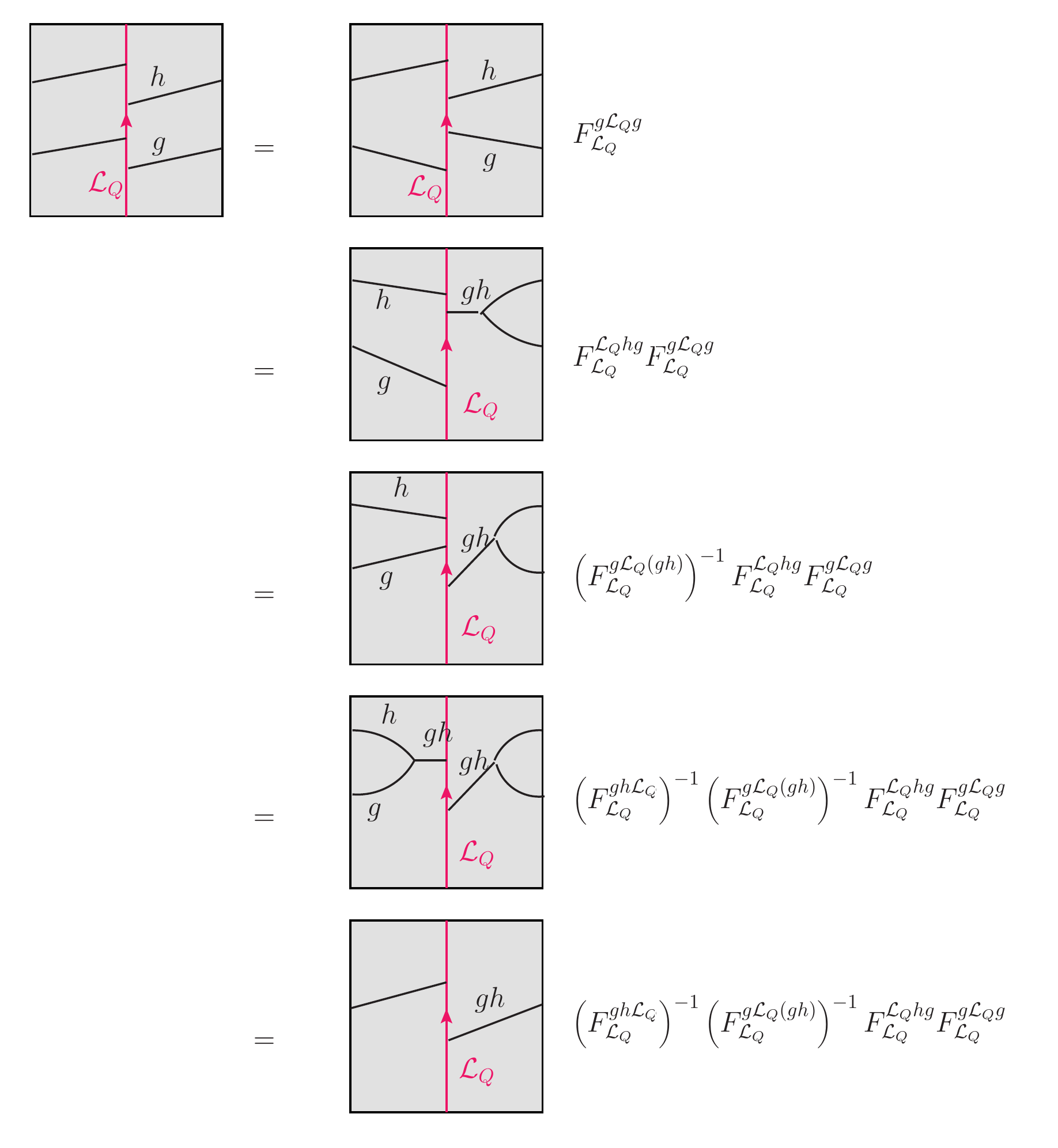}
    \caption{The calculation of the product of $\hat{g}_{\CL_Q}$ and $\hat{h}_{\CL_Q}$ using $F$-moves.}
    \label{fig:central_extension}
\end{figure}

To do this, we first consider the action of $g \in \doubleZ_2\times \doubleZ_2$ on the defect Hilbert space $\scriptH_{\llq}$. Following in the convention in \eqref{eq:general_twisted_partition_function} and suppressing the $1,1$ indices for the fusion channel (since the multiplicity is just $1$ in this case), we denote the operator as $\hat{g}_{\llq}$. This is depicted in Figure \ref{fig:symmetryondefectH}.
Notice that the action of $\doubleZ_2\times \doubleZ_2$ on $\scriptH_{\llq}$ can be twisted by 2-cocycle $\gamma(g,h) \in Z^2(\doubleZ_2\times \doubleZ_2,U(1))$ such that
\begin{equation}
    \hat{h}_{\llq}\cdot \hat{g}_{\llq} = \gamma(h,g)\widehat{hg}_{\llq}.
\end{equation}
We can compute $\gamma$ from the $F$-symbols via following configuration shown in the Figure \ref{fig:central_extension}.
Under a sequence of $F$-move, we relate the phase $\gamma(g,h)$ to products of $F$-symbols as following 
\begin{equation}
    \gamma(h,g) = \left(F^{gh\llq}_{\llq}\right)^{-1}\left(F^{g \llq (gh)}_{\llq}\right)^{-1} F^{\llq hg}_{\llq} F^{g\llq g}_{\llq} = \begin{pmatrix}1 & 1 & 1 & 1 \\ 1 & 1 & -1 & -1 \\ 1 & -1 & 1 & -1 \\ 1 & -1 & -1 & 1\end{pmatrix}.
\end{equation}
We list the eigenvalues of allowed irreducible representations as the following
\begin{equation}\label{eq:proj1}
    (1,1,1,-1),\quad (1,1,-1,1), \quad (1,-1,1,1), \quad (1,-1,-1,-1).
\end{equation}
Notice that the 2-cocycle $\gamma$ above is cohomologically trivial in the group cohomology, which is consistent with the fact that we have $4$ 1-dimensional irreducible representations.\footnote{$\doubleZ_2\times\doubleZ_2$ only has a single 2-dimensional irreducible representation when twisted by the cohomologically non-trivial 2-cocycle.} Yet we will see its importance when deriving the spin selection rule for the intrinsic triality defects in later sections. 

\begin{center}
\begin{figure}[H]
    \centering
    \includegraphics[scale = 0.75]{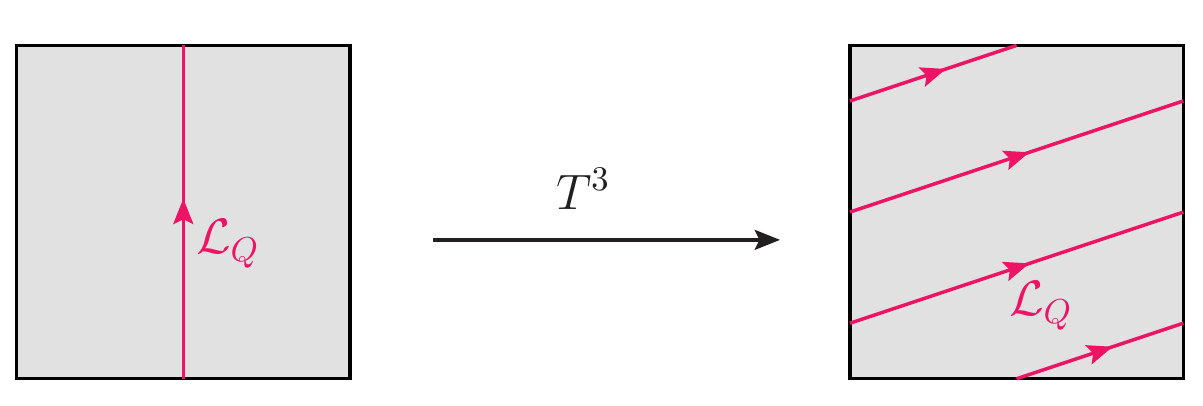}
    \caption{Applying $T^3$ to the twisted partition function $Z_{\llq}(\tau)$. This is equivalent to insert $e^{6\pi i s}$ in the trace over $\mathcal{H}_{\llq}$.}
    \label{fig:triplespin}
\end{figure}
\end{center}

\begin{center}
\begin{figure}
    \centering
    \includegraphics[scale = 0.75]{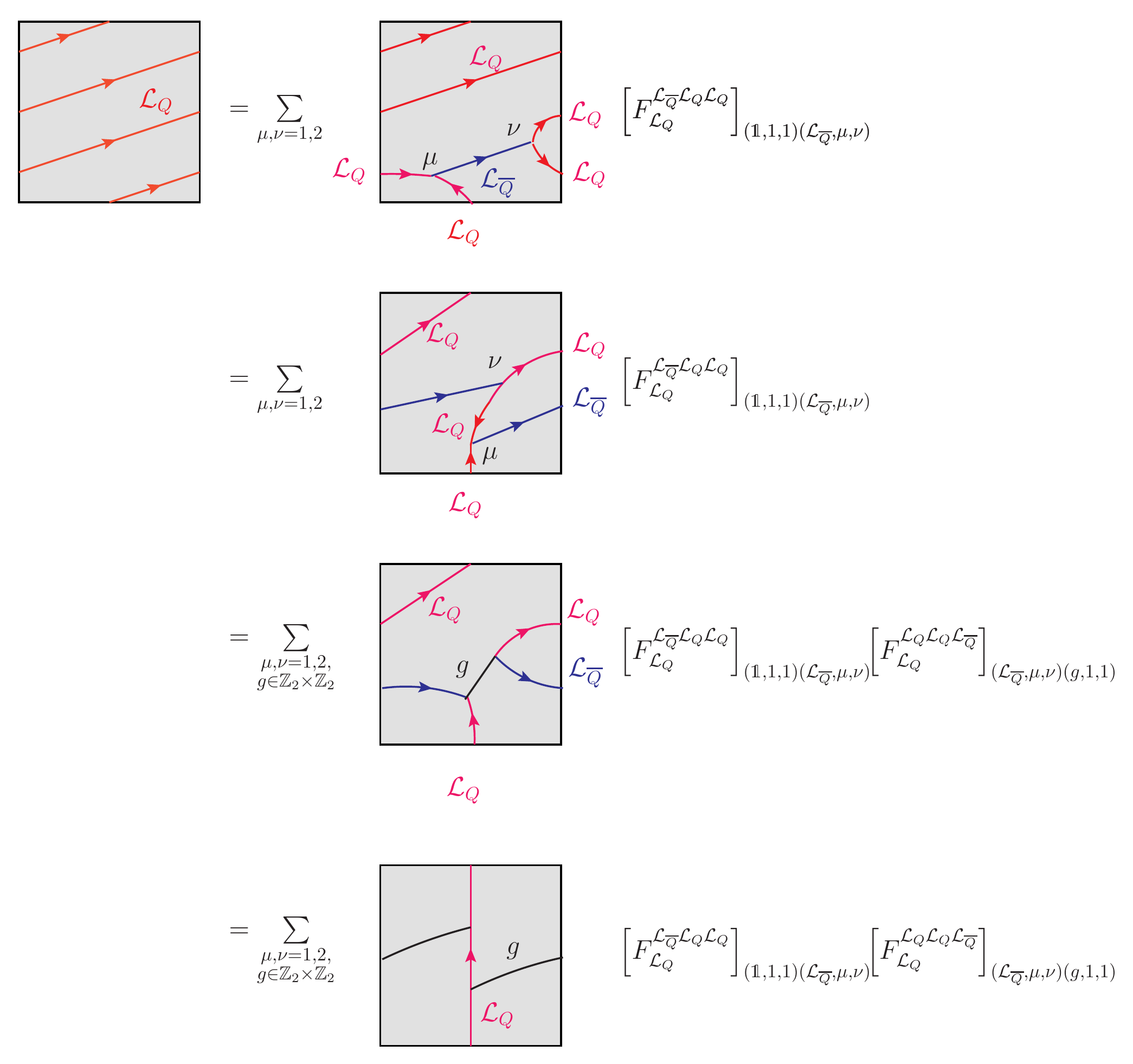}
    \caption{Here, we start with the configuration which computes the $e^{6\pi is}$. Under sequence of $F$-moves, we relate it to the action of $g \in \mathbb{Z}_2\times \mathbb{Z}_2$ on the defect Hilbert space $\scriptH_{\llq}$.}
    \label{fig:spinselection}
\end{figure}
\end{center}

Next, to derive the spin selection rules, we consider twisted partition function $Z_{\llq}(\tau)$ and apply $T$ modular transformation three times, see Figure \ref{fig:triplespin}. This amounts to inserting $e^{6\pi is}$ in the trace over the defect Hilbert space $\scriptH_{\llq}$ where $s$ is the spin of the state. 
We then apply a sequence of $F$-moves to relate $Z_{\llq}(\tau+3)$ to the action of the symmetry $g\in \doubleZ_2 \times \doubleZ_2$ on the defect Hilbert space $\scriptH_{\llq}$, as shown in Figure \ref{fig:spinselection}. We find the following relation:
\begin{equation}
\begin{aligned}
    Z_{\llq}(\tau+3) &= \sum_{\substack{\mu,\nu = 1,2, \\ g \in \doubleZ_2\times \doubleZ_2}} \left[F^{\llqb \llq \llq}_{\llq}\right]_{(\dsi,0,0)(\llq,\mu,\nu)}\left[F^{\llq \llq \llqb}_{\llq}\right]_{(\llqb,\mu,\nu)(g,0,0)}Z_{\llq g}^{\llq}(\tau) \\ 
    &= \alpha Z^{\llq}_{\llq \dsi}(\tau).
\end{aligned}
\end{equation}
This implies the spin $s$ of the states in defect Hilbert space $\scriptH_{\CL_Q}$ satisfies the following relation:
\begin{equation}
    e^{6\pi \ii s} = \alpha,
\end{equation}
which implies,
\begin{equation}\label{eq:spinselectionrule1}
    e^{2\pi \ii s } = \begin{cases} &e^{\frac{2\pi \ii k}{3}}, \quad k = 0,1,2, \quad \text{when} \quad \alpha = 1 \\
    & e^{\frac{2\pi \ii k}{3} + \frac{2\pi \ii}{9}}, \quad k = 0,1,2, \quad \text{when} \quad \alpha = e^{2\pi \ii/3}, \\
    & e^{\frac{2\pi \ii k}{3} - \frac{2\pi \ii}{9}}, \quad k = 0,1,2, \quad \text{when} \quad \alpha = e^{-2\pi \ii/3}.
    \end{cases}
\end{equation}
Now, we provide an alternative derivation of the same spin selection rule by constructing the twisted partition function from the ungauged theory. Consider a CFT $\scriptT$ with $A_4$ global symmetry with the anomaly parameterized by $\omega_0^{2k} \in H^3(A_4,U(1)) \simeq \doubleZ_6$ where $k = 0,1,2$. The $\doubleZ_2^\sigma \times \doubleZ_2^\eta$ is free of anomaly but the $\doubleZ_3$ subgroup generated by $q$ has 't Hooft anomaly. As pointed out before, by gauging a $\doubleZ_2$ symmetry, we get the non-intrinsic triality fusion category with FS indicator $\alpha = e^{2\pi \ii k/3}$. By the fusion rule of the triality defect $\llq$,
\begin{equation}
    \llq \times \llq \times \llq = \llqb \times \llqb \times \llqb = 2 \sum_{g\in \doubleZ_2\times\doubleZ_2} g.
\end{equation}
In the gauged theory $\scriptT/\doubleZ_2$, the twisted sector is odd under the quantum $\doubleZ_2$-symmetry, hence the entire twisted sector is annihilated by the triality defects $\llq$ or $\llqb$. Then, we can construct the twisted partition function $Z^{\llq}_{1,\llqb}(\tau)$ of the gauged theory from the twisted partition function of the original theory as follows,
\begin{equation}
\begin{aligned}
    & (Z_{\scriptT/\doubleZ_2})^{\llq}_{1,\llqb}(\tau) \\
    = & \text{Tr}_{\CH_{\scriptT/\doubleZ_2}}(\mathcal{L}_Q q^{L_0 - 1/24}\oq^{\oL_0-1/24}) \\ = & \text{Tr}_{\CH_{\scriptT/\doubleZ_2,\text{untwisted}}}(\CL_Q q^{L_0 - 1/24} \oq^{\oL_0 - 1/24}) \\ 
    = & \text{Tr}_{\CH_{\scriptT}}((q+\sigma q\sigma)\frac{1+\sigma}{2}q^{L_0 - 1/24}\oq^{\oL_0 - 1/24}) \\ 
    = & \frac{(Z_{\scriptT})^q(\tau) + (Z_{\scriptT})^{\sigma q}(\tau) + (Z_{\scriptT})^{q\sigma}(\tau) + (Z_{\scriptT})^{\sigma q \sigma}(\tau)}{2} \\
    = & (Z_{\scriptT})^q(\tau) + (Z_{\scriptT})^{\sigma q}(\tau),
\end{aligned}
\end{equation}
where we used the above twisted partition function only depending on the conjugacy class. Applying the $S$-modular transformation on both sides, we find
\begin{equation}
    (Z_{\scriptT/\doubleZ_2})^{\llq}_{\llq,1}(\tau) = (Z_{\scriptT})_q(\tau) + (Z_{\scriptT})_{\sigma q}(\tau).
\end{equation}
Then, the spin selection rules of the triality defect $\llq$ are the same as the symmetry defect $q$ and $\sigma q$ which generates $\doubleZ_3$ symmetries in $A_4$. The spin selection rules of $\doubleZ_3$-symmetry defect has been derived in \cite{Chang:2018iay}, which takes the form
\begin{equation}\label{eq:gp_spin_selection_rule2}
    s \in \frac{1}{3}\doubleZ + \frac{k}{9}.
\end{equation}
We then find agreement between \eqref{eq:spinselectionrule1} and \eqref{eq:gp_spin_selection_rule2}.

\subsection{Asymptotic density of states}\label{sec:asymptotic_density_of_states}
In this subsection, we derive several asymptotic density of states for different sectors in the Hilbert space $\scriptH$ for a theory with the group theoretical triality category symmetries.

When acting on the Hilbert space $\scriptH$, the 6 simple lines in the triality fusion category $I,\hat{\sigma}, \eta, \hat{\sigma}\eta, \llq, \llqb$ correspond to 6 operators whose product satisfies the fusion rule \eqref{eq:triality_fusion_rule}, from which we learn they all commute with each other, hence can be simultaneously diagonalized. We enumerate eigenvalues for all $6$ possible 1-dim irreducible representations as follows,
\begin{equation}
\begin{matrix}
    (1,1,1,1,2,2), & (1,1,1,1,2\omega,2\omega^2), & (1,1,1,1,2\omega^2,2\omega), \\
    (1,1,-1,-1,0,0), & (1,-1,1,-1,0,0), & (1,-1,1,-1,0,0).
\end{matrix}
\end{equation}
We now derive a formula of the asymptotic density of states for all the $6$ irreps by relating the above 6 irreps to representations of finite group symmetries in the ungauged theory and utilizing the result in \cite{Pal:2020wwd}, which we reviewed in section \ref{sec:review_asymptotic_density_of_states}. Notice that since the asymptotic density of states for different irreps of $\scriptH$ only depends on the fusion ring structure of the fusion category, we can derive the asymptotic density of states from the simplest case when the fusion category is acquired from gauging $\doubleZ_2^\sigma$ subgroup of $A_4$ with the trivial anomaly. Then, by relating the above $6$ irreps to the representation of the $A_4$ in the ungauged theory, we can use the result in \cite{Pal:2020wwd} to get the result. 
Let's first consider the twisted sector in the gauged theory $\scriptT/\doubleZ_2^{\sigma}$, which is odd under the quantum symmetry $\hat{\sigma}$. In the ungauged theory $\scriptT$, these states correspond to the $\doubleZ_2^{\sigma}$-even states in the defect Hilbert space $\scriptH_{\scriptT,\sigma}$. Since there's no mixed 't Hooft anomaly between $\doubleZ_2^\sigma$ and $\doubleZ_2^\eta$, there are well-defined $\doubleZ_2^\eta$ charges for states in the defect Hilbert space $\scriptH_{\scriptT,\sigma}$, which are the $\doubleZ_2^\eta$ charges for states in the gauged theory. 

Next, let's consider the untwisted sector from the $\doubleZ_2^\sigma$ gauging. This sector is given by the $\doubleZ_2^\sigma$ even states in the ungauged theory with the global $A_4$ symmetry. Here, we can use the fact that the triality line acts on the untwisted sector as $q + \sigma q\sigma$. We will relate the eigenvalues of $q + \sigma q\sigma$ of the states in the $\doubleZ_2^\sigma$-invariant space in each irreducible representation of $A_4$. This group has $4$ irreducible representation, labelled by the dimension $\mathbf{1}, \mathbf{1}_A, \mathbf{1}_B, \mathbf{3}$, where $\mathbf{1}$ is the trivial irrep and $\mathbf{1}_A$ and $\mathbf{1}_B$ are two irreps where $\sigma,\eta$ acts as trivially and $q$ has eigenvalues $\omega$ and $\omega^2$ respectively, which means the states in these two irreps are invariant under the $\doubleZ_2^\sigma$ and have eigenvalues $2\omega$ and $2\omega^2$ under $\llq \simeq q + \sigma q \sigma$ respectively. For the three-dimensional representation $\mathbf{3}$, the representation matrices are given by
\begin{equation}
    U_{\mathbf{3}}(\sigma) = \begin{pmatrix} 1 & 0 & 0 \\ 0 & -1 & 0 \\ 0 & 0 & -1\end{pmatrix}, \quad U_{\mathbf{3}}(\eta) = \begin{pmatrix} -1 & 0 & 0 \\ 0 & 1 & 0 \\ 0 & 0 & -1\end{pmatrix}, \quad U_{\mathbf{3}}(q) = \begin{pmatrix} 0 & 1 & 0 \\ 0 & 0 & 1 \\ 1 & 0 & 0 \end{pmatrix}.
\end{equation}
As one can check, $\doubleZ_2^\sigma$-invariant space is $1$-dimensional, and the operator $\llq \simeq q + \sigma q \sigma$ annihilates this state. 

We summarize the results in the \tabref{tab:spectrum_match}. 
\begin{table}[H]
    \centering
    \begin{tabular}{|c|c|}
        \hline
        Irrep in $\scriptH_{\scriptT/\doubleZ_2^\sigma}$ in the gauged theory $\scriptT/\doubleZ_2$ & Corresponding states in the ungauged theory $\scriptT$  \\
        \hline
        $(1,1,1,1,2,2)$ & the irrep $\mathbf{1}$ of $A_4$ in  $\scriptH_{\scriptT}$ \\
        \hline
        $(1,1,1,1,2\omega,2\omega)$ & the irrep $\mathbf{1}_A$ of $A_4$ in  $\scriptH_{\scriptT}$ \\
        \hline
        $(1,1,1,1,2\omega^2,2\omega^2)$ & the irrep $\mathbf{1}_B$ of $A_4$ in  $\scriptH_{\scriptT}$ \\
        \hline
        $(1,1,-1,-1,0,0)$ & the $\doubleZ_2^\sigma$-even state in the $\mathbf{3}$ of $A_4$ in $\scriptH_{\scriptT}$ \\
        \hline
        $(1,-1,1,-1,0,0)$ & $\doubleZ_2^\eta$-even, $\doubleZ_2^\sigma$-even states in $\scriptH_{\scriptT,\sigma}$ \\
        \hline
        $(1,-1,-1,1,0,0)$ & $\doubleZ_2^\eta$-odd, $\doubleZ_2^\sigma$-even states in $\scriptH_{\scriptT,\sigma}$ \\
        \hline
    \end{tabular}
    \caption{Relating the states in different irreps of triality fusion category $\scriptC(A_4,1,\doubleZ_2^\sigma,1)$ in $\scriptT/\doubleZ_2^\sigma$ to states in the ungauged theory $\scriptT$.}
    \label{tab:spectrum_match}
\end{table}

The last $3$ irreps of $\scriptC(A_4,1,\doubleZ_2^\sigma,1)$ corresponds to the $3$ irreps of $\doubleZ_2^{\hat{\sigma}}\times \doubleZ_2^{\eta}$ in the gauged theory $\scriptT/\doubleZ_2^\sigma$, therefore, we can directly apply the result in \cite{Pal:2020wwd} in the gauged theory and find the asymptotic density of states to be
\begin{equation}
    \rho_{0,(1,1,-1,-1,0,0)}(\Delta) = \rho_{0,(1,-1,1,-1,0,0)}(\Delta) = \rho_{0,(1,-1,-1,1,0,0)}(\Delta) = \frac{1}{4}\rho_0(\Delta),
\end{equation}
where $\rho_0(\Delta)$ is defined in \eqref{eq:asym_dos}.
To determine the asymptotic density of states in irreps $(1,1,1,1,2\omega^k,2\omega^{-k})$ we can simply use the relation in Table \ref{tab:spectrum_match} and apply the result in \cite{Pal:2020wwd}. We then find
\begin{equation}
    \rho_{0,(1,1,1,1,2\omega^k,2\omega^{-k})}(\Delta) = \frac{1}{12}\rho_0(\Delta), \quad k = 0,1,2.
\end{equation}

\subsection{Constraints on RG flow}\label{sec:constraints_on_RG_flows}
To study the constraints on the RG flow, we want to determine if the fusion category symmetry $\scriptC$ is anomalous, in the sense that if it obstructs a $\scriptC$-symmetric trivial gapped phase. \footnote{Here, by $\scriptC$-symmetric gapped phase we mean the Hamiltonian or Lagrangian preserves the $\scriptC$-symmetry while it could be spontaneously broken etc. By $\scriptC$-symmetric trivial gapped phase, we mean there's a unique ground state in this phase.} As pointed out in \cite{Thorngren:2019iar}, module categories $\scriptM$ over $\scriptC$ are in bijection with $\scriptC$-symmetric gapped phases such that the ground states are in bijection with the simple objects in $\scriptM$. Therefore, to check whether a fusion category symmetry $\scriptC$ has a trivial gapped phase is to check if it has a module category $\scriptM$ with a single simple object. Equivalently, one can check if $\scriptC$ admits a fiber functor.

In general, this is not an easy problem. For the case of the group-theoretic fusion category, this is relatively easy because all the indecomposable module categories over $\scriptC(G,\omega,H,\psi)$ can be explicitly constructed as pointed out in \cite{ostrik2002module}. In fact, the indecomposable module categories over $\scriptC(G,\omega,H,\psi)$ and the indecomposable module categories over $\VEC_G^\omega$ are in a canonical bijection. However, as we will see in a concrete example below that this bijection does not preserve the number of simple elements in the module category.\footnote{The authors thank the referee for pointing out a mistake in our draft which leads to this discussion.} Hence, in general, the existence of a module category with a single simple object over $\VEC_G$ does not imply the existence of the a module category with a single simple object over $\scriptC(G,1,H,\psi)$.  Therefore, the fact that $\VEC_G$ has trivial anomaly does not necessarily imply the symmetry $\scriptC(G,1,H,\psi)$ has a symmetric trivial gapped phase.

Physically, this bijection between symmetric gapped phases of $\scriptC(G,\omega,H,\psi)$ and $\VEC_G^\omega$ can be seen as follows. Consider a theory $\scriptT$ with global symmetry $\VEC_G^\omega$ with a relevant operator $O(x)$ preserving the $\VEC_G^\omega$-symmetry driving the $\scriptT$ to some gapped phase $A$. Since $A$ is $\VEC_G^\omega$-symmetric, we can consider gauge the anomaly free subgroup $H$ with discrete torsion $\psi$ to get a $\scriptC(G,\omega,H,\psi)$-symmetric gapped phase denoted as $A/H_\psi$. On the other hand, since the operator $O(x)$ preserve the entire $G$-symmetry, it will survive the $H_\psi$-gauging and remain a local operator in the theory $\scriptT/H_\psi$ which we get from gauging $H_\psi$ in the theory $\scriptT$. $O(x)$ will trigger the RG flow in the theory $\scriptT/H_\psi$ and the theory would flow to the $\scriptC(G,\omega,H,\psi)$-symmetric gapped phase $A/H_\psi$. In this way, we construct a map from the symmetric gapped phases of the theories with $\VEC_G^\omega$-symmetry to the symmetric gapped phases of the theories with $\scriptC(G,\omega,H,\psi)$-symmetry. Similar, we can construct the inverse map by starting with the theories with  $\scriptC(G,\omega,H,\psi)$-symmetry and gauging $\Rep(H_\psi)$ symmetry. The process is summarized as follows:
\begin{equation}
\begin{tikzpicture}[baseline=0,square/.style={regular polygon,regular polygon sides=4},scale=0.6]
\node[black] at (0,3) {$\scriptT$ with $\VEC_G^\omega$-sym};
\draw[black, -stealth, thick] (3,3.15) -- (8,3.15);
\node[black, above] at (5.5,3.15) {gauging $H_\psi$};
\node[black] at (13,3) {$\scriptT/H_\psi$ with $\scriptC(G,\omega,H,\psi)$-sym};
\draw[black, -stealth, thick] (8,3-0.15) -- (3,3-0.15);
\node[black,below] at (5.5,3-0.15) {gauging $\Rep(H_\psi)$};
\draw[black, -stealth, thick] (0,3-0.5) -- (0,-2); 
\node[black, right] at (0,0.5) {RG by $O(x)$};
\node[black, below] at (0,-2) {\Large $\substack{\VEC_G^\omega \text{symmetric} \\ \text{gapped phase} \, A}$};
\draw[black, -stealth, thick] (3,-2.65) -- (8,-2.65);
\node[black, above] at (5.5,-2.65) {gauging $H_\psi$};
\draw[black, -stealth, thick] (8,-2.95) -- (3,-2.95);
\node[black,below] at (5.5,-2.95) {gauging $\Rep(H_\psi)$};
\node[black, below] at (13,-2) {\Large $\substack{\scriptC(G,\omega,H,\psi)- \text{symmetric} \\ \text{gapped phase} \, A/H_\psi}$};
\draw[black, -stealth, thick] (13,2.5) -- (13,-2); 
\node[black, right] at (13,0.5) {RG by $O(x)$};
\end{tikzpicture}.
\end{equation}

All the indecomposable module categories over $\scriptC(G,\omega,H,\psi)$ has been explicitly constructed in \cite{ostrik2002module}. It is parameterized by an anomaly free subgroup $K$ of $G$ (that is, $\omega|_{K\times K \times K} = 1$) together with a 2-cocycle $\psi_K$ and we denote the corresponding module category as $\scriptM(K,\psi_K)$ and the simple objects in $\scriptM(K,\psi_K)$ are parameterized by indecomposable $A(H,\psi)-A(K,\psi_K)$ bimodules\footnote{It is clear from this that the indecomposable module categories over $\scriptC(G,\omega,H,\psi)$ for different $(H,\psi)$ are in canonical bijection as they are parameterized by the same pairs $(K,\psi_K)$. It is also clear that in general, the number of simple objects does not have be the same under the bijection.}, which can be computed using similar techniques as in the section \ref{sec:simplelines}. Specifically, the simple objects are parameterized by a double coset of the form $H g K$ for $g\in G$ and an irreducible projective representation $\rho$ of the little group $H^g := \{(h,k)\in H\times K: hgk = g\}$ twisted by some effective 2-cocycle $\psi_g$. 

Now, we are ready to show that the group theoretical triality fusion categories $\scriptC(A_4,\omega_0^{2k},\doubleZ_2^\sigma,1)$ do not admit a trivial symmetric gapped phases therefore is anomalous. To see this, we only need to show for every possible indecomposable module category contains more than $1$ simple objects. First, we must choose $K$ such that there is only a single double coset of the form $\doubleZ_2^\sigma g K$ and the only possible choices are $K = A_4$ as the only subgroup of $A_4$ has order $\geq 6$ is $A_4$ itself. Notice that $K = A_4$ must be anomaly free which is not true for $k = 1, 2$, therefore, we immediately conclude that $\scriptC(A_4,\omega_0^{2k},\doubleZ_2^\sigma,1)$ for $k = 1, 2$ are anomalous. For $k = 0$, we consider taking the only double coset $\doubleZ_2^\sigma \dsi A_4$ and the little group $H^{\dsi}$ is $\doubleZ_2 \simeq \{(1,1), (\sigma,\sigma)\}$. However, since $H^2(\doubleZ_2, U(1)) = \doubleZ_1$ meaning the effective 2-cocycle $\psi_{\dsi}$ has to be trivial, there has to be two irreducible representations for the little group. This means there are two simple objects in this module category and the ground states are 2-fold degenerate. Hence, we conclude that $\scriptC(A_4,1,\doubleZ_2^\sigma,1)$ is also anomalous. 

It is straight forward to generalize the above discussion \cite{ostrik2002module}. The symmetric trivial gapped phases of the group theoretical fusion category $\scriptC(G,\omega,H,\psi)$ are parameterized by $(K,\psi_K)$ such that
\begin{enumerate}
    \item $K$ is an anomaly free subgroup of $G$;
    \item There is only a single double coset of the form $Hg K$;
    \item The 2-cocycle $\psi_{\dsi} = (\psi|_{H \cap K}) (\psi_K|_{H \cap K})^{-1}$ is non-degenerate which implies there's only a single irreducible projective representation for the little group $H^{\dsi} \simeq H \cap K$.
\end{enumerate}
The existence of the pair $(K,\psi_K)$ satisfying the above three conditions can be viewed as the anomaly free condition for the group theoretical fusion category $\scriptC(G,\omega,H,\psi)$.

\section{Example: $c=1$ Compact boson at Kosterlitz-Thouless point}\label{sec:KT_theory}
In this section, we consider the example of $c=1$ compact boson at the Kosterlitz-Thouless (KT) point and compute the twisted partition functions of triality defect in the $c=1$ compact boson at the KT point. We match the spin selection rule and also show one can not construct a new triality fusion category by combining the triality defect $\llq$ with another generator $\eta$ of the $\doubleZ_3$ symmetry in the KT theory. 

\subsection{A lightning review of $c=1$ compact boson and the triality defect}\label{sec:KT_review}
We first briefly review the $c = 1$ compact boson following the convention in \cite{Thorngren:2021yso}. The theory is described by a scalar field $X$ with period $2\pi R$,
\begin{equation}\label{d-a}
    X \simeq X + 2\pi R.
\end{equation}
It is convenient to define $2\pi$-periodic field $\theta$ and $2\pi$-periodic conjugate momentum $\phi$ and introduce the left and right moving fields $X_{L,R}$,
\begin{equation}\label{d-b}
    \theta = R^{-1} (X_L + X_R), \quad \phi = R(X_L - X_R)/2.
\end{equation}
The global symmetry at a generic radius $R$ is
\begin{equation}\label{d-c}
    G_{bos} = (U(1)^\theta \times U(1)^\phi) \rtimes \mathbb{Z}_2^C,
\end{equation}
where $U(1)^\theta$ and $U(1)^\phi$ are the shifting symmetry of $\theta$ and $\phi$ respectively, and the charge conjugation $C$ flips the sign of $\theta$ and $\phi$ simultaneously.

At a generic radius $R$, the primary local operators in this theory contain vertex operators,
\begin{equation}\label{d-d}
    V_{n,w} = e^{i\left(\frac{n}{R}+\frac{wR}{2}\right)X_L} e^{i\left(\frac{n}{R} - \frac{wR}{2}\right)X_R} = e^{in\theta}e^{iw\phi}, \quad n,w \in \mathbb{Z}
\end{equation}
with the scaling dimension,
\begin{equation}\label{d-e}
    (h,\oh) = \left(\frac{1}{2}\left(\frac{n}{R}+\frac{wR}{2}\right)^2, \frac{1}{2}\left(\frac{n}{R} - \frac{wR}{2}\right)^2\right),
\end{equation}
together with the normal ordered Schur symmetric polynomials in the $U(1)$ currents $j_1 = \partial X_L$ and $\oj_1 = \opartial X_R$ and their derivatives, denoted as,
\begin{equation}\label{d-f}
    j_{n^2} \oj_{m^2} \quad (h,\oh) = (n^2,m^2).
\end{equation}
The spectrum can also be seen from the partition function,
\begin{equation}\label{d-g}
    Z(R) = \frac{1}{|\eta(\tau)|^2}\sum_{n,w\in \mathbb{Z}} q^{\frac{1}{2}\left(\frac{n}{R} +\frac{wR}{2}\right)^2} \oq^{\frac{1}{2}\left(\frac{n}{R} - \frac{wR}{2}\right)^2}.
\end{equation}
For $c=1$ CFT, there are null states in the descendent states when the Virasoro primary state has scaling dimension $h = \frac{n^2}{4}$ for $n \in \mathbb{Z}$. For a generic $h$, there is no null states in the descendent states of a Virasoro primary state and the Virasoro character is given by,
\begin{equation}
    \chi_h(\tau) = \frac{q^{h}}{\eta(\tau)}.
\end{equation}
For the primary state with $h = \frac{n^2}{4}$ with $n \in \mathbb{Z}$, because of the null states, its character takes the form,
\begin{equation}\label{eq:character_deomposition}
    \chi_{h = \left(\frac{n}{2}\right)^2}(\tau) = \frac{q^{\left(\frac{n}{2}\right)^2} - q^{\left(\frac{n}{2} + 1\right)^2}}{\eta(\tau)}.
\end{equation}
At a generic point of the moduli space, terms with $n \neq 0$ or $m \neq 0$ correspond to characters with primaries $V_{n,m}$'s containing no null states. However, the term $\frac{1}{\eta(\tau)\oeta(\otau)}$ with $n = m = 0$ cannot be a character of Virasoro primary (the identity operator) due to the appearance of null states, but should correspond to the sum of characters of primary states and can be seen via the following rewriting:
\begin{equation}\label{d-j}
    \frac{1}{\eta(\tau)\oeta(\otau)} = \frac{1}{\eta(\tau)\oeta(\otau)}\sum_{n,m=0}^\infty (q^{n^2} - q^{(n+1)^2})(\oq^{m^2} - \oq^{(m+1)^2}),
\end{equation}
where each term in the sum is a character for the primary operator with scaling dimension $(h,\oh) = (n^2,m^2)$, corresponding to the primary operator $j_{n^2}\oj_{m^2}$ mentioned above. 

It is worth mentioning at the special radius $R = \sqrt{2}$, the theory becomes $SU(2)_1$, and the global symmetry is enhanced to $SO(4) = \frac{SU(2)_L \times SU(2)_R}{\doubleZ_2}$. We can represent this $SO(4)$ in its vector representation, where the basis is given by 4 operators $(\sin\theta, \cos\theta, \sin\phi, \cos\phi)$ \cite{Thorngren:2021yso}. The charge conjugation is represented as, 
\begin{equation}\label{eq:C_matrix}
    C = \begin{pmatrix} -1 & 0 & 0 & 0 \\ 0 & 1 & 0 & 0 \\ 0 & 0 & -1 & 0 \\ 0 & 0 & 0 & 1 \end{pmatrix}.
\end{equation}
Similarly, the $U(1)_\theta$ and $U(1)_\phi$ can be represented as,
\begin{equation}
    R_\theta(\alpha) = \begin{pmatrix} \cos\alpha & -\sin\alpha & 0 & 0 \\ \sin\alpha & \cos\alpha & 0 & 0 \\ 0 & 0 & 1 & 0 \\ 0 & 0 & 0 & 1\end{pmatrix}, \quad R_\phi(\alpha) = \begin{pmatrix} 1 & 0 & 0 & 0 \\ 0 & 1 & 0 & 0 \\ 0 & 0 & \cos\alpha & -\sin\alpha \\ 0 & 0 & \sin\alpha & \cos\alpha\end{pmatrix}.
\end{equation}
The spectrum of primary operators of the $SU(2)_1$ theory can be derived by decomposing the partition function in terms of characters \eqref{eq:character_deomposition} of irreducible representations of the Virasoro algebra \cite{DiFrancesco:1997nk,Gaberdiel:2001xm}, and the details are presented in the Appendix \ref{App-a}. The $SU(2)_1$ Hilbert space decomposes as,
\begin{equation}\label{eq:SU21HilbertSpace}
    \CH_{SU(2)_1} = \bigoplus_{\substack{ j,\oj \in \frac{1}{2}\mathbb{Z}_{\geq 0}, \\ j + \oj \in \mathbb{Z}}} V_j \otimes V_{\oj} \otimes H^{Vir}_{j^2} \otimes H^{\overline{Vir}}_{\oj^2}, 
\end{equation}
where by $V_j$ ($\oV_{\oj}$) we denote the spin-$j$(spin-$\oj$) representation of $SU(2)_L$ ($SU(2)_R$) and by $H^{Vir}_{j^2}$ we denote the Virasoro representation with $h = j^2$. Notice that here $j$ and $\oj$ label the irrep of $SU(2)_L$ and $SU(2)_R$ symmetry rather than the affine $SU(2)_L$ or $SU(2)_R$, therefore the affine cut-off of $j$ or $\oj$ is not at presence. It is clear from this decomposition that how the $\frac{SU(2)_L\times SU(2)_R}{\mathbb{Z}_2}$ acts on the $\CH_{SU(2)_1}$.

The $\mathbb{Z}_2^C$ symmetry is free of anomaly, so one could consider gauging it. The resulting theories are a class of theories also parameterized by the radius $R$ of the compact boson, and we call the resulting theories the orbifold branch. The spectrum of Virasoro primaries on the $c=1$ orbifold branch consists of two sectors, the untwisted sector which contains $\mathbb{Z}_2^C$ invariant operators of the corresponding compact boson theory, and the twisted sector which contains the $\mathbb{Z}_2^C$ invariant non-local operators ending on the $C$ defect line in the compact boson theory.

The $\mathbb{Z}_2^C$-invariant twisted sector is constructed by acting on the two ground states $\left|\frac{1}{16},\frac{1}{16}\right\rangle_i$ $i = 1,2$ with even powers of the operators $\alpha_{-n}$ and $\overline{\alpha}_{-n'}$ (where now $n,n' \in \frac{1}{2} + \mathbb{Z}_{\geq 0}$) appearing in the mode expansion of the compact boson $\phi$ with twisted boundary condition \cite{Ginsparg:1988ui},
\begin{equation}\label{eq:Hilbertspacetwisted}
    \mathcal{H}_{\text{orbifold, twisted}} = \left\{\alpha_{-n_1}\cdots\alpha_{-n_{l}}\overline{\alpha}_{-n_{l+1}}\cdots \overline{\alpha}_{-n_{2k}}\left|\frac{1}{16},\frac{1}{16}\right\rangle_j: n_i \in \frac{1}{2} + \mathbb{Z}_{\geq 0}\right\}.
\end{equation}
The two ground states $\left|\frac{1}{16},\frac{1}{16}\right\rangle_i$ are denoted as $\sigma_i$ where $i=1,2$.\footnote{We abuse the notation slightly here. These $\sigma_i$ should be distinguished from $\sigma$ appear in the previous section, which denotes an element of the $A_4$ group. The readers should be able to distinguish the two based on context.} The first two excited states are primary states given by $\alpha_{-\frac{1}{2}}\overline{\alpha}_{-\frac{1}{2}}\left|\frac{1}{16},\frac{1}{16}\right\rangle_i$ which both have scaling dimensions $(\frac{9}{16},\frac{9}{16})$ and we will denote the two as $\tau_i$ where $i=1,2$.

The untwisted sector contains,
\begin{equation}\label{d-l}
    V_{n,w}^+ = \frac{V_{n,w} + V_{-n,-w}}{\sqrt{2}}
\end{equation}
which are invariant under the $\mathbb{Z}_2^C$, as well as the $\mathbb{Z}_2^C$ invariant normal-ordered Schur polynomials which are given by,
\begin{equation}\label{d-m}
    j_{n^2} j_{m^2}, \quad \text{with} \quad m - n \in 2\mathbb{Z}.
\end{equation}
At a generic point of the orbifold branch, there is a $D_8 = \langle s,r| s^2 = r^4 = (rs)^2 = 1\rangle$ global symmetry, acting on the untwisted sector as,
\begin{equation}\label{d-n}
    r: (\theta,\phi) \rightarrow (\theta + \pi, \phi + \pi), \quad s: (\theta,\phi) \rightarrow (\theta, \phi + \pi).
\end{equation}
For the operators in the twisted sector, $D_8$ acts as follows. The generator $s$ exchanges two ground states $\sigma_i$ while $r$ acts as
\begin{equation}\label{eq:r_action_1}
    r: (\sigma_1,\sigma_2) \mapsto (i\sigma_1, -i\sigma_2)
\end{equation}
and
\begin{equation}\label{eq:r_action_2}
    r:\alpha_{-n}\mapsto -\alpha_{-n}, \quad r: \overline{\alpha}_{-m}\mapsto -\overline{\alpha}_{-m}, \quad n,m \in \doubleZ_{\geq 0} + \frac{1}{2}.
\end{equation}
This implies $r:(\tau_1,\tau_2) \mapsto (i\tau_1, -i \tau_2)$.
There are two important $D_4$ subgroups of $D_8$:
\begin{equation}
    D_4^A \equiv \langle r^2,s \rangle, \quad D_4^B \equiv \langle r^2, sr\rangle.
\end{equation}
An important result we will use later to determine the action of the triality line $\CL_Q$ on the twisted sector $\mathcal{H}_{KT, \text{twisted}}$ is that the action of $r^2$ acts as $-1$ on the entire twisted sector, which can be seen from \eqref{eq:Hilbertspacetwisted}\eqref{eq:r_action_1}\eqref{eq:r_action_2}. Furthermore, since $r^2$ acts trivially on the entire untwisted sector and acts as $-1$ on the entire twisted sector, we identify as the generator $\hat{C}$ of the quantum $\doubleZ_2$ symmetry from the $\doubleZ_2^C$ gauging.

\subsection{Spectrum of triality defect and twisted partition functions on torus}\label{sec:KT_twisted_partition_function}
The KT theory can be acquired by gauging the $\mathbb{Z}_2^C$ symmetry of the $SU(2)_1$ theory, which locates at the intersection point between the circle branch and the orbifold branch. And in \cite{Thorngren:2021yso}, the triality defect $\mathcal{L}_Q$ has been identified with the element $Q \in \frac{SU(2)_L\times SU(2)_R}{\mathbb{Z}_2}$ global symmetries of the $SU(2)_1$ theory. In the representation of $SO(4)$ we used above, $Q$ can be represented as 
\begin{equation}\label{db-a}
    Q = \left(
\begin{array}{cccc}
 0 & 1 & 0 & 0 \\
 0 & 0 & -1 & 0 \\
 -1 & 0 & 0 & 0 \\
 0 & 0 & 0 & 1 \\
\end{array}
\right).
\end{equation}
However, the symmetry operator $Q$ does not commute with $C$, therefore, in the gauged theory, the reminiscent of the symmetry operator $Q$ is given by the triality line $\mathcal{L}_Q$, related to $Q$ as
\begin{equation}\label{db-b}
    \mathcal{L}_Q = Q + CQC,
\end{equation}
with the fusion rule \cite{Thorngren:2021yso}
\begin{equation}\label{db-c}
    \CL_Q \times \CL_Q = 2 \CL_{\oQ}, \quad \CL_Q \times \CL_{\oQ} = \sum_{g\in D_{4B}} g.
\end{equation}
As one can see, the charge conjugation $C$ corresponds to $\sigma \in A_4$ and $Q$ corresponds to $q \in A_4$ discussed previously and as one can check using the matrix representation above the minimal subgroup of $SO(4)$ containing $C$ and $Q$ is indeed $A_4$.

From the above fusion rule and the irreducible representations of $D_4 = \mathbb{Z}_2\times \mathbb{Z}_2$ given by
\begin{equation}\label{db-d}
    (1,1,1,1), \quad (1,1,-1,-1), \quad (1,-1,1,-1) \quad (1,-1,-1,1),
\end{equation}
we find the action of $\CL_Q$ on a state $|\psi\rangle$ in the KT theory are non-trivial only if $|\psi\rangle$ transforms in the trivial representation of $D_4$.

Now, we determine the action of the triality line $\CL_Q$ on the states in the KT theory. For the states in twisted sector, all twisted sector states transform non-trivially under $\doubleZ_2 = \langle r^2 \rangle$, thus non-trivially under $D_{4B}$ as well. Hence, by the fusion rule \eqref{db-c}, the triality operator $\CL_Q$ must annihilate all the states in the twisted sector. 

The action of $\CL_Q$ on the untwisted sector can be determined by \eqref{db-b}. We simply need to construct the representation matrices of $C$ and $Q$ for $(j,\oj)$ irreducible representation of $SO(4)$ global symmetry in the $SU(2)_1$ theory. And the action of $\scriptL_Q$ on the untwisted sector is simply given by $Q+CQC$ restricted on the $\doubleZ_2^C$-invariant sector of each $(j,\oj)$ irreducible representation of $SO(4)$.

Knowing the action of $\CL_Q$ on the KT theory Hilbert space $\scriptH_{KT}$ allows us to compute the twisted partition function $(Z_{KT})^{\mathcal{L}_Q}$. Since $\CL_Q$ annihilates the twisted sector, the twisted partition function $(Z_{KT})^{\CL_Q} = \text{Tr}_{\scriptH_{KT}}(\CL_Q q^{L_0 - 1/24} \oq^{\oL_0 - 1/24})$ can be reduced to the untwisted sector and expressed as the following sum of the twisted partition function of the $SU(2)_1$ theory,
\begin{equation}\label{db-e}
\begin{aligned}
    (Z_{KT})^{\llq} &= \text{Tr}_{\CH_{KT}}(\mathcal{L}_Q q^{L_0 - 1/24}\oq^{\oL-1/24})\\ &= \text{Tr}_{\CH_{KT,\text{untwisted}}}(\CL_Q q^{L_0 - 1/24} \oq^{\oL - 1/24}) \\ &= \text{Tr}_{\CH_{SU(2)_1}}((Q+CQC)\frac{1+C}{2}q^{L_0 - 1/24}\oq^{\oL_0 - 1/24}) \\ &= \frac{(Z_{SU(2)_1})^Q + (Z_{SU(2)})^{CQ} + (Z_{SU(2)_1})^{QC} + (Z_{SU(2)_1})^{CQC})}{2}.
\end{aligned}
\end{equation}
To evaluate the twisted partition function in $SU(2)_1$ theories, we must first rewrite the partition function in terms of irreps of $\frac{SU(2)_L\times SU(2)_R}{\mathbb{Z}_2}$ global symmetries, as in \cite{DiFrancesco:1997nk,Gaberdiel:2001xm}. As shown in the Appendix \ref{App-a}, this is given by,
\begin{equation}\label{db-f}
    Z_{SU(2)_1}(\tau,\otau) = \sum_{\substack{j,\oj \in \frac{1}{2}\mathbb{Z}_{\geq 0}, \\ j + \oj \in \mathbb{Z}}} (2j+1)(2\oj+1)\chi_{j^2}(\tau) \ochi_{\oj^2}(\otau).
\end{equation}
Since TDL commutes with the stress energy tensor $T(z)$ and $\oT(\overline{z})$, we only need to study its action on the $V_j\otimes \oV_{\oj}$. For this purpose, we can represent the group element $Q,C \in SO(4) = \frac{SU(2)_L\times SU(2)_R}{2}$ as the tensor product of representations of $SU(2)_L$ and $SU(2)_R$, that is,
\begin{equation}
    Q = Q_L \otimes Q_R, \quad C = C_L \otimes C_R,  
\end{equation}
where $Q_L,C_L$ are matrices of a spin-$j$ representation of $SU(2)_L$ and $Q_R,C_R$ are matrices of a spin-$\overline{j}$ representation of $SU(2)_R$. This allows us to compute trace easily since the generic form of the character of $SU(2)$ is well-known. Following the calculation in Appendix \ref{App-b}, we find,
\begin{equation}
\begin{aligned}
    & Tr_{V_j\otimes \oV_{\oj}}Q = Tr_{V_j\otimes \oV_{\oj}}(CQC) = (Tr_{V_j} Q_L) (Tr_{\oV_{\oj}}Q_R) = \frac{\sin((2j+1)\pi/3)}{\sin(\pi/3)}\frac{\sin((2\oj+1)\pi/3)}{\sin(\pi/3)}, \\
    &Tr_{V_j\otimes \oV_{\oj}} CQ = Tr_{V_j\otimes \oV_{\oj}}(QC) = (Tr_{V_j} Q_L C_L) (Tr_{\oV_{\oj}}Q_R C_R) \\
    = & \frac{\sin((2j+1)2\pi/3)}{\sin(2\pi/3)}\frac{\sin((2\oj+1)2\pi/3)}{\sin(2\pi/3)} \\
    = & \frac{\sin((2j+1)\pi/3)}{\sin(\pi/3)}\frac{\sin((2\oj+1)\pi/3)}{\sin(\pi/3)}, \quad \text{for} \quad (j,\oj) \in (\doubleZ_{\geq 0} \oplus \doubleZ_{\geq 0})\cup \left((\frac{1}{2}+\doubleZ_{\geq 0}) \oplus (\frac{1}{2} + \doubleZ_{\geq 0})\right).
\end{aligned}
\end{equation}
The twisted partition function is therefore given by,
\begin{equation}\label{eq:KTLQ1}
    (Z_{KT})^{\CL_Q} = \frac{8}{3|\eta(\tau)|^2}\sum_{\substack{j,\oj \in \frac{1}{2}\mathbb{Z}_{\geq 0}, \\ j + \oj \in \mathbb{Z}}} \sin(\frac{\pi(2j+1)}{3})\sin(\frac{\pi(2\oj+1)}{3}) (q^{j^2}-q^{(j+1)^2})(\oq^{\oj^2} - \oq^{(\oj+1)^2}).
\end{equation}
We can then rewrite the partition function over the familiar sum over the Narain lattice,
\begin{equation}\label{eq:KTLQ2}
    (Z_{KT})^{\CL_Q}(\tau,\otau) = \frac{1}{|\eta(\tau)|^2}\sum_{n,w\in \mathbb{Z}} \bigg(\cos(\frac{2\pi n}{3}) + \cos(\frac{2\pi w}{3})\bigg) q^{(\frac{n+w}{2})^2} \oq^{(\frac{n-w}{2})^2}.
\end{equation}
Using the $S$-modular transformation, we find,
\begin{equation}
    (Z_{KT})_{\CL_Q}(\tau,\otau) = \frac{1}{|\eta(\tau)|^2}\sum_{n,w\in \mathbb{Z}} q^{\frac{(n+w+\frac{1}{3})^2}{4}}\oq^{\frac{(n-w+\frac{1}{3})^2}{4}} + q^{\frac{(n+w-\frac{1}{3})^2}{4}}\oq^{\frac{(n-w+\frac{1}{3})^2}{4}}.
\end{equation}
This twisted partition function computes the states in the defect Hilbert space $\scriptH_{KT,\llq}$ and is consistent as it has integer coefficients in the $q$ and $\oq$ expansion.

Then applying $T$-transformation, we find
\begin{equation}
\begin{aligned}
    & (Z_{KT})_{\CL_Q,\llqb}^{I}(\tau) \\ =& (Z_{KT})_{\CL_Q}(\tau+1) \\ =& \frac{1}{|\eta(\tau)|^2}\sum_{n,w\in \mathbb{Z}} e^{\frac{2\pi i w}{3}}q^{\frac{(n+w+\frac{1}{3})^2}{4}}\oq^{\frac{(n-w+\frac{1}{3})^2}{4}} + e^{\frac{-2\pi in}{3}}q^{\frac{(n+w-\frac{1}{3})^2}{4}}\oq^{\frac{(n-w+\frac{1}{3})^2}{4}}.
\end{aligned}
\end{equation}
This twisted partition function computes the spin of the twisted Hilbert space $\scriptH_{KT,\llq}$, which is given by the phase in front of the $q$ of $\oq$ expansion. And the result is consistent with the spin selection rule derived in \eqref{eq:spinselectionrule1} for the case where the FS indicator $\alpha = 1$. 

Next, we move to compute the twisted partition function of $\CL_{\oQ}$. By the fusion rule $\CL_{\oQ} = \CL_Q \times \CL_Q$, $\CL_{\oQ}$ annihilates the twisted sector in the KT theory and therefore to compute $(Z_{KT})^{\CL_{\oQ}}$, we only need to focus on the untwisted sector. There are two ways to represent the actions of $\CL_{\oQ}$ on the untwisted sector. The first is to consider the action of 
\begin{equation}
    \CL_{\oQ} = Q^2 + CQ^2C
\end{equation}
on the $C$-invariant subspace of $H_{SU(2)_1}$. Alternatively, we may consider using the fusion rule and compute the action of $\CL_{Q}^2$ on the $C$-invariant subspace of $H_{SU(2)_1}$. As a consistency check, one can show the two approaches agree with each other. 

The twisted partition function is given, 
\begin{equation}\label{eq:KTLQB1}
    (Z_{KT})^{\CL_{\oQ}} = \frac{2}{|\eta(\tau)|^2}\sum_{\substack{j,\oj \in \frac{1}{2}\mathbb{Z}_{\geq 0}, \\ j + \oj \in \mathbb{Z}}} \frac{\sin(\frac{(2j+1)2\pi}{3})}{\sin{\frac{2\pi}{3}}}\frac{\sin(\frac{(2\oj+1)2\pi}{3})}{\sin{\frac{2\pi}{3}}}(q^{j^2}-q^{(j+1)^2})(\oq^{\oj^2} - \oq^{(\oj+1)^2}),
\end{equation}
and can be written as a sum over the Narain lattice where $j = \frac{n+w}{2}, \oj = \frac{n-w}{2}$,
\begin{equation}\label{eq:KTLQB2}
    (Z_{KT})^{\CL_{\oQ}}(\tau,\otau) = \frac{1}{|\eta(\tau)|^2}\sum_{n,w\in \mathbb{Z}} \bigg(\cos(\frac{2\pi n}{3}) + \cos(\frac{2\pi w}{3})\bigg) q^{(\frac{n+w}{2})^2} \oq^{(\frac{n-w}{2})^2}
\end{equation}
taken the same form as $(Z_{KT})^{\CL_{Q}}(\tau,\otau)$. 

\subsection{Constructing more triality lines from the known ones}\label{sec:KT_new_trialities}
Now we explore the possibility of constructing more triality line $\CL_Q'$ from the known one via combining the known triality line $\CL_Q$ with the global symmetry $G_{bos}$ at the KT point. 

The most apparent strategy is to take the generator $\eta$ of some $\doubleZ_3 \subset G_{bos} \equiv (U(1)^{\tilde{\theta}}\times U(1)^{\tilde{\phi}})\rtimes \doubleZ_2^{\widetilde{C}}$, and attempt to construct the line operator,
\begin{equation}\label{dc-a}
    \CL_Q' = \CL_Q \eta , \quad \CL_Q' = \eta \CL_Q,
\end{equation}
which has been considered in \cite{Thorngren:2021yso,Thorngren:2019iar,Chang:2018iay} to construct the duality line $N$ with different FS indicator. However, for this to preserve the fusion rule in general, $\CL_Q \times \CL_Q \times \CL_Q = 2\sum_{g\in \doubleZ_2\times \doubleZ_2} g$, $\eta$ has to commute with $\CL_Q$. Indeed, the construction used in \cite{Thorngren:2021yso,Thorngren:2019iar,Chang:2018iay} is to tensor product one theory with duality line $N$ and another theory with anomalous $\doubleZ_2$ global symmetry $\eta$, and consider the operator $N \eta$, where the duality line $N$ in one theory apparently commutes with the operator $\eta$ in another theory. As we will see, however, the candidate $\mathbb{Z}_3$ subgroups are the $\mathbb{Z}_3$ subgroups of $U(1)\times U(1)$, which does not commute with $\CL_Q$, therefore fusion $\CL_Q$ with generators of $\doubleZ_3$ will not lead to new triality lines.

To see this is the case, we consider the action of $\CL_Q \eta$ or $\eta \CL_Q$ on the untwisted sector $\CH_{KT, \text{untwisted}}$ and check whether $(\CL_Q \eta)^3$ or $(\eta \CL_Q)$ only has eigenvalues $0,8$ or not.

For that, we need to understand the origin of the $U(1)^{\thetatilde}\times U(1)^{\phitilde}$ in the KT theory from the $SU(2)_1$ theory. Under the $\doubleZ_2^C$-gauging, the subgroup of the $SO(4)$ global symmetry commute with the $\doubleZ_2^C$ would survive the gauging and remain as the global symmetry of the resulting KT theory. Since the charge conjugation $C$ in the adjoint representation of $SO(4)$ is given by \eqref{eq:C_matrix}, the commutant of $\mathbb{Z}_2^C$ therefore contains
\begin{equation}\label{da-h}
    R_1(\alpha) = \begin{pmatrix} \cos\alpha & 0 & \sin\alpha & 0 \\ 0 & 1 & 0 & 0 \\ -\sin\alpha & 0 & \cos\alpha & 0 \\ 0 & 0 & 0 & 1 \end{pmatrix}, \quad R_2(\beta) = \begin{pmatrix} 1 & 0 & 0 & 0 \\ 0 & \cos\beta & 0 & \sin\beta & \\ 0 & 0 & 1 & 0 \\ 0 & -\sin\beta & 0 & \cos\beta \end{pmatrix}.
\end{equation}
Notice that $C$ is identified as the $\pi$-rotation $R_1(\pi)$. Hence, gauging $\doubleZ_2^C$ would half the radius of $R_1(\alpha)$ which we identify as $U(1)_{\tilde{\theta}}$ (that is, $R_1(\pi)$ acts trivially on every state in the KT theory) and double the radius of $R_2(\alpha)$ which we identify as $U(1)_{\tilde{\phi}}$ (that is $R_2(2\pi)$ acts non-trivially on the twisted sector in the KT theory). 

To check the fusion rule, we only need to consider the action of $\eta \llq$ or $\llq \eta$ on the untwisted sector, as $\eta \llq$ or $\llq \eta$ automatically annihilates the twisted sector therefore satisfies the fusion rule when acting on the twisted sector. The $\doubleZ_3^{\tilde{\theta}} \subset U(1)^{\tilde{\theta}}$ is generated by either $R_1(\pi/3)$ or $R_1(2\pi/3)$ while the $\doubleZ_3^{\tilde{\phi}} \subset U(1)^{\tilde{\phi}}$ is generated by either $R_2(8\pi/3) = R_2(2\pi/3)$ or $R_2(4\pi/3)$ when acting on the untwisted sector. For convenience, we take the generator $\eta_{\tilde{\theta}}$ of $\doubleZ_3^{\tilde{\theta}}$ to be $R_1(2\pi/3)$ and the generator of $\eta_{\tilde{\phi}}$ of $\doubleZ_3^{\tilde{\phi}}$ to be $R_2(2\pi/3)$ as well.

We can check explicitly that on the $(j,\oj) = (3/2,3/2)$ irrep of $SO(4)$ that $\CL_Q$ does not commute with $\eta$ and their product $\CL_Q \eta$ does not lead to triality line. Following the convention in Appendix \ref{App-b}, we construct the matrix of $Q + CQC$ as well as $\eta$ and diagonalize it using $C$ eigenstates as a basis. For $(j,\oj) = (3/2,3/2)$ irrep, the dimension of $C$ invariant states is $5$ and project $Q + CQC$ to this subspace we find,
\begin{equation}\label{dc-b}
    Q + CQC = \left(
\begin{array}{cccccccc}
 0 & 0 & 0 & 0 & 0 & 0 & 0 & 0 \\
 0 & \frac{1}{2} & 0 & -\frac{\sqrt{3}}{2} & -\frac{3}{2} & 0 & \frac{\sqrt{3}}{2} & 0 \\
 0 & 0 & 0 & 0 & 0 & 0 & 0 & 0 \\
 0 & \frac{\sqrt{3}}{2} & 0 & \frac{1}{2} & \frac{\sqrt{3}}{2} & 0 & \frac{3}{2} & 0 \\
 0 & -\frac{3}{2} & 0 & -\frac{\sqrt{3}}{2} & \frac{1}{2} & 0 & \frac{\sqrt{3}}{2} & 0 \\
 0 & 0 & 0 & 0 & 0 & 0 & 0 & 0 \\
 0 & -\frac{\sqrt{3}}{2} & 0 & \frac{3}{2} & -\frac{\sqrt{3}}{2} & 0 & \frac{1}{2} & 0 \\
 0 & 0 & 0 & 0 & 0 & 0 & 0 & 0 \\
\end{array}
\right), 
\end{equation}
and the generator $\eta_{\tilde{\theta}}$ of $\doubleZ_3^{\tilde{\theta}} \subset U(1)^{\tilde{\theta}}$ and the generator $\eta_{\tilde{\phi}}$ of $\doubleZ_3^{\tilde{\phi}} \subset U(1)^{\tilde{\phi}}$ are
\begin{equation}\label{dc-c}
\begin{aligned}
    \eta_{\tilde{\theta}} &= \left(
\begin{array}{cccccccc}
 \frac{7}{16} & \frac{3}{8} & \frac{3 \sqrt{3}}{16} & 0 & -\frac{3}{8} & -\frac{9}{16} &
   0 & -\frac{3 \sqrt{3}}{16} \\
 -\frac{3}{8} & \frac{1}{16} & \frac{\sqrt{3}}{4} & \frac{3 \sqrt{3}}{16} & \frac{9}{16}
   & -\frac{3}{8} & -\frac{3 \sqrt{3}}{16} & 0 \\
 \frac{3 \sqrt{3}}{16} & -\frac{\sqrt{3}}{4} & \frac{1}{16} & \frac{3}{8} & 0 & \frac{3
   \sqrt{3}}{16} & -\frac{3}{8} & -\frac{9}{16} \\
 0 & \frac{3 \sqrt{3}}{16} & -\frac{3}{8} & \frac{7}{16} & \frac{3 \sqrt{3}}{16} & 0 &
   \frac{9}{16} & -\frac{3}{8} \\
 \frac{3}{8} & \frac{9}{16} & 0 & \frac{3 \sqrt{3}}{16} & \frac{1}{16} & \frac{3}{8} &
   -\frac{3 \sqrt{3}}{16} & \frac{\sqrt{3}}{4} \\
 -\frac{9}{16} & \frac{3}{8} & \frac{3 \sqrt{3}}{16} & 0 & -\frac{3}{8} & \frac{7}{16} &
   0 & -\frac{3 \sqrt{3}}{16} \\
 0 & -\frac{3 \sqrt{3}}{16} & \frac{3}{8} & \frac{9}{16} & -\frac{3 \sqrt{3}}{16} & 0 &
   \frac{7}{16} & \frac{3}{8} \\
 -\frac{3 \sqrt{3}}{16} & 0 & -\frac{9}{16} & \frac{3}{8} & -\frac{\sqrt{3}}{4} &
   -\frac{3 \sqrt{3}}{16} & -\frac{3}{8} & \frac{1}{16} \\
\end{array}
\right), \\ \eta_{\tilde{\phi}} &= \left(
\begin{array}{cccccccc}
 -\frac{13}{32} & \frac{15}{32} & -\frac{3 \sqrt{3}}{32} & \frac{3 \sqrt{3}}{32} &
   \frac{15}{32} & -\frac{9}{32} & \frac{9 \sqrt{3}}{32} & \frac{3 \sqrt{3}}{32} \\
 -\frac{15}{32} & -\frac{7}{32} & \frac{7 \sqrt{3}}{32} & -\frac{3 \sqrt{3}}{32} &
   \frac{9}{32} & -\frac{3}{32} & -\frac{9 \sqrt{3}}{32} & \frac{9 \sqrt{3}}{32} \\
 -\frac{3 \sqrt{3}}{32} & -\frac{7 \sqrt{3}}{32} & -\frac{7}{32} & \frac{15}{32} &
   \frac{9 \sqrt{3}}{32} & \frac{9 \sqrt{3}}{32} & -\frac{3}{32} & -\frac{9}{32} \\
 -\frac{3 \sqrt{3}}{32} & -\frac{3 \sqrt{3}}{32} & -\frac{15}{32} & -\frac{13}{32} &
   -\frac{3 \sqrt{3}}{32} & \frac{9 \sqrt{3}}{32} & \frac{9}{32} & \frac{15}{32} \\
 -\frac{15}{32} & \frac{9}{32} & -\frac{9 \sqrt{3}}{32} & -\frac{3 \sqrt{3}}{32} &
   -\frac{7}{32} & -\frac{3}{32} & -\frac{9 \sqrt{3}}{32} & -\frac{7 \sqrt{3}}{32} \\
 -\frac{9}{32} & \frac{3}{32} & \frac{9 \sqrt{3}}{32} & -\frac{9 \sqrt{3}}{32} &
   \frac{3}{32} & \frac{11}{32} & \frac{5 \sqrt{3}}{32} & -\frac{9 \sqrt{3}}{32} \\
 -\frac{9 \sqrt{3}}{32} & -\frac{9 \sqrt{3}}{32} & \frac{3}{32} & \frac{9}{32} & -\frac{9
   \sqrt{3}}{32} & -\frac{5 \sqrt{3}}{32} & \frac{11}{32} & -\frac{3}{32} \\
 \frac{3 \sqrt{3}}{32} & -\frac{9 \sqrt{3}}{32} & -\frac{9}{32} & -\frac{15}{32} &
   \frac{7 \sqrt{3}}{32} & -\frac{9 \sqrt{3}}{32} & \frac{3}{32} & -\frac{7}{32} \\
\end{array}
\right),
\end{aligned}
\end{equation}
and the possible $\eta = \eta^i_\alpha \eta^j_\beta$ where $(i,j)\neq (0,0)$. As one can check explicitly, the product $\CL_Q \eta$ or $\eta \CL_Q$ does not lead to new duality line, as $(\CL_Q \eta)^3$ or $(\eta \CL_Q)^3$ does not have eigenvalues which are either $0$ or $8$. Hence, we conclude we can't build new triality out of the known one $\CL_Q$ from this procedure.

Since the generator $\eta$ of $\doubleZ_3$ and, in fact, elements of $U(1)\times U(1)$ in general, does not commute with $\CL_Q$, we can consider another possible construction, namely to conjugate $\CL_Q$ by an element $h \in U(1)\times U(1)$,
\begin{equation}\label{dc-d}
    \CL_Q' = h^{-1} \CL_Q h
\end{equation}
with the fusion rule,
\begin{equation}\label{dc-e}
    \CL_Q'\times \CL_Q' \times \CL_Q' = 2 \sum_{g\in \doubleZ_2\times \doubleZ_2} h^{-1}g h,
\end{equation}
and $\CL_{\oQ}' = h^{-1} \CL_{\oQ} h$. Notice that we do get a ``new'' triality category under this procedure, since $(h^{-1}gh)^2 = 1$. However, this ``new'' triality defect should be Morita equivalent to the old one.

\section{More Triality Fusion Categories}\label{sec:more_trialities}
One might wonder if there exist more fusion categories besides the ones described previously satisfying the same fusion rule. Indeed, there are another set of $F$-symbols that have been computed in the condensed matter literature \cite{teo2015theory}. It is natural to ask if their $F$-symbols give the same fusion categories as ours and if there are more inequivalent $F$-symbols. We will answer these questions in this section. 

\subsection{The classification of triality fusion category}\label{sec:classifications}
We first review the result in \cite{jordan2009classification} which classifies the fusion category whose simple objects containing $g \in \doubleZ_2\times \doubleZ_2$, $\CL_Q$ and $\CL_{\oQ}$. They satisfy the following fusion relations,
\begin{equation}\label{eq:triality_fusion_rule}
\begin{aligned}
    & g \times \CL_Q = \CL_Q \times g = \CL_Q, \quad g \times \CL_{\oQ} = \CL_{\oQ} \times g = \CL_{\oQ}, \\
    & \CL_Q \times \CL_Q = 2 \CL_{\oQ}, \quad \CL_{\oQ} \times \CL_{\oQ} = 2 \CL_Q, \\
    & \CL_Q \times \CL_{\oQ} = \CL_{\oQ} \times \CL_{Q} = \sum_{g\in \doubleZ_2\times \doubleZ_2} g.
\end{aligned}
\end{equation}
Theorem 1.4 in \cite{jordan2009classification} then implies there are $6$ inequivalent fusion categories with the simple objects satisfying the above fusion relations. Since the Frobenius-Schur indicator is given by an element in $H^3(\doubleZ_3, U(1)) = \doubleZ_3$, the $6$ inequivalent fusion categories organize into $2$ classes each containing $3$ related by choosing different FS indicator. 

Theorem 1.1 in \cite{jordan2009classification} further describes the two classes of fusion categories. The first one is the group theoretic fusion category with different FS indicator $\alpha$ (which are the ones we studied in section \ref{sec:group_theoretical_triality}) while the second class is constructed explicitly in \cite{jordan2009classification} in terms of the classification data (but the $F$-symbols are not explicitly given). We will argue the $F$-symbols computed in \cite{teo2015theory} correspond to the second class as they lead to different fusion categories from ours. Since the second class is not group theoretical fusion category, these triality fusion categories are intrinsic non-invertible in the sense of \cite{Kaidi:2022uux}.

\subsection{F-symbols of intrinsic triality fusion category}\label{sec:instric_triality_F_symbol}
We follow \cite{teo2015theory} to list the $F$-symbols of the triality category. Recall that the triality line satisfies,
\begin{equation}
    \llq \times \llq = 2\llqb,\quad \llq \times \llqb = \sum_{g\in \IZ_2\times \IZ_2} g.
\end{equation}
and invertible symmetry lines $g$ satisfy the fusion rule of $\mathsf{Vec}_{\IZ_2\times\IZ_2}$. We represent invertible symmetry line $g$ by $\IZ_2$-valued vectors $\{(0,0),(1,0),(0,1),(1,1)\}$ (which corresponds to $(\dsi,\hat{\sigma},\eta,\hat{\sigma}\eta)$ in previous notation). In this representation, the triality symmetry is also a $\IZ_2$-valued matrix,
\begin{equation}
    \Lambda_3 = \begin{pmatrix}0 & -1 \\ 1 & -1 \end{pmatrix}=\begin{pmatrix}0 & 1 \\ 1 & 1 \end{pmatrix}.
\end{equation}
The $R$-symbols between different invertible symmetry lines are,
\begin{equation}
    R^{g,h}=(-1)^{g^\intercal \sigma^1 \Lambda_3^2 h},\quad g,h\in \IZ_2\times \IZ_2,
\end{equation}
where $\sigma^i$ are the Pauli $i$ matrices. 
The braiding phase is $\CDS_{g,h}=R^{g,h}R^{h,g}=(-1)^{g^\intercal \sigma^1 h}$. The $F$-symbols can be understood as a representation of the double cover of $A_4$. We choose the 2-d representations as,
\begin{equation}
    \CA_{(0,0)}=\sigma^0,\  \CA_{(1,0)}=\ii \sigma^1,\  \CA_{(0,1)}=-\ii \sigma^2,\  \CA_{(1,1)}=\ii \sigma^3.
\end{equation}
The $F$-symbols consist of a free parameter $\alpha = e^{2\pi k \ii/3}, k = 0,1,2$ which is the Frobenius-Schur indicator. The list of $F$-symbols is given by,
\begin{equation}
    F^{ghk}_{g+h+k}=1,\quad g,h,k \in \IZ_2\times \IZ_2
\end{equation}
\begin{align}
    &F^{\llq g h}_{\llq}=R^{g,(\Lambda_3 h)},\quad F^{gh\llq}_{\llq}=R^{h,(\Lambda_3^2 g)},\quad F^{g\llq h}_{\llq}=\CDS_{(\Lambda_3 g),h},\\
    &{F_{h}^{\llq \llqb g}} = R^{g,\Lambda_3^2 h},\quad  {{F_{h}^{g \llq \llqb }} = R^{g\times h,\Lambda_3^2 g}},\quad {{F_{h}^{\llq g \llqb}} = {{\CDS}_{g,\Lambda_3 h}R^{g,\Lambda_3 g}}}.
\end{align}
When exchanging $\llq \leftrightarrow \llqb$, the $F$-symbols are obtained by replacing $\Lambda_3\leftrightarrow \Lambda_3^2$,
\begin{align}
    &{F^{\llqb g h}_{\llqb}}=R^{g,(\Lambda_3^2 h)},\quad {F^{gh\llqb}_{\llqb}}=R^{h,(\Lambda_3 g)},\quad F^{g\llqb h}_{\llqb}=\CDS_{(\Lambda_3^2 g),h},\\
    &{F_{h}^{\llqb \llq g}} = R^{g,\Lambda_3 h},\quad  {{F_{h}^{g \llqb \llq }} = R^{g\times h,\Lambda_3 g}},\quad {{F_{h}^{\llqb g \llq}} = {{\CDS}_{g,\Lambda_3^2 h}R^{g,\Lambda_3^2 g}}},
\end{align}
and
\begin{align}
    {[F^{\llq\llqb \llq}_{\llq}]_{g,h}} = -\frac{\alpha}{2}\CDS_{\Lambda_3 g,h}R^{h,\Lambda_3 h},\quad {[F^{\llqb\llq \llqb}_{\llqb}]_{g,h}} = -\frac{\alpha^{-1}}{2}\CDS_{\Lambda_3^2 g,h}R^{h,\Lambda_3^2 h}.
\end{align}
Other $F$-symbols are listed in \tabref{tab:teof}.

\begin{table*}[htbp]
\center
\makegapedcells
    \begin{tabular}{|c|c|c|c|c|}
    \hline
    $g$ & $\dsi$ & $\hat{\sigma}$ & $\eta$ & $\hat{\sigma}\eta$ \\
    \hline
    $[F^{\llq \llq g}_{\llqb}]_{(\llqb,\mu,1),(\llq,1,\nu)}$ & $\sigma^0$ & $-\ii \sigma^1$ & $\ii \sigma^2$ & $-\ii \sigma^3$ \\
    \hline
    $[F^{\llq g \llq}_{\llqb}]_{(\llq,1,\mu),(\llq,1,\nu)}$ & $\sigma^0$ & $\ii \sigma^2$ & $-\ii \sigma^3$ & $-\ii \sigma^1$ \\
    \hline
    $[F^{g \llq \llq}_{\llqb}]_{(\llq,1,\mu),(\llqb,\nu,1)}$ & $\sigma^0$ & $\ii \sigma^3$ & $\ii \sigma^1$ & $-\ii \sigma^2$ \\
    \hline
    $[F^{\llqb\llqb g}_{\llq}]_{(\llq,\mu,1),(\llqb,1,\nu)}$ & $\sigma^0$ & $-\ii \sigma^1$ & $\ii \sigma^2$ & $-\ii \sigma^3$ \\
    \hline
    $[F^{\llqb g \llqb}_{\llq}]_{(\llqb,1,\mu),(\llqb,1,\nu)}$ & $\sigma^0$ & $\ii \sigma^3$ & $\ii \sigma^1$ & $-\ii \sigma^2$\\
    \hline
    $[F^{g\llqb\llqb}_{\llq}]_{(\llqb,1,\mu),(\llq,\nu,1)}$ & $\sigma^0$ & $-\ii \sigma^2$ & $\ii \sigma^3$ & $\ii \sigma^1$\\
    \hline 
    $[F^{\llq\llq\llq}_g]_{(\llqb,\mu,1),(\llqb,\nu,1)}$ & $\alpha^{-1}\asym$ & $\ii \alpha^{-1} \sigma^1 \asym$ & $-\ii \alpha^{-1} \sigma^2 \asym$ & $\ii \alpha^{-1} \sigma^3 \asym$ \\
    \hline
    $[F^{\llqb\llqb\llqb}_g]_{(\llq,\mu,1),(\llq,\nu,1)}$ & $\alpha\asymi$ & $\ii \alpha \sigma^1 \asymi$ & $-\ii \alpha\sigma^2 \asymi$ & $\ii \alpha \sigma^3 \asymi$ \\ 
    \hline
    $[F^{\llq \llq \llqb}_{\llq}]_{(\llq,\mu,\nu),(g,1,1)}$ & $\frac{1}{\sqrt{2}}\ii \asym \sigma^1 $ & $\frac{1}{\sqrt{2}} \asym \sigma^0$ & $-\frac{1}{\sqrt{2}}\ii \asym \sigma^3$ & $-\frac{1}{\sqrt{2}}\ii \asym \sigma^2$ \\
    \hline
    $[F^{\llqb\llqb\llq}_{\llqb}]_{(\llq,\mu,\nu),(g,1,1)}$ & $\frac{\alpha}{\sqrt{2}}\ii \asymi \sigma^3$ & $-\frac{\alpha}{\sqrt{2}}\ii \asymi \sigma^2$ & $-\frac{\alpha}{\sqrt{2}}\ii \asymi \sigma^1$ & $-\frac{\alpha}{\sqrt{2}} \asymi \sigma^0$ \\
    \hline
    $[F^{\llqb\llq\llq}_{\llq}]_{(g,1,1),(\llqb,\mu,\nu)}$ & $\frac{\ii \alpha^{-1} \sigma^2}{\sqrt{2}}$ & $\frac{-\alpha^{-1}\sigma^0}{\sqrt{2}}$ & $\frac{-\ii \alpha^{-1}\sigma^1}{\sqrt{2}}$ & $\frac{\ii \alpha^{-1}\sigma^3}{\sqrt{2}}$ \\
    \hline
    $[F^{\llq\llqb\llqb}_{\llqb}]_{(g,1,1),(\llq,\mu,\nu)}$ & $\frac{\ii \sigma^2}{\sqrt{2}}$ & $\frac{-\ii \sigma^1}{\sqrt{2}}$ & $\frac{\ii \sigma^3}{\sqrt{2}}$ & $\frac{- \sigma^0}{\sqrt{2}}$ \\
    \hline
\end{tabular}
\caption{Where $\asym = \exp(\frac{\pi}{3}\frac{\CA_{sym}}{\sqrt{3}})$ and $\CA_{sym}=\sum_{g=\{(1,0),(0,1),(1,1)\}}\CA_g$. $\alpha=e^{\ii 2\pi k/3}$ is the FS indicator.}
\label{tab:teof}
\end{table*}

\subsection{Spin selection rules}\label{sec:intrinsic_spin_selection_rule}
We now derive the spin selection rules from the above $F$-symbols. Repeating the same calculation in section \ref{sec:group_theoretical_spin_selection_rules}, we find,
\begin{equation}
    \gamma(h,g) = \begin{pmatrix} 1 & 1 & 1 & 1 \\ 1 & 1 & -1 & -1 \\ 1 & -1 & 1 & -1 \\ 1 & -1 & -1 & 1\end{pmatrix},
\end{equation}
with the same four 1-dimensional irreducible representations as in the non-intrinsic or group theoretical case,
\begin{equation}\label{eq:proj2}
    (1,1,1,-1),\quad (1,1,-1,1), \quad (1,-1,1,1), \quad (1,-1,-1,-1).
\end{equation}
Then, consider the same calculation in Figure \ref{fig:spinselection} and plug in the $F$-symbols for the intrinsic triality defects, we find,
\bee\label{eq:T32}
Z_{\CL_Q}(\tau+3) &= \sum_{\substack{\mu,\nu = 1,2,\\ g\in \mathbb{Z}_2\times \mathbb{Z}_2}}\left[F^{\llqb \llq \llq}_{\llq}\right]_{(\dsi,1,1)(\llq,\mu,\nu)}\left[F^{\llq \llq \llqb}_{\llq}\right]_{(\llqb,\mu,\nu)(g,1,1)}Z_{\llq g}^{\llq}(\tau) 
\\ &= \frac{\alpha}{2} \sum_{g\in \mathbb{Z}_2\times \mathbb{Z}_2} Z_{\llq g}^{\llq}(\tau). 
\eee
where $\alpha = e^{2\pi \ii k/3}, k = 0,1,2$ is the FS indicator. This implies, 
\begin{equation}
    e^{6\pi \ii s} = \frac{\alpha}{2}\sum_{g\in \mathbb{Z}_2\times \mathbb{Z}_2}\widehat{g}_{\llq}.
\end{equation}
Using the eigenvalues of the irreducible representations of $\mathbb{Z}_2\times \mathbb{Z}_2$ with phase $\gamma(g,h)$ given by \eqref{eq:proj2}, we have,
\be
    e^{6\pi \ii s}=\pm \alpha.
\ee
Notice that 2-cocycle $\gamma(g,h)$ ensures the eigenvalues lead to consistent result in \eqref{eq:T32}, as the result on the right-hand side has to be a phase, which is not true for the eigenvalues of the usual irreducible representations of $\doubleZ_2\times \doubleZ_2$.
We find the allowed spin is given by, 
\begin{equation}
e^{2\pi \ii s} = \begin{cases} e^{k\pi \ii/3}, \quad k = 0,1,2,3,4,5, \quad \alpha = 1, \\ e^{\frac{2\pi \ii}{9} + \frac{k\pi \ii}{3}}, \quad k = 0,1,2,3,4,5,  \quad \alpha = e^{2\pi \ii/3}, \\ e^{-\frac{2\pi \ii}{9} + \frac{k\pi \ii}{3}}, \quad k = 0,1,2,3,4,5, \quad \alpha = e^{4\pi \ii/3}.\end{cases}
\end{equation}
\subsubsection{Determine the triality fusion category from spin selection rules}
We now argue that one can determine the triality fusion category from the spins of the defect Hilbert space $\scriptH_{\llq}$. It is clear that we can determine the FS indicator from the spins that appear in $\scriptH_{\llq}$. 

We now argue that we can distinguish between intrinsic and non-intrinsic triality fusion categories from the spins. For example, considering the case when the FS indicator $\alpha = 1$, then the allowed spin $s$ for the non-intrinsic triality fusion categories satisfies $s \in \frac{1}{3}\doubleZ$ while the allowed spin $s$ for the intrinsic triality fusion categories satisfies $s \in \left( \frac{1}{3}\doubleZ \right) \cup \left(\frac{1}{3}\doubleZ + \frac{1}{2}\right)$. To distinguish the two cases, we only need to show the additional spin where $s \in \frac{1}{3}\doubleZ + \frac{1}{2}$ must appear.  

To see this, we can show all the allowed irreducible representations of $\doubleZ_2\times\doubleZ_2$ in \eqref{eq:proj2} has to appear using the technique in \cite{Pal:2020wwd}. Specifically, consider the following partition function
\begin{equation}
    \frac{1}{4}\sum_{g\in \doubleZ_2\times\doubleZ_2} \chi_\alpha(g) Z^{\llq}_{\llq,g}(\tau = \ii\beta)
\end{equation}
where $\chi_\alpha$ is the character associated with the irreducible representation $\alpha$, which can be seen from \eqref{eq:proj2}. For each choice of $\chi_\alpha$, we keep only the contribution from the particular irreducible representation $\alpha$. We only need to show this is non-zero for any $\alpha$. 

For this, we consider applying the $S$-modular transformation and get,
\begin{equation}
    \frac{1}{4}\sum_{g\in \doubleZ_2\times\doubleZ_2} \chi_\alpha(g) Z^{\llq}_{\llq,g}(\tau = \ii\beta) = \frac{1}{4}\sum_{g\in\doubleZ_2\times \doubleZ_2}\chi_\alpha(g) \Tr_{\scriptH_g}\left(\widehat{\llqb}\right)_{g,+} e^{-\frac{4\pi^2}{\beta}\left(H - \frac{c}{12}\right)}.
\end{equation}
Considering the high-temperature limit $\beta \rightarrow 0$, we can see in the dual channel on the right-hand side, the partition sum is dominated by the ground state in each defect Hilbert space $\scriptH_g$. As long as the $\doubleZ_2\times \doubleZ_2$ acts faithfully in the theory, the ground state in $\scriptH_g$ for $g\neq 1$ has positive energy, hence the R.H.S. is dominated by the vacuum state in $\scriptH_1$, which implies
\begin{equation}
    \frac{1}{4} \sum_{g\in\doubleZ_2\times\doubleZ_2} \chi_\alpha(g) Z^{\llq}_{\llq,g}(\tau = \ii\beta) \xrightarrow{\beta \rightarrow 0} \frac{1}{2} e^{\frac{3\pi^2c}{\beta}} > 0, 
\end{equation}
where in the last step we used $\chi_\alpha(1) = 1$ for every $\alpha$ as in \eqref{eq:proj2} and $\llqb$ acts on vacuum state as its quantum dimension $2$. Since in the high-temperature limit the partition sum over a particular fixed irreducible representation is positive, we know each irreducible representation must appear. This then implies the spin $s$ such that $s \in \frac{1}{3}\doubleZ + \frac{1}{2}$ has to appear since it comes from the states with irreducible representation $(1,-1,-1,-1)$. 

Hence, we can distinguish the different triality fusion categories by the spins that appeared in the defect Hilbert space $\scriptH_{\llq}$.

To conclude this subsection, we briefly comment on when the spin selection rule should be saturated.\footnote{The authors thank Yifan Wang for mentioning this example which leads to this discussion.} For illustration, let's consider the three-state Potts model. This RCFT contains two $\doubleZ_3$ self-duality lines $N,N'$ \cite{Chang:2018iay} with the fusion rules:
\begin{equation}
    N^2 = (N')^2 = I + \eta + \oeta,
\end{equation}
where $\eta$ generates the $\doubleZ_3$ global symmetries. The spins of the defect Hilbert space satisfy,
\begin{equation}
    \scriptH_N\,: \, s \in \frac{1}{2} \doubleZ + \left\{\frac{1}{8},-\frac{1}{24}\right\}, \quad \scriptH_{N'}\, : \, s \in \frac{\doubleZ}{2} + \left\{ -\frac{1}{8}, \frac{1}{24}\right\}.
\end{equation}
The spin selection rule is derived in \cite{Chang:2018iay} from the relation,
\begin{equation}
    e^{4\pi \ii s}\langle \psi'|\psi\rangle = \frac{1}{\sqrt{3}}\langle \psi'|1 + \hat{\eta}_- + \hat{\bar{\eta}}_-|\psi\rangle = \frac{1+\omega^a +\omega^b}{\sqrt{3}}
\end{equation}
by requiring that $\frac{1+\omega^a +\omega^b}{\sqrt{3}}$ where $a,b = 0,1,2$ and $\omega = e^{2\pi \ii/3}$ is a phase which takes the form,
\begin{equation}
    s \in \frac{1}{2}\doubleZ \pm \left\{\frac{1}{24}, \frac{1}{8}\right\}.
\end{equation}
At first glance, the spin selection rule is not saturated. This is because the spin selection rule is not derived from the eigenvalues of the (projective) representation. For instance, let's consider the irrep of $\doubleZ_2\times \doubleZ_2$ which leads to $e^{4\pi \ii s} = e^{-\frac{2\pi \ii}{12}}$ in the defect Hilbert space ${\scriptH}_{\scriptN}$, say $(1, \omega^2, 1)$. From this, we can derive the 2-cocycle $\gamma(g,h)$,
\begin{equation}
    \gamma(g,h) = \begin{pmatrix} 1 & 1 & 1 \\ 1 & \omega & \omega^2 \\ 1 & \omega^2 & \omega \end{pmatrix}.
\end{equation}
The allowed irreducible representations twisted by this 2-cocycle are given by,
\begin{equation}
    (1, \omega^2, 1), \quad (1, 1, \omega^2), \quad (1,\omega, \omega),
\end{equation}
where the first two lead to spin such that $e^{4\pi \ii s} = e^{-\frac{2\pi \ii}{12}}$ and the last one leads to the spin such that $e^{4\pi \ii s} = e^{\frac{2\pi \ii}{4}}$. By the same argument, each irreducible representation has to appear and this spin selection rule must be saturated, which indeed is the case. Similarly, if we consider the defect Hilbert space of $\scriptH_{\scriptN'}$, the 2-cocycle $\gamma'(g,h)$ is now given by
\begin{equation}
    \gamma'(g,h) = \begin{pmatrix} 1 & 1 & 1 \\ 1 & \omega^2 & \omega \\ 1 & \omega & \omega^2 \end{pmatrix}
\end{equation}
and the allowed irreducible representations are given by
\begin{equation}
    (1,\omega,1), \quad (1,1,\omega), \quad (1,\omega^2,\omega^2),
\end{equation}
where the first two leads to spin such that $e^{4\pi \ii s} = e^{2\pi \ii/12}$ and the last one leads to spin such that $e^{4\pi \ii s} = e^{-2\pi \ii/4}$. Similarly, this spin selection rule is also saturated. One might wonder why there could be two different 2-cocycles arising from the same $\doubleZ_3$-duality category. This is because when deriving the crossing kernels, even after fixing the FS-indicator $\epsilon = 1$, one needs to choose the $\omega$ to be either $e^{ 2\pi \ii/3}$ or $e^{-2\pi \ii/3}$. Two different choices relate to each other by relabeling $\eta$ as $\oeta$. Such relabelling will change $\gamma(g,h)$ to $\gamma'(g,h)$ as well. In the case of two duality lines $N$ and $N'$, both form a $\doubleZ_3$-duality category with the same $\doubleZ_3$ symmetry, hence there's no way we can relabel the $\eta$ in one $\doubleZ_3$ fusion category without doing the same relabeling for the other. Therefore, the choices of $\omega = e^{\pm 2\pi \ii s/3}$ matter here.

\newpage
\section*{Acknowledgements}
We thank Ken Intriligator, John McGreevy, Yifan Wang, Justin Kulp, Justin Kaidi, Sridip Pal, Meng Cheng, Yi-Zhuang You for useful conversation. Z.S. is especially grateful for David Jordan and Pavel Etingof for patiently answer basic questions on tensor categories. We also thank Yi-Zhuang You for sharing his Mathematica code for computing group cohomologies. Z.S. is supported from the US Department of Energy (DOE) under cooperative research agreement DE-SC0009919,  Simons Foundation award No. 568420 (K.I.) and the Simons Collaboration on Global Categorical Symmetries.

\appendix
\section{Details on the compact boson partition function}\label{App-a}
We show that the compact boson partition function, which is derived using \eqref{eq:SU21HilbertSpace}, can be rewritten as the familiar sum over lattice. We start with breaking $Z_{SU(2)_1}$ into the partition sum over irreps with integer spins $j,\oj$ denoted as $Z_{SU(2)_1,I}$ and the partition sum over irreps with half-integer spins $j,\oj$ denoted as $Z_{SU(2)_1,II}$.
\begin{equation}
\begin{aligned}
    &|\eta(\tau)|^2 Z_{SU(2)_1,I} \\ =& \sum_{\substack{j=0,1,\cdots \\ \oj = 0,1,\cdots}} (2j+1)(2\oj+1) (q^{j^2} - q^{(j+1)^2})(\oq^{\oj^2} - \oq^{(\oj+1)^2}) \\ =& \sum_{\substack{j=0,1,\cdots \\ \oj = 0,1,\cdots}} \big[(2j+1)(2\oj+1) q^{j^2} \oq^{\oj^2}\big] - \bigg(\sum_{\substack{j=0,1,\cdots,\\ \oj = 0,1,\cdots}}\big[(2j+1)(2\oj -1) q^{j^2} \oq^{\oj^2} \big] - \sum_{j = 0,1,\cdots} (2j+1)(-1) q^{j^2}\bigg) \\ & - \bigg( \sum_{\substack{j=0,1,\cdots \\ \oj = 0,1,\cdots}} \big[ (2j-1)(2\oj+1) q^{j^2} \oq^{\oj^2} \big] - \sum_{\oj = 0,1,\cdots}(2\oj+1)(-1) \oq^{\oj^2}\bigg) \\ & + \bigg(\sum_{\substack{j = 0,1,\cdots, \\ \oj = 0,1,\cdots}}[(2j-1)(2\oj-1)q^{j^2}\oq^{\oj^2}]  - \sum_{j = 0, 1,\cdots} \big[(2j-1)(-1)q^{j^2}\big] - \sum_{\oj = 0,1,\cdots}\big[(2\oj-1)(-1)\oq^{\oj^2}\big] + 1 \bigg) \\ =& \bigg(\sum_{\substack{j = 0,1,\cdots, \\ \oj = 0,1,\cdots}}4 q^{j^2} \oq^{\oj^2}\bigg) - \bigg(\sum_{j=0,1,\cdots}2 q^{j^2}\bigg) - \bigg(\sum_{\oj = 0,1,\cdots}2 \oq^{\oj^2}\bigg) + 1
\end{aligned}
\end{equation}
and similarly,
\begin{equation}
    |\eta(\tau)|^2 Z_{SU(2)_1,II} = \sum_{\substack{j = \frac{1}{2}, \frac{3}{2},\cdots, \\ \oj = \frac{1}{2}, \frac{3}{2}, \cdots}} 4 q^{j^2} \oq^{\oj^2}.
\end{equation}
It is then straightforward to see from the above that summing over two parts lead to the familiar sum over $2d$ Narain lattice,
\begin{equation}
    |\eta(\tau)|^2 (Z_{SU(2)_1,I} + Z_{SU(2)_1,II}) = \sum_{n,w\in\mathbb{Z}}q^{(\frac{n+w}{2})^2} \oq^{(\frac{n-w}{2})^2}.
\end{equation}
The twisted partition functions \eqref{eq:KTLQ1} and \eqref{eq:KTLQB1} can be rewritten as a sum over Narain lattice \eqref{eq:KTLQ2} and \eqref{eq:KTLQB2} using the same method.

\section{Group theory convention}\label{App-b}
We normalize the generators $T^i$ ($i=1,2,3$) of $su(2)$ Lie algebra such that,
\begin{equation}\label{App-b-a}
    [T^i,T^j] = \epsilon^{ijk}T^k.
\end{equation}
We match the generators of the vector representation of $SO(4)$ with $SU(2)_L\times SU(2)_R$ as,
\begin{equation}\label{App-b-b}
    T^1_L = \left(
\begin{array}{cccc}
 0 & 0 & 0 & \frac{1}{2} \\
 0 & 0 & \frac{1}{2} & 0 \\
 0 & -\frac{1}{2} & 0 & 0 \\
 -\frac{1}{2} & 0 & 0 & 0 \\
\end{array}
\right), \quad T^2_L = \left(
\begin{array}{cccc}
 0 & 0 & \frac{1}{2} & 0 \\
 0 & 0 & 0 & -\frac{1}{2} \\
 -\frac{1}{2} & 0 & 0 & 0 \\
 0 & \frac{1}{2} & 0 & 0 \\
\end{array}
\right), \quad T^3_L = \left(
\begin{array}{cccc}
 0 & \frac{1}{2} & 0 & 0 \\
 -\frac{1}{2} & 0 & 0 & 0 \\
 0 & 0 & 0 & \frac{1}{2} \\
 0 & 0 & -\frac{1}{2} & 0 \\
\end{array}
\right), 
\end{equation}
and,
\begin{equation}\label{App-b-c}
    T^1_R = \left(
\begin{array}{cccc}
 0 & 0 & 0 & \frac{1}{2} \\
 0 & 0 & -\frac{1}{2} & 0 \\
 0 & \frac{1}{2} & 0 & 0 \\
 -\frac{1}{2} & 0 & 0 & 0 \\
\end{array}
\right), \quad T^2_R = \left(
\begin{array}{cccc}
 0 & 0 & -\frac{1}{2} & 0 \\
 0 & 0 & 0 & -\frac{1}{2} \\
 \frac{1}{2} & 0 & 0 & 0 \\
 0 & \frac{1}{2} & 0 & 0 \\
\end{array}
\right), \quad T^3_R = \left(
\begin{array}{cccc}
 0 & \frac{1}{2} & 0 & 0 \\
 -\frac{1}{2} & 0 & 0 & 0 \\
 0 & 0 & 0 & -\frac{1}{2} \\
 0 & 0 & \frac{1}{2} & 0 \\
\end{array}
\right).
\end{equation}
The symmetry operator $Q \in SO(4)$ can be written as,
\begin{equation}\label{App-b-d}
    Q = R^\theta\bigg(\frac{\pi}{2}\bigg) R^{\phi'}\bigg(\frac{\pi}{2}\bigg) = \left(
\begin{array}{cccc}
 0 & 1 & 0 & 0 \\
 0 & 0 & -1 & 0 \\
 -1 & 0 & 0 & 0 \\
 0 & 0 & 0 & 1 \\
\end{array}
\right),
\end{equation}
where,
\begin{equation}\label{App-b-e}
    R^\theta\bigg(\frac{\pi}{2}\bigg) = \left(
\begin{array}{cccc}
 0 & 1 & 0 & 0 \\
 -1 & 0 & 0 & 0 \\
 0 & 0 & 1 & 0 \\
 0 & 0 & 0 & 1 \\
\end{array}
\right), \quad R^{\phi'}\bigg(\frac{\pi}{2}\bigg) = \left(
\begin{array}{cccc}
 0 & 0 & 1 & 0 \\
 0 & 1 & 0 & 0 \\
 -1 & 0 & 0 & 0 \\
 0 & 0 & 0 & 1 \\
\end{array}
\right).
\end{equation}
Notice that we can write,
\begin{equation}\label{App-b-f}
    R^\theta\bigg(\frac{\pi}{2}\bigg) = \exp\bigg(\frac{\pi}{2}(T_L^3 + T_R^3)\bigg), \quad R^{\phi'}\bigg(\frac{\pi}{2}\bigg) = \exp\bigg(\frac{\pi}{2}(T_L^2 - T_R^2)\bigg).
\end{equation}
This allows us to break $Q$ into tensor product of $Q_L \in SU(2)_L$ and $Q_R \in SU(2)_R$ where,
\begin{equation}\label{App-b-g}
    \begin{aligned}
        & Q_L = \exp(\frac{\pi}{2} t_L^3) \exp(\frac{\pi}{2} t_L^2) = \left(
\begin{array}{cc}
 \frac{e^{-\frac{\ii\pi }{4}}}{\sqrt{2}} & -\frac{e^{-\frac{\ii\pi
   }{4}}}{\sqrt{2}} \\
 \frac{e^{\frac{\ii\pi }{4}}}{\sqrt{2}} & \frac{e^{\frac{\ii\pi
   }{4}}}{\sqrt{2}} \\
\end{array}
\right) , \\ & Q_R = \exp(\frac{\pi}{2}t_R^3) \exp(-\frac{\pi}{2} t_R^2) = \left(
\begin{array}{cc}
 \frac{e^{-\frac{\ii\pi }{4}}}{\sqrt{2}} & \frac{e^{-\frac{\ii\pi
   }{4}}}{\sqrt{2}} \\
 -\frac{e^{\frac{\ii\pi }{4}}}{\sqrt{2}} & \frac{e^{\frac{\ii\pi
   }{4}}}{\sqrt{2}} \\
\end{array}
\right),
\end{aligned}
\end{equation}
where in the adjoint representation of $SU(2)$, $t^i = -\frac{\ii}{2}\sigma^i$ satisfying $[t^i,t^j] = \epsilon^{ijk} t^k$.

The charge conjugation $C$ of $c = 1$ compact boson is chosen to be $R^{\phi'}(\pi)$ which can also be written as a tensor product $C = C_L\otimes C_R$ where,
\begin{equation}\label{App-b-h}
    C_L = - C_R = -\ii\sigma^2.
\end{equation}
Notice that since $Q_L, Q_R, C_L, C_R$ are all $SU(2)$ elements, their trace over a spin$-j$ representation is nothing but the character $\chi_j(\phi)$ of $SU(2)$ for some $\phi$, given by,
\begin{equation}\label{App-b-i}
    \chi_j(\phi) = \frac{\sin((2j+1)\phi)}{\sin\phi}.
\end{equation}
And the $\phi$ can be solved by matching the $j = \frac{1}{2}$ result known from the above representation. We find,
\begin{equation}\label{App-b-j}
\begin{aligned}
    & Tr_{V_j} (Q_L) = \frac{\sin((2j+1)\pi/3)}{\sin(\pi/3)}, \quad Tr_{V_j}(Q_L C_L) = \frac{\sin((2j+1)2\pi/3)}{\sin(2\pi/3)}, \\ & Tr_{\oV_{\oj}} (Q_R) = \frac{\sin((2\oj+1)\pi/3)}{\sin(\pi/3)}, \quad Tr_{\oV_{\oj}} (Q_R C_R) = \frac{\sin((2\oj+1)2\pi/3)}{\sin(2\pi/3)}.
\end{aligned}
\end{equation}
Next, we list the group theory result for $\overline{Q} = Q^2$. The decomposition is given by,
\begin{equation}\label{App-b-k}
    \oQ_L = \left(
\begin{array}{cc}
 -\frac{1}{2}-\frac{\ii}{2} & -\frac{1}{2}+\frac{\ii}{2} \\
 \frac{1}{2}+\frac{\ii}{2} & -\frac{1}{2}+\frac{\ii}{2} \\
\end{array}
\right), \quad \oQ_R = \left(
\begin{array}{cc}
 -\frac{1}{2}-\frac{\ii}{2} & \frac{1}{2}-\frac{\ii}{2} \\
 -\frac{1}{2}-\frac{\ii}{2} & -\frac{1}{2}+\frac{\ii}{2} \\
\end{array}
\right).
\end{equation}
We then have,
\begin{equation}\label{App-b-l}
\begin{aligned}
    & Tr_{V_j}(\oQ_L) = Tr_{V_j}(\oQ_L C_L) = \frac{\sin((2j+1)2\pi/3)}{\sin(2\pi/3)}, \\ & Tr_{\oV_{\oj}}(\oQ_R) = Tr_{\oV_{\oj}}(\oQ_R C_R) = \frac{\sin((2\overline{j}+1)2\pi/3)}{\sin(2\pi/3)}.
\end{aligned} 
\end{equation}
We will also need to construct the representation matrices of the group $SO(4)$ on the $(j,\oj)$ irrep as well, which can be done via the tensor product of the representation matrices of the group $SU(2)_L$ and $SU(2)_R$. The spin-$j$ representation matrices of the $SU(2)$ can be acquired by exponentiate the generators $S_{(j)}^i$ where $j$ labels the spin and $i = 1, \cdots, 3$. The $S_{(j)}^i$ is constructed from,
\begin{equation}\label{App-b-m}
    [S_{(j)}^+]_{mn} = \delta_{m+1,n} \sqrt{j(j+1) - (j + 1 - m)(j - m)},
\end{equation}
and,
\begin{equation}\label{App-b-n}
    S_{(j)}^1 = \frac{S_{(j)}^+ + (S_{(j)}^+)^\dagger}{2i}, \quad S_{(j)}^2 = -\frac{S_{(j)}^+ - (S_{(j)}^+)^\dagger}{2}, \quad S^3_{(j)} = [S_{(j)}^1, S_{(j)}^2].
\end{equation}
As one can check,
\begin{equation}\label{App-b-o}
    [S^k_{(j)}, S^l_{(j)}] = \epsilon^{klm} S^m_{(j)}.
\end{equation}
\section{Detail on the basis in $\Hom_A(M\otimes_A N,L)$}
In this section, we list our choices of basis in $\Hom_A(M\otimes_A N,L)$. Notice that if $\phi \in \Hom_A(M\otimes_A N,L)$, then,
\begin{equation}\label{eq:Hom_A_Condition}
    a_1\phi(m\otimes_A n)a_2 = \phi((a_1 m)\otimes_A (n a_2)), \quad \forall a_i \in A, m\in M, n\in N.
\end{equation}
Let's consider the example $H_A(M_{\eta}^-\otimes_A M_{\oQ}, M_{\oQ})$. If we choose, 
\begin{equation}
    \phi_{M_\eta^-\otimes_A M_{\oQ} \rightarrow M_{\oQ}}(m_\eta^-\otimes_A m_{(134)}) = m_{(243)},
\end{equation}
then acting $\sigma$ on left or right or both side on $\phi_{M_\eta^-\otimes_A M_{\oQ} \rightarrow M_{\oQ}}(m_\eta^-\otimes_A m_{(134)})$ allows us to determine,
\begin{equation}
    \phi_{M_\eta^-\otimes_A M_{\oQ} \rightarrow M_{\oQ}}: \begin{pmatrix} m_\eta^- \otimes_A m_{(134)} \\ m_\eta^- \otimes_A m_{(123)} \\ m_\eta^- \otimes_A m_{(142)}  \\ m_\eta^- \otimes_A m_{(243)} \end{pmatrix} \mapsto \begin{pmatrix} m_{(243)} \\  m_{(142)} \\ - m_{(123)} \\ - m_{(134)}\end{pmatrix}. 
\end{equation}
Hence, we will only list the action of $\phi \in \Hom_A(M\otimes_A N,L)$ on a single element which allows one to determine its action on the rest of the elements as follows,
\begin{equation}
\begin{aligned}
    & \phi_{M^+_I \otimes_A M_{HgH}^\rho \rightarrow M_{HgH}^\rho}: m_1^+ \otimes_A m_g^\rho \mapsto m_g^\rho, \quad \phi_{M_{HgH}^{\rho} \otimes_A M_1^+ \rightarrow M_{HgH}^{\rho}}: m_g^\rho\otimes_A m_1^+ \mapsto m_g^\rho, \\
    & \phi_{M_I^-\otimes_A M_J^+ \rightarrow M_J^-}: m_1^-\otimes_A m_\mu^+ \mapsto m_\mu^-, \quad 
    \phi_{M_I^-\otimes_A M_J^- \rightarrow M_J^+}: m_1^-\otimes_A m_\mu^- \mapsto m_\mu^+, \\
    & \phi_{M_J^+\otimes_A M_J^+\rightarrow M_I^+}: m_\eta^+\otimes_A m_\eta^+ \mapsto m_1^+, \quad \phi_{M_J^-\otimes_A M_J^+\rightarrow M_I^-}: m_\eta^-\otimes_A m_\eta^+ \mapsto m_1^-, \\
    & \phi_{M_J^+\otimes_A M_J^-\rightarrow M_I^-}: m_\eta^+\otimes_A m_\eta^- \mapsto m_1^-, \quad \phi_{M_J^-\otimes_A M_J^-\rightarrow M_I^-}: m_\eta^-\otimes_A m_\eta^- \mapsto m_1^+, \\
    & \phi_{M_Q\otimes_A M_I^- \rightarrow M_Q}: m_{(143)}\otimes_A m_1^- \mapsto m_{(143)}, \quad \phi_{M_I^-\otimes_A M_Q \rightarrow M_Q}: m_1^- \otimes_A m_{(143)} \mapsto m_{(143)}, \\
    & \phi_{M_Q\otimes_A M_J^+ \rightarrow M_Q}: m_{(143)}\otimes_A m_\eta^+ \mapsto m_{(234)}, \quad \phi_{M_J^+\otimes_A M_Q \rightarrow M_Q}: m_\eta^+\otimes_A m_{(143)}\mapsto m_{(124)}, \\
    & \phi_{M_Q\otimes_A M_J^- \rightarrow M_Q}:m_{(143)}\otimes_A m_\eta^- \mapsto m_{(234)}, \quad \phi_{M_J^-\otimes_A M_Q \rightarrow M_Q}:m_\eta^- \otimes_A m_{(143)}\mapsto m_{(124)}, \\
    & \phi_{M_{\oQ}\otimes_A M_I^- \rightarrow M_{\oQ}}: m_{(134)}\otimes_A m_1^- \mapsto m_{(134)}, \quad \phi_{M_I^-\otimes_A M_{\oQ} \rightarrow M_{\oQ}}: m_1^-\otimes_A m_{(134)} \mapsto m_{(134)}, \\
    & \phi_{M_{\oQ}\otimes_A M_J^+ \rightarrow M_{\oQ}}: m_{(134)}\otimes_A m_\eta^+ \mapsto m_{(142)}, \quad \phi_{M_J^+\otimes_A M_{\oQ} \rightarrow M_{\oQ}}:m_{\eta}^+\otimes_A m_{(134)} \mapsto m_{(243)}, \\
    & \phi_{M_{\oQ}\otimes_A M_J^- \rightarrow M_{\oQ}}:m_{(134)}\otimes_A m_\eta^- \mapsto m_{(142)}, \quad \phi_{M_J^-\otimes_A M_{\oQ} \rightarrow M_{\oQ}}: m_\eta^- \otimes_A m_{(134)} \mapsto m_{(243)}, \\
\end{aligned}
\end{equation}
and,
\begin{equation}
\begin{aligned}
    & \phi_{M_Q\otimes_A M_Q\rightarrow M_{\oQ},1}: m_{(143)}\otimes_A m_{(143)} \mapsto m_{(134)}, \quad \phi_{M_Q\otimes_A M_Q\rightarrow M_{\oQ},2}: m_{(143)}\otimes_A m_{(234)} \mapsto m_{(142)}, \\
    & \phi_{M_{\oQ}\otimes_A M_{\oQ}\rightarrow M_Q,1}: m_{(134)}\otimes_A m_{(134)} \mapsto m_{(143)}, \quad \phi_{M_{\oQ}\otimes_A M_{\oQ}\rightarrow M_Q,2}: m_{(134)}\otimes_A m_{(142)} \mapsto m_{(234)}, \\
    & \phi_{M_Q\otimes_A M_{\oQ}\rightarrow M_I^+}: m_{(132)}\otimes_A m_{(123)} \mapsto \frac{1}{\sqrt{2}}m_1^+, \quad \phi_{M_Q\otimes_A M_{\oQ}\rightarrow M_I^-}: m_{(132)}\otimes_A m_{(123)} \mapsto \frac{1}{\sqrt{2}}m_1^-, \\
    & \phi_{M_Q\otimes_A M_{\oQ}\rightarrow M_J^+}: m_{(132)}\otimes_A m_{(243)} \mapsto \frac{1}{\sqrt{2}}m_\eta^+, \quad \phi_{M_Q\otimes_A M_{\oQ}\rightarrow M_J^-}: m_{(132)}\otimes_A m_{(243)} \mapsto \frac{1}{\sqrt{2}} m_\eta^-, \\
    & \phi_{M_{\oQ}\otimes_A M_{Q}\rightarrow M_I^+}: m_{(134)}\otimes_A m_{(143)} \mapsto \frac{1}{\sqrt{2}}m_1^+, \quad \phi_{M_{\oQ}\otimes_A M_{Q}\rightarrow M_I^-}: m_{(134)}\otimes_A m_{(143)} \mapsto \frac{1}{\sqrt{2}}m_1^-, \\
    & \phi_{M_{\oQ}\otimes_A M_{Q}\rightarrow M_J^+}: m_{(142)}\otimes m_{(132)} \mapsto \frac{1}{\sqrt{2}} m_\eta^+, \quad \phi_{M_{\oQ}\otimes_A M_{Q}\rightarrow M_J^-}: m_{(142)}\otimes m_{(132)} \mapsto \frac{1}{\sqrt{2}} m_\eta^-.
\end{aligned}
\end{equation}
The $\phi(m\otimes_A n)$'s which can not be determined using \eqref{eq:Hom_A_Condition} are set to zero.

\bibliographystyle{utphys}
\bibliography{orbifold}

\end{document}